\documentclass{cernyrep} 
\pdfoutput=1
\usepackage{texnames}
\usepackage[T1]{fontenc}
\usepackage[bookmarks, colorlinks=true, linktoc=page, pdftex, linkcolor=red, citecolor=red, urlcolor=red]{hyperref}
\usepackage{cite}
\usepackage{lineno}
\usepackage{import}
\usepackage{subcaption}
\begin{document}
\title{Heavy ions at the Future Circular Collider}
 
\author{
A.~Dainese\,$^{1}$,\,
U.A.~Wiedemann\,$^{2}$\,
(editors),\,
N.~Armesto\,$^{3}$,\, 
D.~d'Enterria\,$^{2}$,\, 
J.M.~Jowett\,$^{2}$,\, 
J.-P.~Lansberg\,$^{4}$,\, 
J.G.~Milhano\,$^{5,2}$,\, 
C.A.~Salgado\,$^{3}$,\, 
M.~Schaumann\,$^{2}$,\, 
M.~van~Leeuwen\,$^{6,7}$\, 
(section editors),\,
J.L.~Albacete\,$^{8}$,\, 
A.~Andronic\,$^{9}$,\, 
P.~Antonioli\,$^{10}$,\, 
L.~Apolin\'ario\,$^{5}$,\, 
S.~Bass\,$^{11}$,\, 
A.~Beraudo\,$^{12}$,\, 
A.~Bilandzic\,$^{13}$,\, 
S.~Borsanyi\,$^{14}$,\, 
P.~Braun-Munzinger\,$^{9}$,\, 
Z.~Chen\,$^{15}$,\, 
L.~Cunqueiro Mendez\,$^{16}$,\, 
G.S.~Denicol\,$^{17}$,\, 
K.J.~Eskola\,$^{18}$,\, 
S.~Floerchinger\,$^{19}$,\, 
H.~Fujii\,$^{20}$,\, 
P.~Giubellino\,$^{12}$,\, 
C.~Greiner\,$^{21}$,\, 
J.F.~Grosse-Oetringhaus\,$^{2}$,\, 
C.-M.~Ko\,$^{22}$,\, 
P.~Kotko\,$^{23}$,\, 
K.~Krajcz\'ar\,$^{2,24}$,\, 
K.~Kutak\,$^{25}$,\, 
M.~Laine\,$^{26}$,\, 
Y.~Liu\,$^{27}$,\, 
M.P.~Lombardo\,$^{28}$,\, 
M.~Luzum\,$^{29,3}$,\, 
C.~Marquet\,$^{30}$,\, 
S.~Masciocchi\,$^{9}$,\, 
V.~Okorokov\,$^{31}$,\, 
J.-F.~Paquet\,$^{32,33}$,\, 
H.~Paukkunen\,$^{3,18,34}$,\, 
E.~Petreska\,$^{30,3}$,\, 
T.~Pierog\,$^{35}$,\, 
M.~Ploskon\,$^{36}$,\, 
C.~Ratti\,$^{37}$,\, 
A.H.~Rezaeian\,$^{38}$,\, 
W.~Riegler\,$^{2}$,\, 
J.~Rojo\,$^{39}$,\, 
C.~Roland\,$^{24}$,\, 
A.~Rossi\,$^{40,1}$,\,  
G.P.~Salam\,$^{2}$,\, 
S.~Sapeta\,$^{25,2}$,\, 
R.~Schicker\,$^{19}$,\, 
C.~Schmidt\,$^{41}$,\, 
J.~Stachel\,$^{19}$,\, 
J.~Uphoff\,$^{21}$,\, 
A.~van~Hameren\,$^{25}$,\, 
K.~Watanabe\,$^{42}$,\, 
B.-W.~Xiao\,$^{42}$,\, 
F.~Yuan\,$^{36}$,\, 
D.~Zaslavsky\,$^{42}$,\, 
K.~Zhou\,$^{21,15}$,\, 
P.~Zhuang\,$^{15}$  
\vspace*{2cm}
}

\maketitle 

\begin{abstract}
The Future Circular Collider (FCC) Study is aimed at assessing the physics potential and the technical feasibility
of a new collider
with centre-of-mass energies, in the hadron--hadron collision mode,
seven times larger than the nominal LHC energies.
Operating such machine with heavy ions is an option that is being
considered in the accelerator design studies. 
It would provide, for example, Pb--Pb and p--Pb collisions at
$\sqrt{s_{\mathrm{\scriptscriptstyle NN}}}=39$ and 63~TeV, respectively, per nucleon--nucleon
collision, with integrated luminosities above 30~nb$^{-1}$ per month for Pb--Pb.
This is a report by the working group on heavy-ion physics of the FCC Study. 
First ideas on the physics opportunities with heavy ions at the FCC are presented, 
 covering the physics of the Quark--Gluon Plasma, of gluon saturation,
 of photon-induced collisions, as well as connections with other fields of high-energy physics.
\end{abstract}

\newpage

\noindent
$^{1}$ INFN - Sezione di Padova, Italy \\
$^{2}$ CERN, Geneva, Switzerland \\
$^{3}$ Instituto Galego de F\'{\i}sica de Altas Enerx\'{\i}as, Universidade de Santiago de Compostela, Galicia-Spain \\
$^{4}$ IPNO, Univ. Paris-Sud, CNRS/IN2P3, Universit\'e Paris-Saclay, Orsay, France\\
$^{5}$ CENTRA, Instituto Superior T\'ecnico, Universidade de Lisboa, Portugal \\
$^{6}$ Utrecht University, Netherlands \\
$^{7}$ NIKHEF, Netherlands \\
$^{8}$ Granada University, Spain \\
$^{9}$ GSI, Darmstadt, Germany \\
$^{10}$ INFN - Sezione di Bologna, Italy \\
$^{11}$ Duke University, Durham, USA \\
$^{12}$ INFN - Sezione di Torino, Italy \\
$^{13}$ Niels Bohr Institute, University of Copenhagen, Denmark \\
$^{14}$ Wuppertal University, Germany \\
$^{15}$ Tsinghua University, Beijing, China \\
$^{16}$ Munster University, Germany \\
$^{17}$ Brookhaven National Lab, Upton, USA \\
$^{18}$ University of Jyv\"askyl\"a, Finland\\
$^{19}$ University of Heidelberg, Germany \\
$^{20}$ Tokyo University, Japan \\
$^{21}$ Frankfurt University, Germany \\
$^{22}$ Texas A\&M University, College Station, USA \\
$^{23}$ Pennsylvania State University, University Park, USA \\
$^{24}$ Massachusetts Institute of Technology , Cambridge, USA \\
$^{25}$ Institute of Nuclear Physics PAN, Krakow, Poland \\
$^{26}$ AEC, ITP, University of Bern, Switzerland \\
$^{27}$ Tianjin University, China \\
$^{28}$ INFN - Laboratori Nazionali di Frascati, Italy \\
$^{29}$ Universidade de Sao Paulo, Brasil \\
$^{30}$ Ecole Polytechnique, CNRS, Universit\'e Paris-Saclay, Palaiseau, France \\
$^{31}$ National Research Nuclear University MEPhI, Moscow, Russia\\
$^{32}$ McGill University, Montr\'eal, Canada \\
$^{33}$ Stony Brook University, Stony Brook, USA  \\
$^{34}$ Helsinki Institute of Physics, University of Helsinki, Finland\\
$^{35}$ KIT, IKP, Karlsruhe, Germany \\
$^{36}$ Lawrence Berkeley National Lab, Berkeley, USA \\
$^{37}$ University of Houston, USA \\
$^{38}$ Universidad Tecnica Federico Santa Maria, Valparaiso, Chile \\
$^{39}$ University of Oxford, UK \\
$^{40}$ University of Padova, Italy \\
$^{41}$ Bielefeld University, Germany \\
$^{42}$ Central China Normal University, Wuhan, China \\

\tableofcontents
 
\newpage 
 
\newcommand{\met}{E\!\!\!\!/_T} 
\def\iab{ab$^{-1}$}
\def\ifb{fb$^{-1}$}
\def\ipb{pb$^{-1}$}
\def \gsim{\mathrel{\vcenter
     {\hbox{$>$}\nointerlineskip\hbox{$\sim$}}}}
\def \lsim{\mathrel{\vcenter
     {\hbox{$<$}\nointerlineskip\hbox{$\sim$}}}}

\newcommand{\qty}[2]{\ensuremath{#1~\mathrm{#2}}}
\newcommand{\Eb}{\ensuremath{E_b}}
\newcommand{\etev}[1]{\ensuremath{\Eb=\qty{#1}{TeV}}}

\newcommand{\pp}{\ensuremath{\rm pp}\xspace}
\newcommand{\ppbar}{\ensuremath{\rm p\overline{p}}\xspace}
\newcommand{\pa}{{p--A}\xspace}
\newcommand{\pA}{{p--A}\xspace}
\newcommand{\pPb}{{p--Pb}\xspace}
\newcommand{\dAu}{{d--Au}\xspace}
\newcommand{\pb}{{Pb--Pb}\xspace}
\newcommand{\AuAu}{{Au--Au}\xspace}
\newcommand{\CuCu}{{Cu--Cu}\xspace}
\newcommand{\CuAu}{{Cu--Au}\xspace}
\newcommand{\UU}{{U--U}\xspace}
\newcommand{\AAcoll}{{AA}\xspace}
\newcommand{\PbPb}{{Pb--Pb}\xspace}

\newcommand{\snn}{\ensuremath{\sqrt{s_{\mathrm{\scriptscriptstyle NN}}}}\xspace}
\newcommand{\sqrtsNN}{\ensuremath{\sqrt{s_{\mathrm{\scriptscriptstyle NN}}}}\xspace}
\newcommand{\s}{\ensuremath{\sqrt{s}}\xspace}

\newcommand{\jpsi}{\ensuremath{J/\psi}\xspace}
\newcommand{\chic}{\ensuremath{\chi_c}\xspace}
\newcommand{\chib}{\ensuremath{\chi_b}\xspace}
\newcommand{\psiP}{\ensuremath{\psi\text{(2S)}}\xspace}
\newcommand{\doubleRatioPsi}{\ensuremath{\left.(N_{\psiP}/N_{\jpsi})_{\mathrm{Pb-Pb}}/(N_{\psiP}/N_{\jpsi})_{\mathrm{pp}}\right.\xspace}}
\newcommand{\ups}{\ensuremath{\Upsilon}\xspace}
\newcommand{\upsa}{\ensuremath{\Upsilon\text{(1S)}}\xspace}
\newcommand{\upsb}{\ensuremath{\Upsilon\text{(2S)}}\xspace}
\newcommand{\upsc}{\ensuremath{\Upsilon\text{(3S)}}\xspace}
\newcommand{\upsbc}{\ensuremath{\Upsilon\text{(2S+3S)}}\xspace}
\newcommand{\upsabc}{\ensuremath{\Upsilon\text{(1S+2S+3S)}}\xspace}
\newcommand{\upsn}{\ensuremath{\Upsilon\text{(nS)}}\xspace}
\newcommand{\doubleRatioUps}{\ensuremath{\left.(N_{\upsb}/N_{\upsa})_{\mathrm{Pb-Pb}}/(N_{\upsb}/N_{\upsa})_{\mathrm{pp}}\right.\xspace}}

\newcommand{\qqbar}{\ensuremath{{q\overline{q}}}\xspace}
\newcommand{\QQbar}{\ensuremath{{Q\overline{Q}}}\xspace}
\newcommand{\Qcal}{\ensuremath{\Phi\xspace}}
\newcommand{\ccbar}{\ensuremath{{c\overline{c}}}\xspace}
\newcommand{\bbbar}{\ensuremath{{b\overline{b}}}\xspace}

\newcommand{\Dmeson}[1]{\ensuremath{\mathrm{D}^{#1}}\xspace}
\newcommand{\Bmeson}[1]{\ensuremath{\mathrm{B}^{#1}}\xspace}
\newcommand{\Dzero}{\Dmeson{0}}
\newcommand{\Dplus}{\Dmeson{+}}
\newcommand{\Dstarplus}{\Dmeson{*+}}
\newcommand{\Ds}{\ensuremath{\mathrm{D}^{+}_{s}}\xspace}
\newcommand{\kaon}{\ensuremath{\mathrm{K}}\xspace}
\newcommand{\hfe}{\ensuremath{HF \to e^{\pm}}\xspace}
\newcommand{\hfm}{\ensuremath{HF \to \mu^{\pm}}\xspace}

\newcommand{\lambdac}{\ensuremath{\Lambda_c}\xspace}
\newcommand{\etac}{\ensuremath{\eta_c}\xspace}
\newcommand{\lambdab}{\ensuremath{\Lambda_b}\xspace}
\newcommand{\etab}{\ensuremath{\eta_b}\xspace}
\newcommand{\lambdacplus}{\ensuremath{\Lambda_c^+}\xspace}
\newcommand{\etacplus}{\ensuremath{\eta_c^+}\xspace}
\newcommand{\chicplus}{\ensuremath{\Xi_c^+}\xspace}
\newcommand{\sigmacplus}{\ensuremath{\Sigma_c^+}\xspace}

\newcommand{\lambdabplus}{\ensuremath{\Lambda_b^+}\xspace}
\newcommand{\etabplus}{\ensuremath{\eta_b^+}\xspace}
\newcommand{\proton}{\mathrm{p}\xspace}

\newcommand{\ee}{\ensuremath{e^+e^-}\xspace}
\newcommand{\mumu}{\ensuremath{\mu^+\mu^-}\xspace}

\newcommand{\eminus}{\ensuremath{e^{-}}\xspace}
\newcommand{\eplus}{\ensuremath{e^{+}}\xspace}
\newcommand{\electron}{\ensuremath{e}\xspace}
\newcommand{\muminus}{\ensuremath{\mu^{-}}\xspace}
\newcommand{\muplus}{\ensuremath{\mu^{+}}\xspace}
\newcommand{\cquark}{\ensuremath{c}\xspace}
\newcommand{\bquark}{\ensuremath{b}\xspace}
\newcommand{\uquark}{\ensuremath{u}\xspace}
\newcommand{\dquark}{\ensuremath{d}\xspace}
\newcommand{\squark}{\ensuremath{s}\xspace}
\newcommand{\gluon}{\ensuremath{g}\xspace}

\newcommand{\raa}{\ensuremath{R_{\mathrm{AA}}}\xspace}
\newcommand{\RAA}{\ensuremath{R_{\mathrm{AA}}}\xspace}
\newcommand{\rpa}{\ensuremath{R_{\mathrm{pA}}}\xspace}
\newcommand{\spa}{\ensuremath{S_{\mathrm{pA}}}\xspace}
\newcommand{\rppb}{\ensuremath{R_{\mathrm{pPb}}}\xspace}
\newcommand{\rdau}{\ensuremath{R_{\mathrm{dAu}}}\xspace}
\newcommand{\rcp}{\ensuremath{R_{\mathrm{CP}}}\xspace}
\newcommand{\rfb}{\ensuremath{R_{\mathrm{FB}}}\xspace}

\newcommand{\MeV}{\ensuremath{\text{~MeV}}\xspace}
\newcommand{\GeV}{\ensuremath{\text{~GeV}}\xspace}
\newcommand{\TeV}{\ensuremath{\text{~TeV}}\xspace}
\newcommand{\mev}{\ensuremath{\text{MeV}}}
\newcommand{\gev}{\ensuremath{\text{GeV}}}
\newcommand{\tev}{\ensuremath{\text{TeV}}}
\newcommand{\MeVc}{\ensuremath{\text{~MeV}/c}\xspace}
\newcommand{\GeVc}{\ensuremath{\text{~GeV}/c}\xspace}
\newcommand{\GeVcc}{\ensuremath{\text{~GeV}/c^2}\xspace}
\newcommand{\lumi}{\ensuremath{\mathcal{L}}\xspace}
\newcommand{\mbinv}{\ensuremath{\text{~mb}^\text{$-$1}}\xspace}
\newcommand{\mubinv}{\ensuremath{~\mu\text{b}^\text{$-$1}}\xspace}
\newcommand{\nbinv}{\ensuremath{\text{~nb}^\text{$-$1}}\xspace}
\newcommand{\Lunits}{\rm cm^{-2}s^{-1}}


\newcommand{\dd}{\ensuremath{\mathrm{d}}}
\newcommand{\la}{\langle}
\newcommand{\ra}{\rangle}
\newcommand{\av}[1]{\ensuremath{\left\langle #1 \right\rangle}}
\newcommand{\bk}{\vec{k}}
\newcommand{\bq}{\vec{q}} 
\newcommand{\sabs}{\ensuremath{\sigma_\mathrm{abs}}\xspace}
\newcommand{\muf}{\ensuremath{\mu_F}\xspace}
\newcommand{\mur}{\ensuremath{\mu_R}\xspace}

\newcommand{\beq}{\begin{equation}}
\newcommand{\eeq}{\end{equation}}
\newcommand{\beqn}{\begin{eqnarray}}
\newcommand{\eeqn}{\end{eqnarray}}
\newcommand{\noi}{\noindent}
\newcommand{\beqa}{\begin{eqnarray}}
\newcommand{\eeqa}{\end{eqnarray}}
\def\lsim{\raise0.3ex\hbox{$<$\kern-0.75em\raise-1.1ex\hbox{$\sim$}}}
\def\gsim{\raise0.3ex\hbox{$>$\kern-0.75em\raise-1.1ex\hbox{$\sim$}}}

\newcommand{\Npart}{\ensuremath{\mathrm{N_{part}}}\xspace}
\newcommand{\Ncoll}{\ensuremath{\mathrm{N_{coll}}}\xspace}
\newcommand{\taa}{\ensuremath{T_{\mathrm{AA}}}\xspace}
\newcommand{\dEdx}{\ensuremath{{\mathrm{d}}E/{\mathrm{d}}x}\xspace}
\newcommand{\vtwo}{\ensuremath{v_{2}}\xspace}
\newcommand{\pt}{\ensuremath{p_{\mathrm{T}}}\xspace}
\newcommand{\mt}{\ensuremath{m_{\mathrm{T}}}\xspace}
\newcommand{\kt}{\ensuremath{k_{\mathrm{T}}}\xspace}
\newcommand{\invYield}[1]{\ensuremath{\frac{\dd^2 N^{#1}}{2 \pi \pt \dd \pt \dd y}}}
\newcommand{\invYieldEta}[1]{\ensuremath{\frac{\dd^2 N^{#1}}{2 \pi \pt \dd \pt \dd \eta}}}
\newcommand{\invYieldEtaN}[1]{\ensuremath{\frac{\dd^2 N^{#1}}{\pt \dd \pt \dd \eta}}}
\newcommand{\iaa}{\ensuremath{I_{\mathrm{AA}}}\xspace}
\newcommand{\ylab}{\ensuremath{y_{\mathrm{lab}}}\xspace}
\newcommand{\ycm}{\ensuremath{y}\xspace}
\newcommand{\xf}{\ensuremath{x_{\mathrm{F}}}\xspace}
\newcommand{\dsdpt}{\ensuremath{{\rm d}\sigma/{\rm d}p_{\mathrm{T}}}\xspace}
\newcommand{\dsdy}{\ensuremath{{\rm d}\sigma/{\rm d}y}\xspace}
\newcommand{\meanpt}{\ensuremath{\langle p_{\mathrm{T}} \rangle}\xspace}
\newcommand{\tc}{\ensuremath{\tau_{\rm c}}\xspace}
\newcommand{\tf}{\ensuremath{\tau_{\rm f}}\xspace}
\newcommand{\tcross}{\ensuremath{\tau_{\rm cross}}\xspace}

\newcommand{\sqrts}{\sqrt{s}}
\newcommand{\sqrtsnn}{\sqrtsNN}

\newcommand{\alphaS}{\alpha_{\rm s}}
\newcommand\cO{{\cal O}}
\newcommand\cN{{\cal N}}
\newcommand\bR{{\cal B}}

\newcommand{\Sig}{\mathcal{S}}

\newcommand{\LumiInt}{\mathcal{L}_{\mbox{\rm \tiny{int}}}}
\newcommand{\pn}{{\rm{p-n}}}
\newcommand{\pN}{{\rm{p-N}}}
\newcommand{\TpA}{T_{_{\rm pA}}}
\newcommand{\TpPb}{T_{_{\rm pPb}}}

\newcommand{\pythia}{{\sc pythia}}
\newcommand{\mcfm}{{\sc mcfm}}
\newcommand{\ttbar}    {t\bar{t}}
\newcommand{\ttbars}    {\ensuremath{t,\bar{t}}}
\newcommand{\MET}{\ensuremath{{E\!\!\!/}_{_{\rm T}}}}

\newcommand{\CHECK}[1]{\textbf{\color{red}#1}}

\def\ttt#1{\texttt{\scriptsize #1}}

\newcommand{\fig}[1]{\figurename~\ref{#1}}
\newcommand{\tab}[1]{\tablename~\ref{#1}}
\newcommand{\sect}[1]{Section~\ref{#1}}
\newcommand{\eq}[1]{Equation~\ref{#1}}
\newcommand{\ci}[1]{Ref.~\cite{#1}}
\newcommand{\cis}[1]{Refs.~\cite{#1}}

\newcommand{\figs}{Figures\xspace}
\newcommand{\tabs}{Tables\xspace}
\newcommand{\sects}{Sections\xspace}
\newcommand{\eqs}{Equations\xspace}


\def\mean#1{\ensuremath{\left<#1\right>}}
\def\ttt#1{\texttt{\scriptsize #1}}

\providecommand{\epem}{e^{+}e^{-}}
\providecommand{\pp}{pp}
\providecommand{\pA}{p--A}
\providecommand{\pPb}{p--Pb}
\providecommand{\AaAa}{A--A}
\providecommand{\PbPb}{Pb--Pb}
\providecommand{\gaga}{\gamma\,\gamma}
\providecommand{\gp}{\gamma\,p}
\providecommand{\gA}{\gamma\,A}

\providecommand{\mgg}{\rm m_{\gamma\gamma}}
\providecommand{\sigmagg}{\sigma_{\gaga\to\gaga}}

\newcommand{\sqrtsgp}{\sqrt{s_{_{\gamma p}}}}
\newcommand{\sgg}{s_{_{\gamma\,\gamma }}}
\newcommand{\sqrtsgg}{\sqrt{s_{_{\gamma\,\gamma }}}}

\newcommand{\ABgaga}{A\,B\,$\xrightarrow{\gaga}$ A\,$\gaga$\,B}
\newcommand{\ABglgl}{A\,B\,$\xrightarrow{g\,g}$ A\,$\gaga$\,B}
\newcommand{\pPbgaga}{$p \,$Pb$\,\xrightarrow{\gaga} p \,\gaga \,$Pb}
\newcommand{\pPbgagaX}{$p \,$Pb$\,\xrightarrow{\gaga} X \,\gaga \,$Pb}
\newcommand{\pAgaga}{$p \,A\,\xrightarrow{\gaga} p \,\gaga \,A$}
\newcommand{\pAgagaX}{$p \,A\,\xrightarrow{\gaga} X \,\gaga \,A$}

\newcommand{\ABgagaX}{$A \,B\,\xrightarrow{\gaga} X \,\gaga \,B$}

\providecommand{\madgraph}{{\sc MadGraph}}
\providecommand{\superchic}{{\sc SuperChic}}

\newcommand{\Pom} {I\!P} 

\newcommand*{\cm}{c.m.\@\xspace}

\renewcommand\arraystretch{1.2}



\newpage 
\section{Executive summary}
\label{sec:HI_intro}

A five-year international design study called Future Circular Collider (FCC) has been launched by CERN in February 2014~\cite{FCCweb,FCCkickoff}. 
The main goal is to assess the feasibility and physics potential of a hadron collider with a centre-of-mass energy $\sqrt s$ of 100~TeV for pp collisions in a new 80--100~km tunnel near Geneva. 
The starting date is targeted for 2035--40.
Operating such machine with heavy ions is part of the accelerator design studies. 

For a centre-of-mass energy $\sqrt{s}= 100$~TeV for pp collisions, the relation $\sqrtsNN= \sqrt {s} \sqrt{Z_1 Z_2 / A_1 A_2}$ 
gives the energy per nucleon--nucleon collision of $\sqrtsNN = 39$~TeV for Pb--Pb ($Z=82$, $A=208$) and 63~TeV for p--Pb collisions. 
The present estimate of the integrated luminosity for Pb--Pb collisions results in about 33~nb$^{-1}$ per month of running, which is more than an order of magnitude larger than
the current projection for the future LHC runs~\cite{rliup,LHCHIRun3}.

The increase in the centre-of-mass energy and integrated luminosity
with respect to the LHC opens new opportunities for physics with heavy
ions. 
This report summarises the projected machine performance and the physics opportunities for a nuclear beam programme at the FCC. 
We point out the existence of an ongoing design study by the Chinese community for a machine similar to the FCC but with 
smaller circumference and centre-of-mass energy~\cite{cepc}. The hadronic machine is called SppC and the centre-of-mass energy for Pb--Pb collisions would be $\sqrtsNN\sim 20$--$30$~TeV. A report on heavy-ion studies at SppC was recently published and includes several projections and ideas on high-energy nuclear physics in the multi-TeV domain~\cite{SppCHI}.

At the time of writing this report, the physics community still looks ahead to more than one decade of experimentation with nuclear beams at the LHC. 
Our understanding of most of the measurements discussed in this report is likely
to evolve significantly in the coming years in the light of future LHC data and further advances in theory. In this sense, many of the basic motivations for a heavy-ion programme
at the FCC are the basic motivations for continuing the heavy-ion programme at the LHC or they arise naturally from it. 

There is by now ample historical 
evidence that an order of magnitude increase in energy or luminosity of heavy-ion collisions advances significantly our understanding of the nature
of the hot and dense QCD matter produced in these collisions, denoted Quark--Gluon Plasma (QGP), and that it can lead to unexpected discoveries. While unexpected discoveries, by their very nature, cannot be anticipated in a
working group report (despite being one major motivation for exploring a previously-uncharted energy range with nuclear beams), we focus here mainly on those 
fundamental questions about the nature of QCD matter at high temperature and density for which we expect qualitative
advances from the FCC. Our study is not exhaustive, but it aims at supporting with a selected set of arguments and proposed measurements the following main
motivations for a heavy-ion programme at the FCC:
\begin{enumerate}
	\item {\bf FCC provides novel access to QCD thermodynamics and QCD equilibration processes}\\
	Substantially increasing the centre-of-mass energy leads to the creation of initially denser and hotter systems that expand for a longer duration and over a larger volume, thereby
	developing stronger collective phenomena. Beyond expected quantitative gains, this may bring novel qualitative phenomena into experimental reach. For instance, FCC energies target an 
	interesting transition region in energy density above which charm quarks start counting towards the thermal degrees of freedom, thus playing a novel role in QCD equilibration processes. 
         Also, the $\sqrt{s}$-dependent increase in event multiplicity combined with sufficient integrated luminosity will allow for the systematic study of flow-like features in smaller collision systems 
         (including pp and pA collisions), and it will facilitate the characterisation of important signatures of collectivity 
on the level of single events rather than event samples only. This opens novel opportunities for understanding the equilibration processes that lead to
         hydrodynamization and thermalization in the non-abelian quantum field theory QCD. These opportunities are discussed in Section~\ref{sec:HI_qgpsoft}.\\
	\item {\bf FCC allows for an unprecedented characterisation of dense QCD matter with hard processes}\\
	In heavy-ion collisions,  hadronic high-transverse-momentum ($\pt$) processes are known to show strong medium-induced modifications, often referred to as jet quenching, up to the highest transverse momenta 
	$O(100~\gev)$ explored at the LHC so far. These jet quenching measurements characterize transport properties of the dense QCD matter through which the hard partons
	propagate and they allow one to follow experimentally how a probe that is initially far out-of-equilibrium evolves towards equilibrium. 
As detailed in Section~\ref{sec:HI_hardprobes}, the increase in energy and integrated luminosity at FCC will provide 
much larger abundance of hard processes than at LHC, as well as the access to qualitatively-novel hard probes that are measurable at FCC only. A remarkable example is represented by high-momentum (thus, high boost) $t \to W \to \qqbar$ decay chains that are promising probes of the time evolution of the QGP density and of the role of colour coherence.
A possible sizeable secondary production of charm quarks in scatterings between quark and gluon constituents of the hot QCD medium could represent a novel observable sensitive to the  medium temperature evolution. Also the yields and kinematic distributions of heavy quarkonium bound states carry information about properties of the produced QCD matter, since quarkonia states are expected to dissociate above critical energy densities (that depend on the binding energy of the state), and since they are expected to form in secondary processes (depending on the density of heavy quarks in the system).
         \item {\bf FCC explores saturated parton densities in a previously-uncharted, ultra-dense kinematic domain}\\
         In the incoming nuclear wave-functions, parton densities increase strongly with decreasing momentum fraction $x$. At any given $\sqrt{s}$, the nuclear parton densities are larger 
          than those in the proton due to geometric enhancement.  On general grounds, this growth at small-$x$ is expected to saturate
         once parton densities reach non-perturbative values of parametric order $\sim 1/\alpha_s$. In the context of heavy-ion collisions, the study of saturated QCD is of fundamental interest 
         mainly because it fixes the initial conditions for the collective dynamics. For instance, the accuracy with which properties of dense QCD matter can be constrained  in a heavy-ion programme
         is expected to depend ultimately on the accuracy with which one characterises the incoming nuclear wave functions at small $x$. More generally, saturated QCD is of fundamental interest as 
         it is a qualitatively-novel kinematic regime where QCD scale dependence is governed by non-linear evolution equations and where bulk properties of QCD may become amenable to perturbative 
         calculations. As discussed in Section~\ref{sec:HI_smallx}, the higher centre-of-mass energy of FCC allows one to explore a wide previously-uncharted kinematic range in $\log Q^2$ and
         $\log 1/x$ within which saturation physics is expected to manifest itself.  A proton--nucleus collision programme at the FCC is needed to explore this opportunity fully.
Such programme would be complementary to that of an electron--hadron collider.
Among the most promising observables, we quote here photon production and photon--hadron correlations at forward rapidity, which are sensitive to the small-$x$ and small-$Q^2$ region where saturation is expected to set in, heavy quarkonium production in photon--nucleus collisions (so called ultra-peripheral heavy-ion collisions),  as well as very heavy objects like $W$, $Z$ and top, which can provide strong constraints on the modification of the parton density functions in nuclei at small $x$ and large $Q^2$. 
\end{enumerate}

So far, the heavy-ion working group did not study detailed detector requirements for an experimental programme with nuclear beams at the FCC-hh. From an experimental viewpoint, it remains in particular to be investigated to what extent the physics opportunities of a heavy-ion programme at the FCC can be exploited with a general purpose detector for pp collisions. Without addressing this question in detail, the
physics opportunities discussed in the present report allow one to identify some general prerequisites for the detector design:
\begin{enumerate}
\item To fully exploit the opportunities for physics with soft probes, one requires a detector with excellent charged-hadron identification to measure low-$\pt$ pions, kaons, protons and light nuclei, their abundance, spectra, flow and correlations, as well as low-$\pt$ charm and beauty
mesons and baryons. Such identification capability could be provided by measurements of specific energy deposition in silicon trackers, time-of-flight, Cherenkov radiation, or a combination of these. 
\item Track reconstruction capability down to low $\pt$, ideally starting from few hundred MeV/$c$, is mandatory for all the aforementioned measurements. This capability requires to minimize the material thickness of the inner tracker and is may be limited by the large values of magnetic field ($4$--$8$~T) that are considered for pp-dedicated detectors at the FCC-hh. Therefore, it would be interesing to assess the feasibility of a general-purpose detector that can be operated also with reduced magnetic field of $\approx 1$~T.
\item To fully exploit the opportunities for physics with hard probes, the basic requirements should match those for the pp programme at the FCC, that is hadronic and electromagnetic large-acceptance calorimeters with excellent energy resolution at high-$\pt$, and excellent detection capabilities for the leptonic decay products of hard processes. These detector specifications need to persist for the higher event multiplicities of heavy-ion collisions. 
\item To fully exploit the opportunities for saturation physics, one requires a detector with excellent forward coverage for charged particles, photons and jets, ideally  up to $\eta\approx 6$. 
\end{enumerate}

In addition to the three key motivations for a heavy-ion beam programme at the FCC listed above, the present document will summarise further opportunities. It is structured as follows.
The FCC-hh machine parameters and projected performance for heavy-ion running are presented in Section~\ref{sec:HI_machine}. The opportunities for studying hot and dense QCD matter
with soft and hard observables are discussed in Sections~\ref{sec:HI_qgpsoft} 
and \ref{sec:HI_hardprobes}, respectively. In Section~\ref{sec:HI_smallx}, the potential for studying gluon saturation and
nuclear-modified PDFs is presented including observables in hadronic proton--nucleus and nucleus--nucleus collisions and in photon-induced ultra-peripheral collisions. In a final Section~\ref{sec:HI_otherhep}, we turn then
to contributions to other sectors of high-energy physics, such as 
searches for new particles in photon--photon scattering processes induced with very large rate by the strong electro-magnetic fields of incident Pb nuclei, 
and the physics with fixed-target collisions using FCC proton or heavy-ion beams.

\newpage 
\section{Heavy-ion performance of FCC-hh \footnote{Editors: J.M.~Jowett, M.~Schaumann}}
\label{sec:HI_machine}

At an early stage in the study of the future hadron collider, FCC-hh, 
a fairly complete study of its potential as a heavy-ion  collider 
was published~\cite{Schaumann:2015fsa};  we take this as a reference for the following.  
Collisions of lead nuclei with each other (Pb--Pb) and 
with protons (p--Pb) were considered, as at the LHC.  
The performance projections were based on a 
very conservative injection scenario, in which the LHC 
was used as the final injector synchrotron and  the 
parameters of the injected beam in an LHC cycle were based on 
those obtained in the 
2013 p--Pb run of the LHC.  
This could be comfortably realised by  simply maintaining the 
present source and injector chain of the LHC at the 
performance levels of LHC Run-1.  
With these parameters, 
and the assumption of a single heavy-ion experiment,  
Ref.~\cite{Schaumann:2015fsa} showed that the optimum operating 
cycle for the FCC-hh was to inject one LHC fill (filling only a fraction of the FCC-hh ring) and immediately ramp and collide.   
The time required to efficiently exhaust the beams in collisions corresponded 
closely to the time required to refill and ramp the LHC again for the
next fill, so that the optimum injection scheme was to fill just a quarter of the ring with a single bunch train from the LHC.  

In Ref.~\cite{Schaumann:2015fsa}, it was also shown that the FCC-hh will enter a 
new, highly-efficient operating regime, in which a large fraction of the
injected intensity can be converted  to useful integrated luminosity.  
Thanks to strong synchrotron radiation damping, the beam emittances shrink rapidly and 
compensate the rapid decay of initial luminosity seen at lower-energy colliders.   
The luminosity may even increase during a fill until the  beams are exhausted. 
Not only is this natural beam cooling twice as fast for heavy ions as for protons, 
it can also be more fully exploited since the lower overall bunch charges 
do not lead, for example, to high beam--beam tune-shifts.

In fact, the first heavy-ion run of LHC Run-2, in 2015, has   shown that
the present LHC and injector complex is 
already capable of higher performance, giving approximately a factor of 
2.4 in luminosity beyond what is assumed in Ref.~\cite{Schaumann:2015fsa}.  
Further gains are expected after the LHC Long Shutdown 2 (from 2021).
Since the publication of Ref.~\cite{Schaumann:2015fsa}, the design work 
on FCC-hh has mainly focussed  
on its performance as a proton--proton collider. 
However, some important developments carry over into increased expectations for heavy-ion performance.   
In particular, measures envisaged to shorten the LHC cycle~\cite{FCCturnaround} mean that the optimum   scheme is to fill 
the entire FCC ring using up to 4 LHC injection cycles, boosting the peak and integrated luminosities by a further factor approaching 4.  
The effect of this is shown in Fig.~\ref{intLPbPb}, which shows the luminosity that would be integrated in an ideal 30-days run at full performance with perfect efficiency (no down time or other interruptions).   
Note that, for simplicity of comparison with~\cite{Schaumann:2015fsa} 
we have maintained the assumption of a single experiment taking data. 
For $n_\mathrm{exp}$ heavy-ion experiments 
(with similar configurations) the integrated luminosity per experiment 
will go down because of luminosity sharing but not as fast as
$1/n_\mathrm{exp}$ (the total luminosity, summed over experiments, will be somewhat increased).

\begin{figure}[t]
\centering
\includegraphics[width=0.49\textwidth]{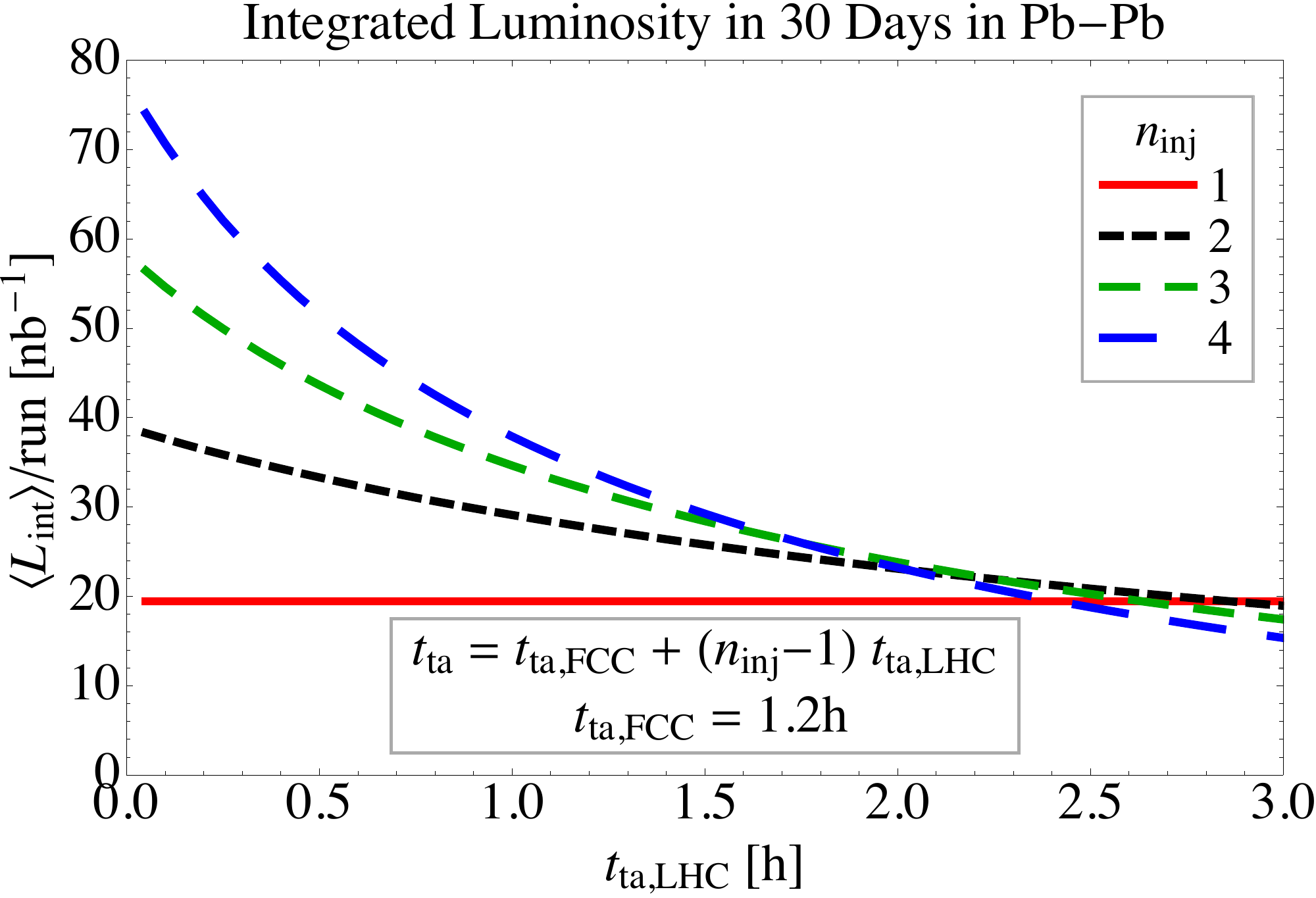}
\includegraphics[width=0.49\textwidth]{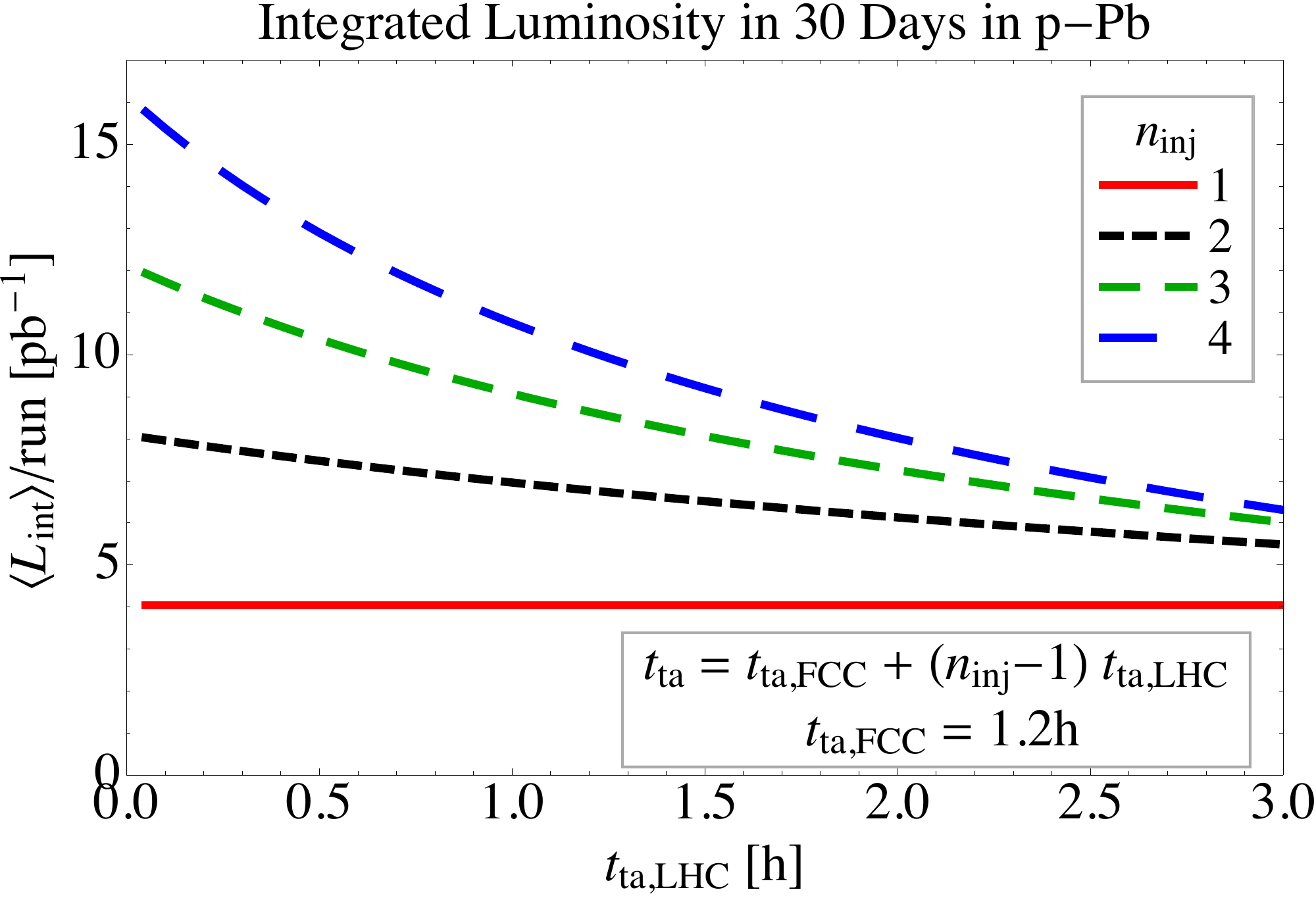}
\caption{\label{intLPbPb}
Integrated luminosity in a 30-days Pb--Pb (left) or p--Pb
run (right) for different numbers of LHC injections, and as
a function of the LHC cycle time (this updates Figs.~7(c) and 11(c) of Ref.\protect{\cite{Schaumann:2015fsa}}).}
\end{figure}     


\begin{table*}
  \centering
          \caption[Heavy-ion collider parameters]{ \label{t_fcc_summary}
          Selected beam and performance parameters for the FCC-hh in
          Pb--Pb and p--Pb modes.  This table is an update of parts of
          Table~VIII of Ref.~\protect{\cite{Schaumann:2015fsa}} which provides further detail.} 
    \begin{tabular}{lc|c|cc}
\hline
  & Unit   & FCC Injection &
    \multicolumn{2}{c}{ FCC Collision} \\
    \hline
     Operation mode					&                                          & Pb    	& Pb--Pb         & p--Pb           \\
    
    \hline
     Beam energy 					& [TeV]                                     & 270   	& 4100          & 50              \\
     $\sqrtsNN$&[TeV] 						&- 			&39.4           & 62.8 \\
     No. of bunches per LHC injection & -										& 518		&518			& 518 \\
     No. of bunches in the FCC		& -                                         & 2072      & 2072          & 2072         \\
     No. of particles per bunch 	& [$10^8$]                           		& 2.0       & 2.0           & 164           \\
     Transv. norm. emittance 		&  [$\mu$m]                            		& 1.5       & 1.5           & 3.75          \\
     
     Number of IPs in collision     & -         								& -         &1               & 1     \\
     Crossing-angle 				& [$\mu$rad]                                & -         &\multicolumn{2}{c}{0}  \\
     Initial luminosity 			& [$10^{27}\text{cm}^{-2}\text{s}^{-1}$] 	& -         & 24.5            & 2052            \\
     Peak luminosity 				& [$10^{27}\text{cm}^{-2}\text{s}^{-1}$]  	& -         & 57.8            & 9918             \\
     Integrated luminosity per fill & [$\mu$b$^{-1}$]           				& -         & 553             & 158630           \\
     Average luminosity 			& [$\mu$b$^{-1}$]                      		& -         & 92           	  & 20736            \\
     Time in collision				& [h]										& - 		& 3				  &6		\\
     Assumed turnaround time 		& [h]                                       & -          & 1.65           & 1.65        \\
     Integrated luminosity/run 		& [$\mathrm{nb}^{-1}$]                                       & -          & 33          & 8000    \\   
\hline
    \end{tabular}%
\end{table*}%

Table~\ref{t_fcc_summary} summarises key parameters for
Pb--Pb and p--Pb  operation at \etev{50Z} in the FCC-hh.  
In the case of p--Pb operation the Pb beam is assumed to be the
same as for Pb--Pb, so the corresponding column only quotes the
proton beam parameters.    
The calculated luminosity values assume an optimised theoretical turnaround time of \qty{9}{min} per LHC cycle and an additional preparation time in the FCC-hh of \qty{1.2}{h} per FCC-hh filling, as quoted in~\cite{FCCturnaround}. 
It was assumed that the first LHC beam is already prepared during preparation time of the FCC-hh, so that a total turnaround time sums up to \qty{1.65}{h}. This represents a theoretical minimum. 
In reality early beam aborts and other faults will increase this 
time and somewhat reduce the integrated luminosity.  

The final values for the integrated luminosity in a typical annual 
one-month run assume an LHC cycle time of 9~min 
and are reduced by a ``performance efficiency factor'' of 50\% 
to allow for set-up time, down-time and other deviations from the 
idealised running described in Fig.~\ref{intLPbPb} 
(a similar factor is applied in HL-LHC performance projections). 
The resulting integrated luminosity values are of 33~nb$^{-1}$ for
Pb--Pb and  8~pb$^{-1}$ for p--Pb collisions. The previous estimates~\cite{Schaumann:2015fsa}
were of 8~nb$^{-1}$ and 1~pb$^{-1}$, respectively.
For the moment, no studies of upgrades to the heavy-ion injectors 
(source, linac, accumulation ring, PS and SPS synchrotrons) have been 
performed.
If upgrades to these machines can be envisaged by the time of 
FCC-hh operation, then still higher 
luminosities are likely to be available. 


Heavy-ion operation will certainly also require certain adaptations of the
FCC-hh main ring, e.g., special absorbers in key locations 
for the high flux of modified ions 
from the bound-free pair-production process 
at the interaction points.  
Collimation of the heavy-ion beams will also be a serious issue and
require further absorbers or, possibly, the application of new collimation 
technologies such as bent crystals or electron lenses.  
The potential of these technologies is under study at the LHC.


\clearpage 
\section{QGP studies: bulk properties and soft observables \footnote{Editor: U.A.~Wiedemann}}
\label{sec:HI_qgpsoft}

A central goal of a heavy-ion programme at a hadron collider is to explore how collective properties emerge from the fundamental 
fields of Quantum Chromodynamics and their non-abelian interactions. So-called ``soft observables'', that is particles at
low transverse momentum, are important in this context since they are the experimentally accessible decay products of the medium 
that is formed during the collision, and since they provide the most direct signals of collective behaviour. Any substantial
increase in the centre-of-mass energy benefits this research programme by significant quantitative gains. In particular, the QGP phase 
in Pb--Pb collisions at $\sqrtsNN=39$~TeV is expected to have larger volume, lifetime, energy density and temperature than Pb--Pb collisions
at LHC energy. Also, the enlarged spatio-temporal extension of the created system is expected to be accompanied by larger
collective effects, and the increased multiplicity per event increases the statistical precision with which statements about collectivity can be made.
In the present section, we summarise basic expectations for the bulk properties and soft observables in heavy-ion collisions at FCC-hh, and we 
provide some examples for the physics opportunities arising from them. Our discussion touches also fundamental questions whose understanding is currently
evolving rapidly. For instance, recent discoveries at the LHC emphasise the need for understanding signatures of collectivity across system size,
 including the study of proton--proton and proton--nucleus collisions. This addresses the fundamental question of what is the smallest length and time
  scale for QCD thermalization or hydrodynamization in  Quantum Chromodynamics, and how this scale depends on energy density.
While there is no doubt that, due to the higher event multiplicities reached in hadronic collisions, experiments at the FCC will greatly contribute to this 
question, there is also little doubt that the motivation for this experimental programme will be refined significantly in the coming years in an interplay
between theory and further analyses of LHC data. A similar comment applies to the increased charm production in heavy-ion collisions at the FCC, 
and its impact on our understanding of QCD thermalization and hadronization from the QCD fireball.

\subsection{Global characteristics of Pb--Pb collisions}
\label{sec:HI_global}

Extrapolating measurements of charged particle multiplicity, transverse energy and femtoscopic correlations 
at lower energies~\cite{Aamodt:2010pb,Abbas:2013bpa,Aamodt:2011mr,Adam:2015ptt,Chatrchyan:2012mb,Adam:2016thv}, one can obtain estimates
for the growth of global event characteristics from LHC to FCC. In particular, up to the top LHC energy, the growth of charged hadron 
event multiplicity per unit rapidity in Pb--Pb collisions is consistent with a weak power-law, 
\begin{equation}
    \frac{{\rm d}N_{\rm ch}}{{\rm d}\eta}\Bigg|_{\eta=0} \propto (\sqrtsNN)^{0.3}\, .
\end{equation}
As can be seen from Table~\ref{tab:PbPb}, this amounts to an increase of a factor $\sim 1.8$ from top LHC to FCC energy. In the absence of a 
fully quantitative theory for soft physics observables such as event multiplicities in hadronic collisions, it is difficult to assign uncertainties to such
an estimate. From past experience with such extrapolations, and from the increased lever arm available to extrapolate now to FCC, we note
simply that all current considerations favour an $\mathcal{O}(2)$ increase of multiplicity from LHC to FCC, while a multiplicity increase by a factor 3 or larger
would be a big surprise. Also, while event multiplicity increases significantly with $\sqrtsNN$ in all models of particle production, this increase is 
tamed in models that account for non-linear QCD saturation physics as a mechanism that regulates the density of incoming parton distributions.
For instance, in the CGC-rcBK model~\cite{ALbacete:2010ad}, one obtains ${\rm d}N_{\rm ch}/{\rm d}\eta = 2700$--$2900$ which is about 25\% lower than the value obtained 
from scaling the multiplicity with $ (\sqrtsNN)^{0.3}$. This illustrates the typical uncertainties in such extrapolations. 

\begin{table}[h]
\caption{Global properties measured in central Pb--Pb collisions (0--5\% centrality class) at $\sqrtsNN=2.76$~TeV and extrapolated to 5.5 and 39~TeV.
The measurements at 2.76~TeV~\cite{Aamodt:2010pb,Abbas:2013bpa,Aamodt:2011mr,Adam:2015ptt,Chatrchyan:2012mb,Adam:2016thv} are reported for comparison only and without experimental uncertainties.}
\small
\begin{center}
\begin{tabular}{lccc}
\hline
Quantity & Pb--Pb 2.76~TeV & Pb--Pb 5.5~TeV & Pb--Pb 39~TeV \\
\hline
${\rm d}N_{\rm ch}/{\rm d}\eta$ at $\eta=0$ & 1600 & 2000 & 3600 \\
Total $N_{\rm ch}$ & 17000 & 23000 & 50000 \\
${\rm d}E_{\rm T}/{\rm d}\eta$ at $\eta=0$ & 1.8--2.0~TeV & 2.3--2.6~TeV & 5.2--5.8~TeV \\
Homogeneity volume $$ & 5000~fm$^3$  & 6200~fm$^3$ & 11000~fm$^3$ \\
Decoupling time & 10~fm/$c$ &  11~fm/$c$ & 13~fm/$c$ \\
$\varepsilon$ at $\tau=1$~fm/$c$ & 12--13~GeV/fm$^3$  & 16--17~GeV/fm$^3$ & 35--40~GeV/fm$^3$ \\
\hline
\end{tabular}
\end{center}
\label{tab:PbPb}
\end{table}

 
\begin{figure}[!t]
\begin{center}
\includegraphics[width=0.49\textwidth]{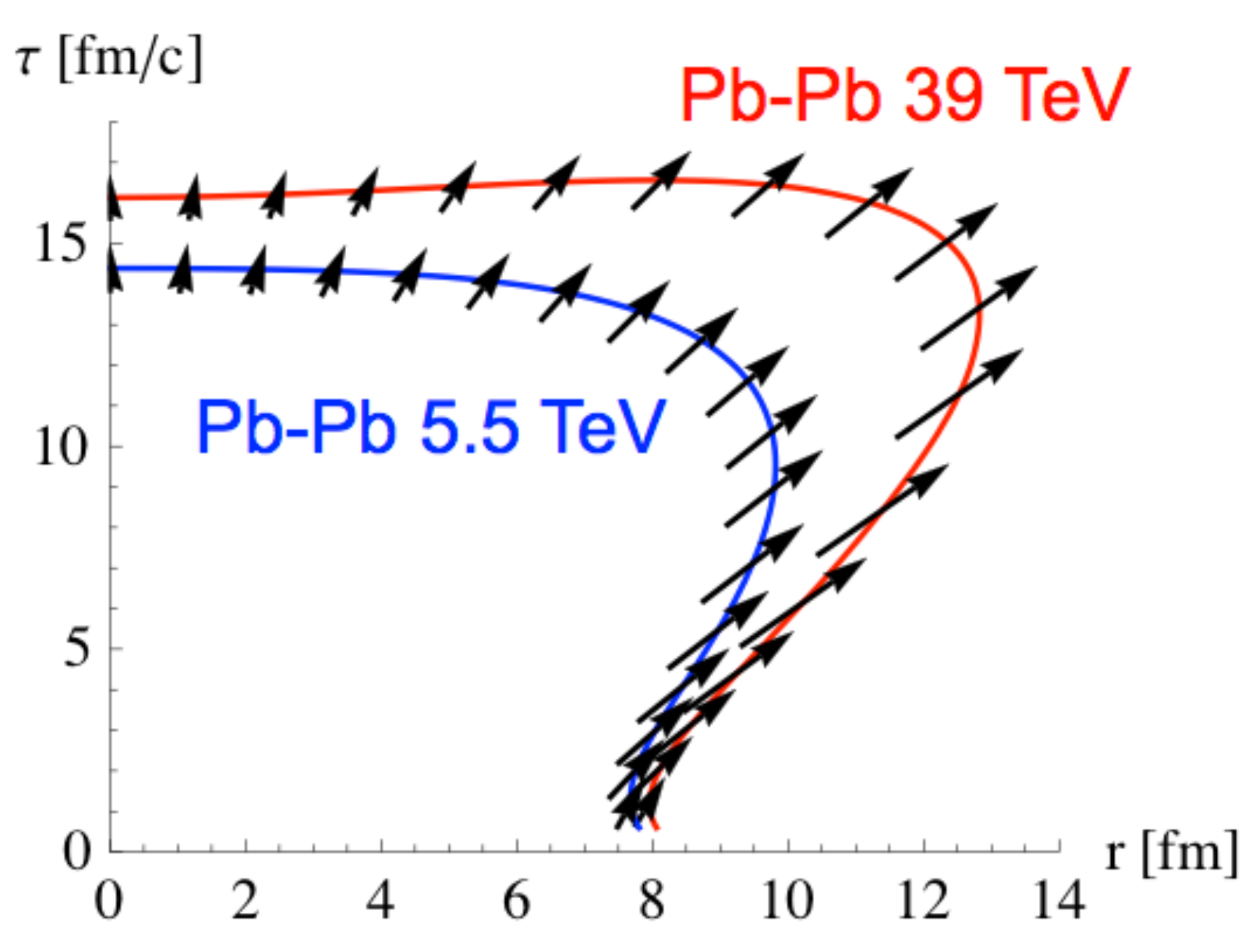}
\includegraphics[width=0.49\textwidth]{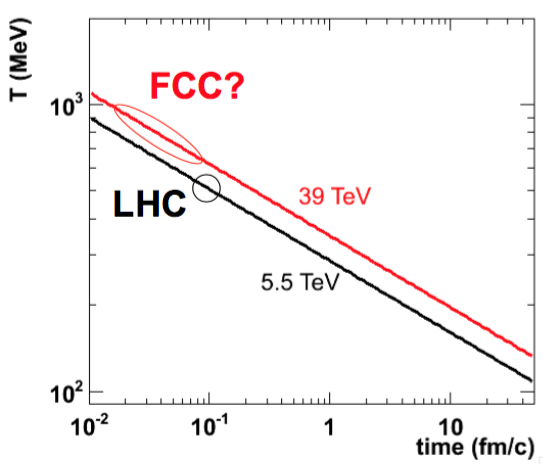}
\caption{Left: space-time profile at freeze-out from hydrodynamical calculations for central Pb--Pb collisions at $\sqrtsNN=5.5$~TeV and $39$~TeV.
Right: time evolution of the QGP temperature as estimated on the basis of the Bjorken relation and the Stefan-Boltzmann equation (see text for details).}
\label{fig:freezeout}
\end{center}
\end{figure}

Fluid dynamic simulations of heavy-ion collisions are sensitive to the initial conditions from which the system is evolved dynamically, 
and they are sensitive to the thermodynamic properties of hot QCD matter as encoded in the the equation of state and in QCD transport properties. 
The increased event multiplicity at FCC energy is of prime importance for the fluid dynamic expansion, since it constrains a central characteristic of the 
initial conditions, namely the entropy density at initial time. More precisely, for a general viscous dynamics, the second law of thermodynamics implies
that the final multiplicity puts an upper bound on the initial entropy. However, the QCD matter produced in heavy-ion collisions shows very small dissipative
properties at TeV energies and is thus expected to follow a close to isentropic expansion: the initial entropy density is then fixed by the final event 
multiplicity. The $\sqrt{s}$-dependence of fluid dynamic simulations of heavy-ion collisions thus results mainly from the increase in event multiplicity with $\sqrt{s}$. 
To illustrate the impact of the expected multiplicity increase from LHC to FCC,  we have run a simplified fluid dynamic simulation for a central Pb--Pb collision. The radial dependence 
of the energy density in the initial conditions was chosen to be determined as the smooth nuclear transverse overlap function of two Wood-Saxon profiles, neglecting any
possible energy dependence and fluctuations. Using a standard parametrisation of a realistic QCD equation of state and minimal dissipative properties (shear viscosity to entropy density ratio $\eta/s = 1/4\pi$), we
show in Fig.~\ref{fig:freezeout} (left) results for the freeze-out hypersurfaces of central Pb--Pb collisions at different collision energies. This figure quantifies the naive expectation 
that the denser system created at higher collision energy has to expand to a larger volume and for a longer time before reaching the freeze-out temperature at which 
decoupling to hadrons sets in. In this way, Fig.~\ref{fig:freezeout} (left) confirms the qualitative expectation that the freeze-out volume should increase proportional to event multiplicity, 
since the system is expected to decouple at a freeze-out temperature that does not depend significantly on $\sqrtsNN$ already at top RHIC and LHC energies. While this overall volume is not directly measurable, the homogeneity volume over which bosons interfere 
constructively is measurable via femptoscopic interferometry measurements, and experimental data up to LHC energy indicate that this volume increases with  $\propto {\rm d}N_{\rm ch}/{\rm d}\eta$, too, see
Table~\ref{tab:PbPb}. Fig.~\ref{fig:freezeout} (left) is also in accordance with the parametric expectation that the decoupling time grows  $\propto ({\rm d}N_{\rm ch}/{\rm d}\eta)^{1/3}$.  

The arrows overlaid with the freeze-out hypersurface in Fig.~\ref{fig:freezeout} (left) indicate the transverse flow of the fluid element at decoupling. This provides quantitative support for 
the qualitative expectation that in a larger and more long-lived system, collective effects can grow stronger. In particular, the figure illustrates that the radial flow field
is expected to increase substantially from LHC to FCC, as indicated by the length of the arrows.

In general, the global event characteristics listed in Table~\ref{tab:PbPb} determine the spatio-temporal extent of the ``cauldron'' in which QCD matter is evolved, and they
constrain the thermodynamic conditions that apply after thermalization. The measured transverse energy per unit rapidity ${\rm d}E_{\rm T}/{\rm d}\eta$ (see Table~\ref{tab:PbPb})
is of particular importance since it constrains the initial energy density. This is most easily illustrated by an estimate, obtained from back-extrapolating ${\rm d}E_{\rm T}/{\rm d}\eta$
under the assumption of free-streaming, i.e., under the assumption that the system makes minimal work. The resulting Bjorken relation 
$\varepsilon(\tau) = \frac{1}{c\, \tau} \frac{1}{\pi R_A^2} {\rm d}E_{\rm T}/{\rm d}\eta$ assumes then that the energy density of the system at very early times is determined 
by the energy ${\rm d}E_{\rm T}/{\rm d}\eta$ contained in a volume given by the transverse overlap area $\propto \frac{1}{\pi R_A^2}$ times the longitudinal extent reached
at time $\tau$. According to this Bjorken estimate, the energy density decreases initially like $\varepsilon(\tau) \sim 1/\tau$, and the temperature evolves as the fourth root.
One may estimate the pre-factor of this relations, for instance by using the Stefan-Boltzmann limit of the QCD equation of state which yields 
 $T(t)= [\varepsilon(t)\,(30/\pi^2)/n_{\rm d.o.f.}]^{1/4}$, where $n_{\rm d.o.f.}=47.5$ is the number of degrees of freedom for a system with gluons and three quark flavours.
The energy density is expected to increase by a factor of two from LHC to FCC, reaching a value of 35--40~GeV/fm$^3$ at the time of 1~fm/$c$. 
 In Fig.~\ref{fig:freezeout} (right), we have plotted the time-dependence of the QGP plasma temperature for Pb--Pb collisions at the LHC and at the FCC. 
We caution that the present use of the Bjorken estimate is subject to several uncertainties. On a quantitative level, one may note e.g. that the QCD equation of state differs from
that of a Stefan-Boltzmann gas and that at sufficiently late times (say $\tau > 1$~fm/$c$), the transverse expansion of the system makes the energy density decay faster than
$1/\tau$. On a qualitative level, we note that the Bjorken estimate constrains the energy within a given initial volume without specifying whether (and if so, at what initial time) 
this energy density is thermalised. While the curves in Fig.~\ref{fig:freezeout} (right) extend to arbitrarily early (logarithmic) times, they are only meaningful at times larger than
a thermalization time-scale that is estimated to be $\mathcal{O}(0.1\, {\rm fm}/c)$ at LHC energies and that is expected to decrease with increasing energy density.
At least in principle, this time-scale is calculable from kinetic theory, but so far only rough estimates can be given. Figure~\ref{fig:freezeout} (right) shows that while the increase 
at a given time is a modest 30\% when going from LHC to FCC, the thermalization time of the system is expected to be significantly smaller. 
One may reach initial temperatures as large as $T_0\approx800$--$1000$~MeV in case that the thermalization time decreases to $\mathcal{O}(0.02\, {\rm fm}/c)$ at FCC energies.

\subsection{Collective phenomena from heavy-ion to pp collisions}

One of the most important characterisations of flow-like phenomena in heavy-ion collision is the study of the azimuthal dependence of particle production. For a single inclusive hadron spectrum,
this can be done for instance  by measuring the azimuthal harmonics $v_n$ in a Fourier decomposition\footnote{This decomposition takes into account only the even terms (cosines), which are far dominant over the odd terms (sines).} in the azimuthal angle $\varphi$ with respect to the nucleus--nucleus reaction plane orientation $\Psi$,
\begin{equation}
	\frac{\dd N_{\rm ch}}{\pt\, \dd\pt\, \dd\eta\, \dd\varphi} = \frac{1}{2\pi} \frac{\dd N_{\rm ch}}{\pt\, \dd\pt\, \dd\eta} \, \left[1 + 2 \sum_{n=1}^{\infty} v_n(\pt,\,\eta)\, \cos(n(\varphi-\Psi)) \right] \, .
\end{equation}
Since heavy nuclei are not uniform spheres but distributions of nucleons, the spatial profile of the initial nucleon--nucleon collisions, which determines the so-called initial conditions of a heavy-ion collision, varies on an event-by-event basis and it has a complex structure that is best characterised with a symmetry plane for each harmonic $v_n$, rather than with a single reaction plane. Therefore, $\Psi$ is replaced in the above expansion by a set of symmetry planes $\Psi_n$.
The azimuthal orientations of $\Psi_n$ need to be extracted from the same set of data as the harmonic flow coefficients $v_n$, and 
effects which only contribute to few-particle correlations and are invariant to symmetry plane orientations (so-called non-flow effects) need to be disentangled from flow-effects,
which typically involve all produced particles. This is done by measuring $v_n$ from multi-particle azimuthal correlations. For the description of the various techniques,
developed to this end, we refer to the literature~\cite{Ollitrault:2009ie,Voloshin:2008dg,Bilandzic:2013kga}.

At FCC energies, the two-fold larger multiplicity in central Pb--Pb collisions
may open up the possibility to carry out flow measurements on an event-by-event basis and to become sensitive to dependencies of transport coefficients that are very difficult to address at the LHC. For example, the different azimuthal coefficients $v_n$ are sensitive to the 
various possibilities for the temperature dependence of shear viscosity to entropy density ratio, $\eta/s(T)$, and this sensitivity becomes stronger with increasing multiplicity and for higher harmonics. This is illustrated in Fig.~\ref{fig:flow}.
 The results are based on the theoretical framework described in Refs.~\cite{Gale:2012rq,Niemi:2014wta,Denicol:2014ywa}. We note that at the time of writing this report, neither the theoretical possibilities for disentangling different conceivable temperature dependences of $\eta/s(T)$ at FCC, nor the opportunities of getting insights into this question from further LHC data are fully explored. In particular, 
recent work studying various soft hadron observables in the EbyE EKRT model context indicates~\cite{Niemi:2015voa,Niemi:2015qia} that an analysis including Pb--Pb data up to the top LHC energies in combination with RHIC Au--Au data can disentangle between different temperature dependencies for  $\eta/s(T)$. While there is a sound qualitative argument that the higher centre-of-mass energy at 
FCC will lead to an improved handle on the temperature dependence of transport coefficients, the expected quantitative gains at FCC are likely to require updating in the light of these ongoing efforts at the
LHC.

\begin{figure}[t]
\begin{center}
\includegraphics[width=0.49\textwidth]{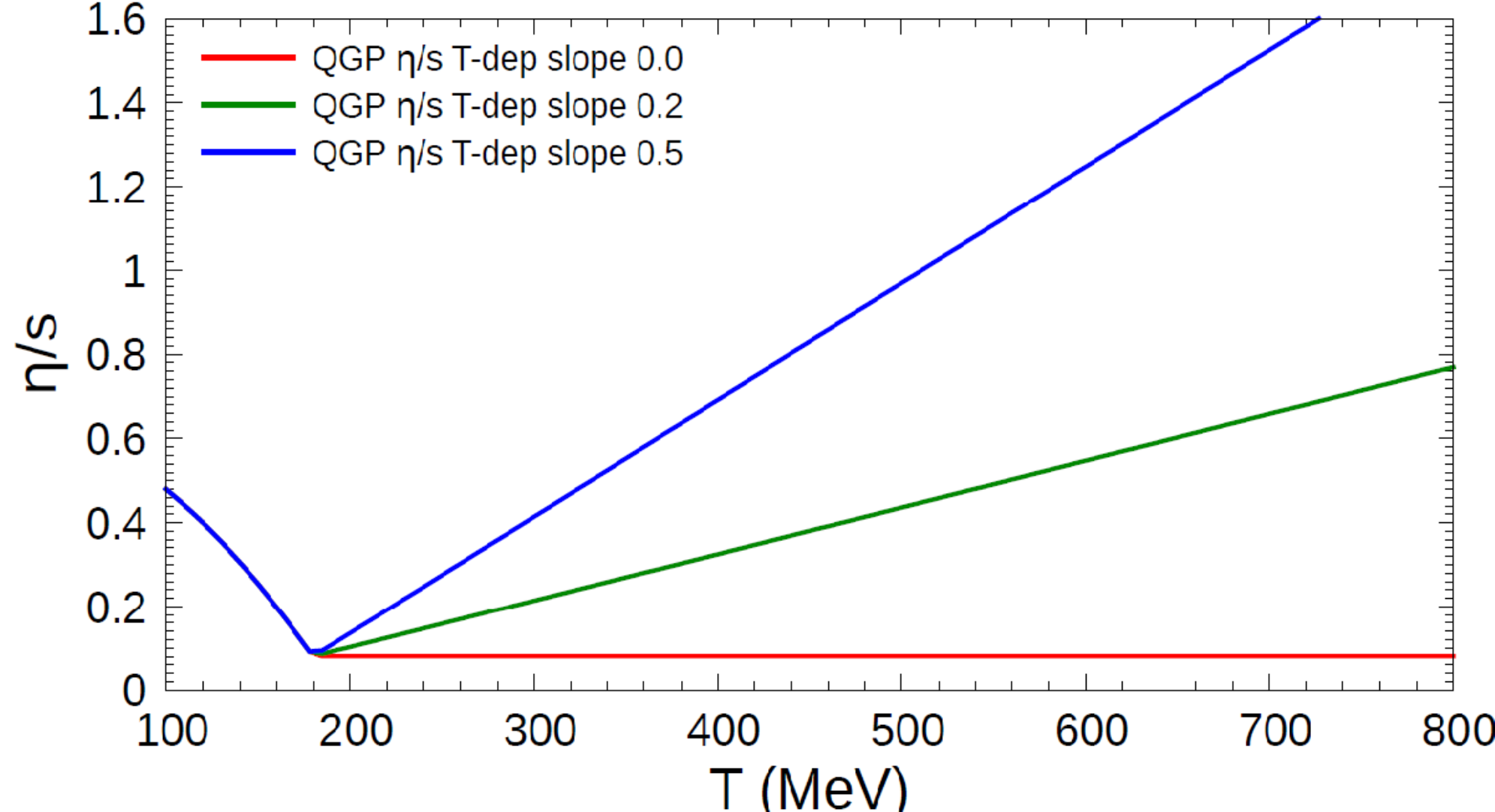}
\hfill
\includegraphics[width=0.46\textwidth]{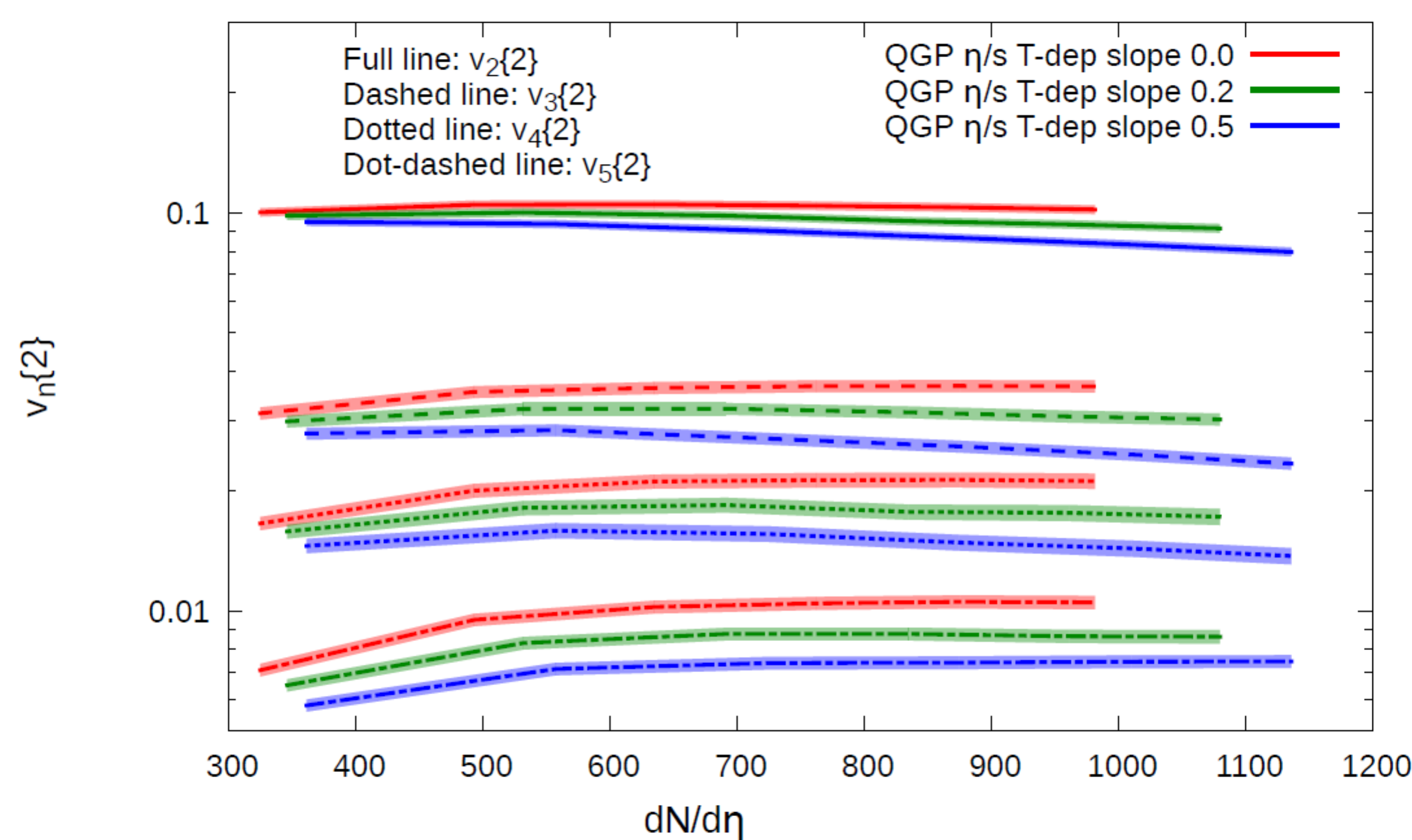}
\caption{Left: parametrisations for the evolution of the ratio $\eta/s$ versus temperature.
Right: results for different $v_n\{2\}$ versus multiplicity from viscous hydrodynamics calculations for different temperature dependencies of $\eta/s$ shown on the left. The results are based on the theoretical framework described in Refs.~\cite{Gale:2012rq,Niemi:2014wta,Denicol:2014ywa}.}
\label{fig:flow}
\end{center}
\end{figure}

In recent years, surprisingly, small and dense systems probed in high multiplicity p--A and pp collisions were found to display flow-like phenomena. In particular, p--Au, d--Au and $^3$He--Au collisions at RHIC, as well as p--Pb and high-multiplicity pp collisions at the LHC have been shown to feature similar ridge-like structures, $v_2$ anisotropy  and, in some of the systems, including high-multiplicity pp collisions, even $v_3$ anisotropy as seen in collisions between large nuclei~\cite{Khachatryan:2010gv,Chatrchyan:2013nka,ABELEV:2013wsa,Abelev:2012ola,Aad:2014lta,Adamczyk:2015xjc,Adare:2015ctn}. The hints for a collective nature of the azimuthal anisotropy seen in p--Pb collisions were reinforced by measuring its higher-order cumulants. New pp data from ATLAS and CMS indicate that the long-range ridge correlation may well extend all the way down to minimum bias collisions (if not below), though differences in the analysis techniques warrant further scrutiny. These recent findings raise fundamental questions about whether the flow-like patterns in small and dense systems are only similar in appearance to
what one observes in heavy-ion collisions, or whether the idea of a minimal scale for the onset of collective phenomena needs to be revisited. 

\begin{figure}[!t]
\begin{center}
\includegraphics[width=0.5\textwidth]{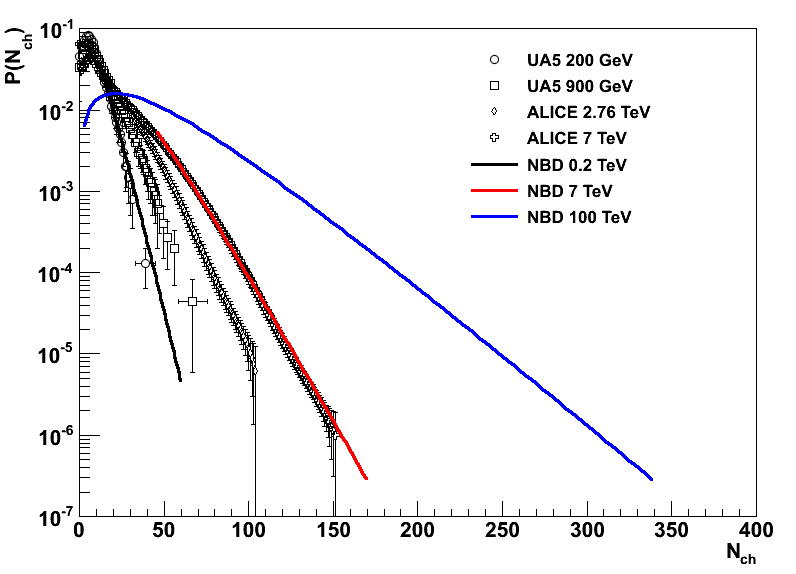}
\caption{Extrapolation of the charged-particle multiplicity distribution ($|\eta| < 1.5$) in pp collisions to $\sqrt s=100$~TeV, based on a negative binomial distribution.}
\label{fig:ppMultReach}
\end{center}
\end{figure}

In general, the experimental study of flow phenomena in hadronic collisions aims at disentangling flow effects from non-flow contributions. Simple non-flow effects can arise for instance from resonance decays or 
(mini)jets. The energy-momentum constraints of these elementary particle production processes lead to patterns in the two-particle correlations that share at least qualitative commonalities
with the patterns measured in the harmonic flow coefficients $v_n$, in the ideal case of zero non-flow correlations.
However, a collective phenomenon is shared by many if not all particles in the event, rather than by the few
particles associated to the same microscopic production process. The ability to disentangle collective correlations from confounding factors, therefore, increases with event multiplicity. This is
seen most explicitly in the so-called cumulant analysis of flow coefficients, via which one tests whether an assumed collective effect persists with equal strength when searched for in 
multi-particle correlations involving higher number of particles. It is an important limitation of the current discussion of the system size dependence of flow at the LHC that the multiplicities in pp collisions are still too small 
to allow for the same cumulant analyses of flow coefficients that are standard in heavy-ion collisions. The increased event multiplicity in pp collisions at the FCC will overcome this
problem. 

Fig.~\ref{fig:ppMultReach} presents an extrapolation of the multiplicity distribution in $|\eta| < 1.5$ to $\sqrt s=100$~TeV. The extrapolation was obtained by fitting multiplicity distributions from 0.2 to 7~TeV with a negative binomial distribution and extrapolating the parameters. To achieve a good description of the high-multiplicity tail of the distribution, the lowest 30\% of the multiplicity range was excluded from the fit. Including the low-multiplicity range leads to a worse fit result and a wider extrapolated multiplicity distribution. Therefore, the presented extrapolation can be seen as a lower limit of the possible reach. The high-multiplicity tail of event-distributions in proton-proton collisions will become accessible at the FCC up to multiplicities of at least 300 charged particles. This makes it feasible to apply statistically demanding analysis techniques for the identification of flow-like phenomena, such as higher-order cumulant analyses, across system size, including the smallest pp collision system.

\subsection{Effect of the charm quark on the QGP equation of state}
\label{sec:HI_charmlattice}

The fluid dynamic interpretation of ultra-relativistic heavy-ion collisions relies on a gradient expansion around local thermodynamic equilibrium. It is thus based entirely on
basic thermodynamic properties such as the equation of state (EOS, i.e.\,the temperature dependence of the pressure $P(T)/T^4$), the velocity of sound and dissipative transport coefficients. As thresholds for particle production are crossed
at higher temperature, the number of relevant thermodynamical degrees of freedom and thus the equation of state changes characteristically.  Remarkably, while still relatively
little is known from first principles about the dependence of the EOS on the quark mass, LO and NLO calculations in finite temperature field theory~\cite{Laine:2006cp}
indicate that charm quarks start contributing to the EOS for temperature as low as 350 MeV (see Fig.~\ref{fig:EOSwithCharmLaine0603048}). This is also supported by first exploratory lattice studies~\cite{Cheng:2007wu,
DeTar:2010xm,Borsanyi:2012vn}. Given that the initial temperature of the systems produced at the FCC will be significantly higher, see Fig.~\ref{fig:freezeout} (left), one
may expect that the system temperature increase could lead to a sizeable production of secondary charm and anti-charm quark ($c\overline c$) pairs from partonic
interactions during the hydrodynamical evolution of the system. However, in heavy-ion collisions, reaching thermal charm abundances will depend on kinetic and chemical
equilibration mechanisms. Their effectiveness depends on the competition between the chemical equilibration rate $\Gamma_{\rm chem}$ and the expansion rate. 
A perturbative analysis allows to relate this equilibration rate to the ratio of quark number susceptibilities of massive ($\chi_f$) and massless ($\chi_0$) quarks,
$\Gamma_{\rm chem} \simeq \textstyle\frac{2\pi\alpha_s^2 T^3}{9\, M_c^2}\left( \textstyle\frac{7}{6} + N_f\right) \textstyle\frac{\chi_f}{\chi_0}$,~\cite{Bodeker:2012gs}.
Fixing these susceptibilities with lattice data~\cite{Ding:2010yz,Borsanyi:2012vn}, and using $\alpha_s = 0.3$ and $M_c = 1.5$ GeV, one finds 
\begin{eqnarray}
	\Gamma_{\rm chem}^{-1} &>& 60\, {\rm fm/c}\qquad \hbox{for}\quad T \simeq 400\, {\rm GeV}\, ,\nonumber \\
	\Gamma_{\rm chem}^{-1} &>& 10\, {\rm fm/c}\qquad \hbox{for}\quad T \simeq 600\, {\rm GeV}\, .\nonumber	
\end{eqnarray}
We note that further theoretical developments are likely to refine these estimates prior to the start of the FCC. At face value, the numbers listed above indicate that FCC targets
an interesting transition region in energy density above which charm quarks start counting towards the thermal degrees of freedom. While charm abundances are likely to lie 
below chemical equilibration values at high temperatures, charm is expected to participate in experimentally accessible kinetic and chemical equilibration processes that can 
be accounted for in the context of kinetic theory. Predictions on thermal charm production at FCC energies will be discussed in Section~\ref{sec:HI_hf}.

\begin{figure}[!t]
\begin{center}
\includegraphics[width=0.5\textwidth]{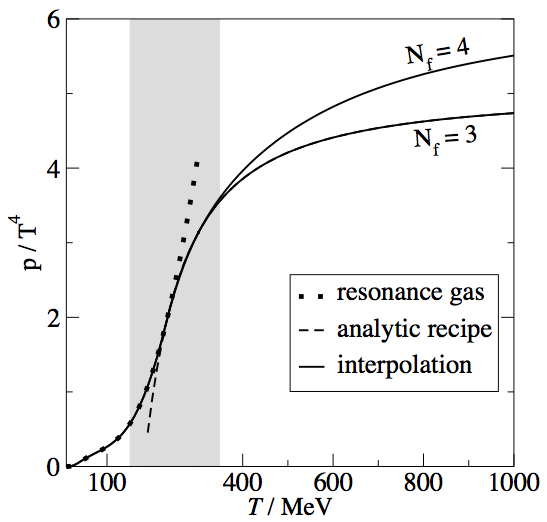}
\caption{For basic thermodynamic observables such as the pressure plotted here, the charm quark plays a visible role at very low temperature $T/m_c \ll 1$.  The present plot is for a perturbative
calculation that accounts for gluonic contributions up to the highest known order $\mathcal{O}(g^6 \ln (1/g))$ and that treats the change in quark masses to order $\mathcal{O}(g^2)$. Figure taken from Ref. ~\cite{Laine:2006cp}. }
\label{fig:EOSwithCharmLaine0603048}
\end{center}
\end{figure}

\subsection{Hadrochemistry}

\begin{figure}[!t]
\begin{center}
\includegraphics[width=0.7\textwidth]{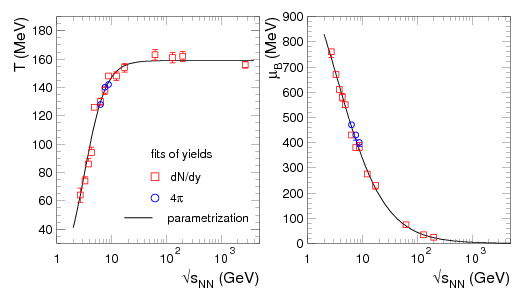}
\caption{The energy dependence of the thermal model parameters, temperature $T$ and baryon chemical potential $\mu_B$, obtained from fitting the statistical model of hadron production \cite{Andronic:2005yp} to identified hadronic yields measured in heavy-ion collisions. The figure is updated to include the most recent LHC results and it was taken from \cite{Andronic:2014zha}.}
\label{fig:pbmthermal}
\end{center}
\end{figure}

In heavy-ion collisions, the relative abundances of different hadronic species are well-described in terms of the grand canonical partition function over the full
hadronic mass spectrum. Hadronic yields are then given in terms of only two free parameters, the temperature $T$ and baryo-chemical potential $\mu_B$ of the system at
decoupling (as well as a volume parameter). A thermally equilibrated QCD system that expands and cools to the limiting temperature of a hadron gas is a system
in which all hadrons are produced with thermal abundance. Therefore, the limiting temperature $T$ and baryo-chemical potential $\mu_B$ obtained from thermal fits to 
hadronic abundances is of interest since it is thought of as tracing the QCD phase boundary. The energy-dependence of both fit parameters, shown in Fig.~\ref{fig:pbmthermal} 
shows that hadrochemical measurements at the FCC are expected to lie far within a plateau in which no further energy evolution occurs: the baryo-chemical potential in
central Pb--Pb collisions vanishes almost at top LHC energies and hence no baryon excess is expected at mid-rapidity at the FCC. Also, a limiting temperature of $\sim 160$ GeV
is reached at LHC energy already and no further evolution is expected. The simple implication of having reached these plateau values of the thermal model of hadron production 
at LHC is that all ratios of thermal hadronic abundances are expected to remain unchanged between LHC and FCC energies. While these measurements are not expected
to reveal surprises at the FCC, a confirmation of this well-tested statistical baseline for soft thermal hadron production could help to define a particularly clean baseline
on top of which dynamical mechanisms of kinetic and chemical equilibration of rarer processes (e.g. in the charm sector) could be established.


\clearpage 
\section{QGP studies: hard probes \footnote{Editors: A.~Dainese, D.~d'Enterria, J.G.~Milhano, C.A.~Salgado}}

\label{sec:HI_hardprobes}

\subsection{Jet quenching}
\label{sec:HI_quenching}

\subsubsection{Current understanding of jet quenching}
\label{sec:HI_jetquenchingintro}

The modification of jet properties in heavy-ion collisions with respect to the proton--proton case, what is commonly referred to as jet quenching, results from the interaction of jet constituents with the QGP that they traverse. 
Jet quenching was discovered at RHIC without full jet reconstruction via the strong reduction of the number of intermediate-$\pt$ hadrons~\cite{Adcox:2004mh,Arsene:2004fa,Back:2004je,Adams:2005dq}. Robust jet reconstruction, above the large and fluctuating background characteristic of heavy-ion collisions, was first carried out at the LHC~\cite{Aad:2010bu,Chatrchyan:2011sx,Abelev:2013kqa} as the combined result of a higher centre-of-mass collision energy, much improved detector capabilities and novel reconstruction techniques~\cite{Cacciari:2010te}. This has given access to a range of measurable jet properties from which the jet-QGP dynamics, and ultimately QGP properties, can be inferred. 

Over the last few years, as several jet properties were measured in heavy-ion collisions~\cite{Aad:2012vca,Aad:2013sla,Aad:2014wha,Aad:2014bxa,Aad:2015bsa,Chatrchyan:2012nia,Chatrchyan:2012gw,Chatrchyan:2012gt,Chatrchyan:2013kwa,Chatrchyan:2013exa,Chatrchyan:2014ava,Khachatryan:2015lha,Khachatryan:2016erx,Adam:2015ewa,Adam:2015doa,Adam:2015mda}, the theoretical understanding of jet--QGP interactions has evolved from the early descriptions of single parton energy loss~\cite{Baier:1994bd,Baier:1996sk,Baier:1996kr,Baier:1998yf,Zakharov:1996fv,Zakharov:1997uu,Wiedemann:2000za,Gyulassy:2000er,Guo:2000nz,Wang:2001ifa,Arnold:2000dr,Arnold:2001ms,Arnold:2002ja} towards an overall understanding of how full jets are modified by the QGP (for a review see~\cite{Mehtar-Tani:2013pia} and references therein). Several important results underlie this emergent picture of in-medium jets. First, that while the hard structure of a jet remains mostly driven by vacuum-like physics, soft jet constituents are strongly affected by the QGP experiencing large broadening effects which ultimately decorrelate them from the jet direction~\cite{CasalderreySolana:2010eh,Qin:2010mn}. Second, that the QGP presence strongly modifies intra-jet coherence properties \cite{MehtarTani:2010ma} leading to a breakdown of angular ordering for radiation induced by transverse momentum exchanges between jet and QGP. Finally, that the QGP-induced radiation pattern of a jet is driven by the number of objects within the jet that can be resolved by the QGP~\cite{CasalderreySolana:2012ef}.  

Extraction of QGP properties from jet observables relies ultimately on the availability of event generators~\cite{Armesto:2009fj,Zapp:2013vla,Lokhtin:2008xi,Casalderrey-Solana:2014bpa} that  accurately model  the interaction of jets with realistic implementations of a hydrodynamically evolving QGP. From an experimental point of view, jet observables provide versatile probes of the different energy scales as well as the space-time picture of the medium. Present LHC data show clearly the potential of these probes with higher statistics. 
The increase in energy, the abundance of probes, especially those involving electroweak bosons together with jets, and the qualitatively new processes available (e.g. boosted jets, see below) make of the FCC-hh the best-suited next machine for a deeper understanding of this physics.

\subsubsection{Hard cross sections at FCC-hh energies}
\label{sec:HI_xsections}

The large increase in energy and luminosity from the LHC to the FCC provides new tools to study the matter created in the collisions of heavy ions.
In Fig.~\ref{fig:hardXsectHIC}, cross sections for different processes
and different energies are computed with  MCFM~\cite{Campbell:2010ff}
at the highest available order. Ratios with respect to the cross
sections at top LHC energy for Pb--Pb collisions ($\sqrtsNN=5.5$~TeV) are also shown for an easier comparison of the available increases. While the increases in $Z$ or beauty production are a factor $\sim 10$ with the expected energy increase from the LHC to the FCC, these figures are much larger for top production (an increase of a factor of $\sim 80$) or the $Z$+jet with an increase of $\sim 20$. Although not shown in Fig.~\ref{fig:hardXsectHIC}, large yields of other processes of interest will be also available, as charm production, heavy quarkonia or jets in the TeV mass region.

The large increase in the top cross section, along with the larger
luminosities expected for the FCC-hh, make the case of top observables
one of the main qualitative differences with respect the LHC. It will,
in particular, allow the study of boosted $W$'s coming from the decay
of the top quarks, a unique probe of colour singlet objects traversing
a medium, even in the hadronic channel (see
Section~\ref{sec:HI_boostedtops}). A rough estimate of the rates,
based on an integrated luminosity of $\sim 30$~nb$^{-1}$ for one month  Pb--Pb run gives several million $t\overline t$ pairs, which is enough for several interesting measurements, in particular with boosted tops and $W$'s.

The large yields in $Z+$jets (several tens of millions) will also allow to study the jet quenching process with excellent calibration of the jet energy. In principle, the measurement of  the energy lost by the jet in $Z$+jet would provide a good experimental measurement of the distribution of the parton energy losses in hot QCD matter.

\begin{figure}[!t]
\begin{center}
\includegraphics[width=0.49\textwidth]{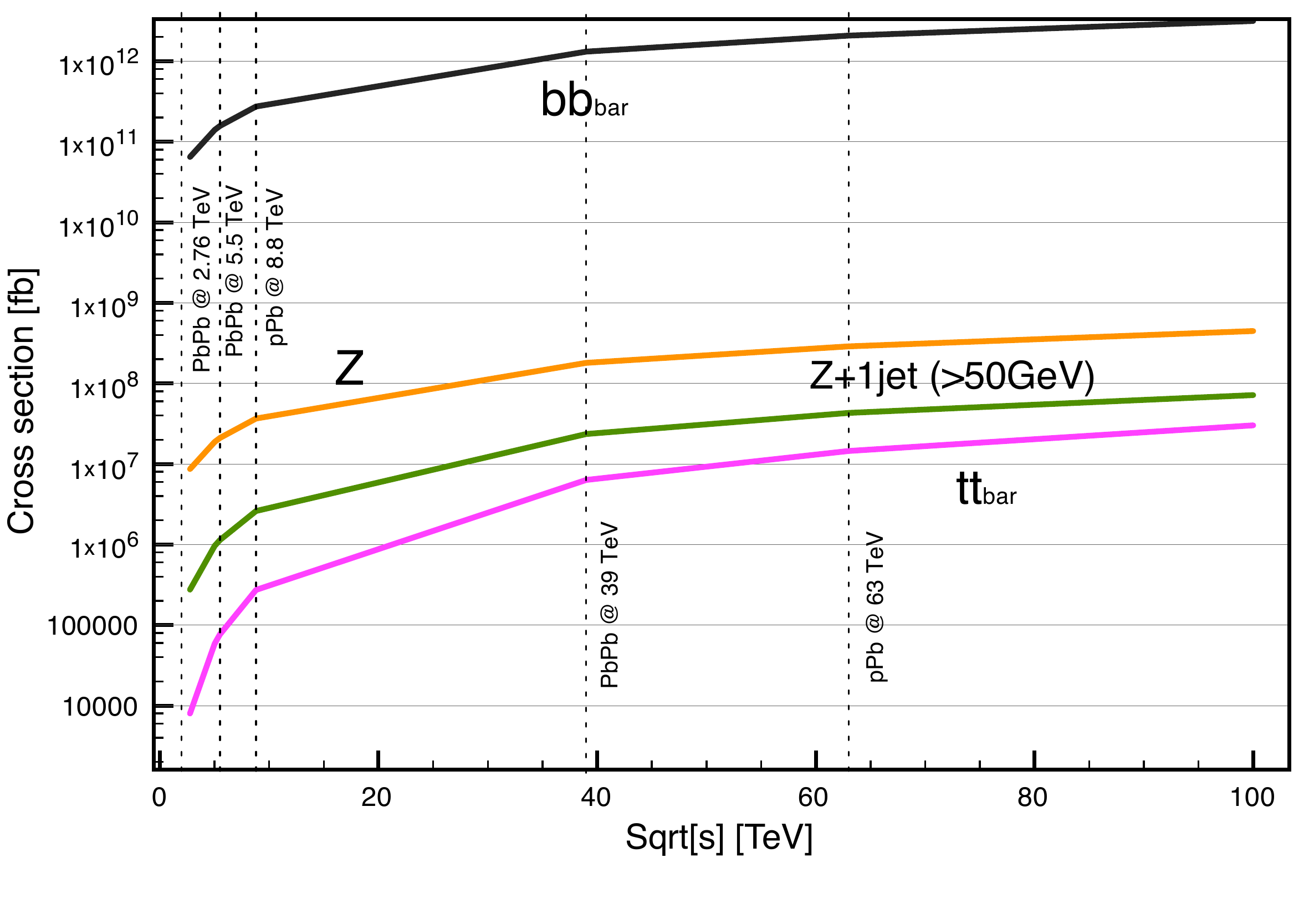}
\hfill
\includegraphics[width=0.49\textwidth]{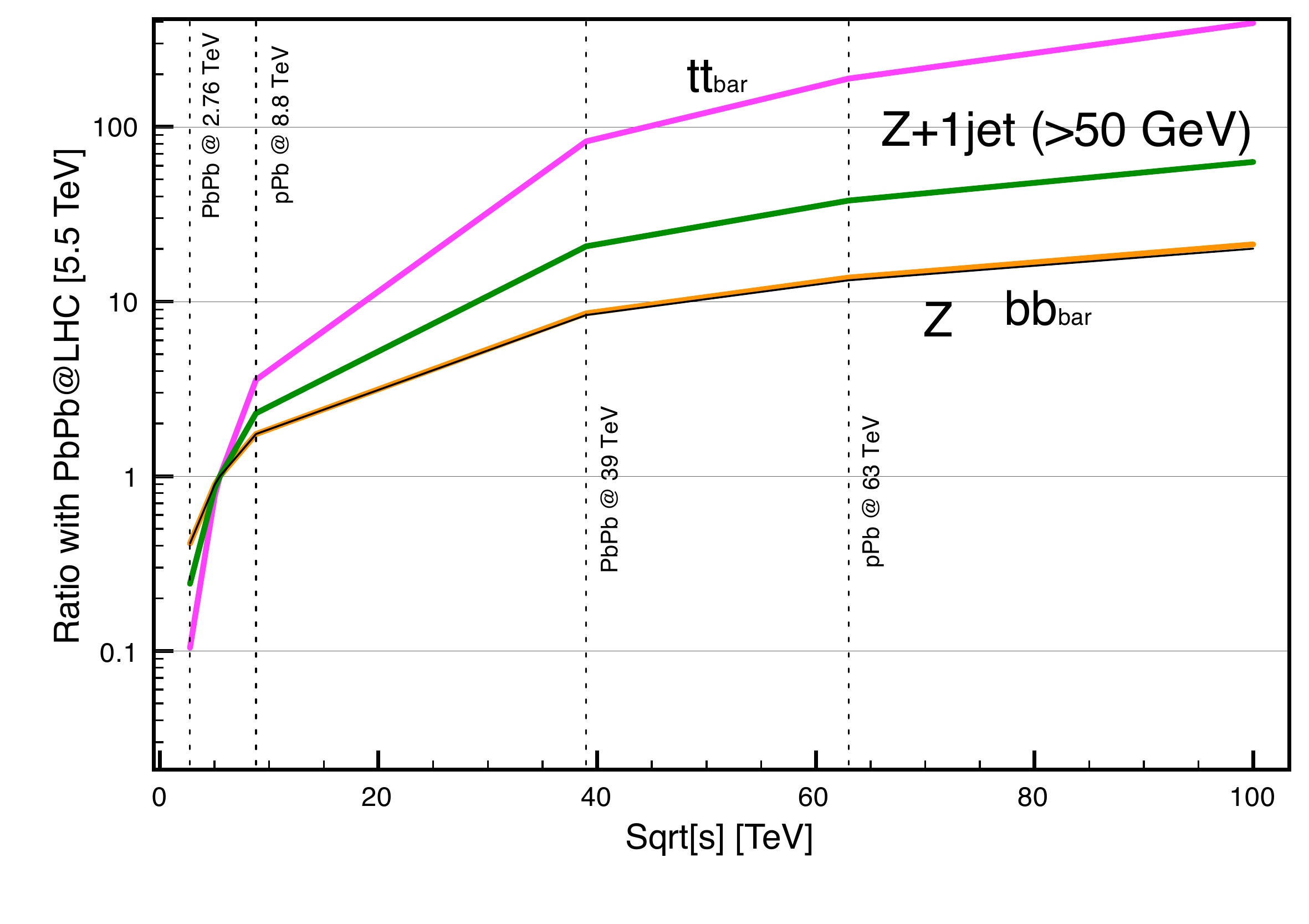}
\caption{Left:$\sqrt s$-dependence of the cross sections for hard processes of
  interest for a heavy-ion programme, calculated with MCFM~\cite{Campbell:2010ff}
at the highest available order. Right: increase factors with respect to
the cross sections at top LHC energy for Pb--Pb collisions.}
\label{fig:hardXsectHIC}
\end{center}
\end{figure}

\subsubsection{Top-quark production in p--Pb and Pb--Pb collisions}

The motivations for measurements of top quarks in heavy-ion collisions at FCC are multifold. For example, in
p--Pb collisions the cross sections efficiently probe the nuclear
gluon PDFs in a wide range in momentum fraction
$x$ at high scale $Q\sim m_{\rm t}$~\cite{d'Enterria:2015jna} (see Section~\ref{sec:HI_npdf}). In Pb--Pb
collisions, the top-quark observables are sensitive to the energy-loss of heavy quarks~\cite{Baskakov:2015nxa}
and by selecting boosted (very high-$p_{\rm T}$) top quarks one could also probe the QGP medium at slightly
later times (though still close to its formation stages) as the decays of boosted top quarks get Lorentz
time dilated (see Section~\ref{sec:HI_boostedtops}). The corresponding measurements at the LHC will be limited by the smaller production cross sections,
while at FCC energies the production cross sections are significantly higher. This is illustrated in
Fig.~\ref{fig:txsec}, which shows the energy dependence of the total top-pair and single-top cross
sections at NLO (computed with MCFM~\cite{Campbell:2010ff}) for pp, p--Pb and
Pb--Pb collisions. The large differences
between the pp and p--Pb (Pb--Pb) curves are due to scaling by $A$ ($A^2$). The effects of nuclear
modifications in PDFs (here EPS09~\cite{Eskola:2009uj}) are at the level of a few percent (see
Section~\ref{sec:HI_npdf}). 

The top quarks decay almost exclusively to $b$ quark and $W$ boson and, in a heavy-ion environment, it is the
leptonic decays of $W$ that can be best resolved from the backgrounds. The estimated measurable yields (using
nominal per-year luminosities from Section~\ref{sec:HI_machine}) with realistic analysis cuts ($b$-jets: anti-$k_{\rm T}$ algorithm with
$R=0.5$, $\pt > 30$~GeV/$c$, $|\eta| < 5$; charged leptons: $R_{\rm
  isol} = 0.3$, $\pt > 20$~GeV/$c$, $|\eta| < 5$; 
neutrinos: $\MET > 40$~GeV) and conservative 50\% efficiency for $b$-jet tagging are shown in
Table~\ref{tab:tyields}. 

\begin{figure}[!t]
\center
\includegraphics[width=0.49\columnwidth,height=7.cm]{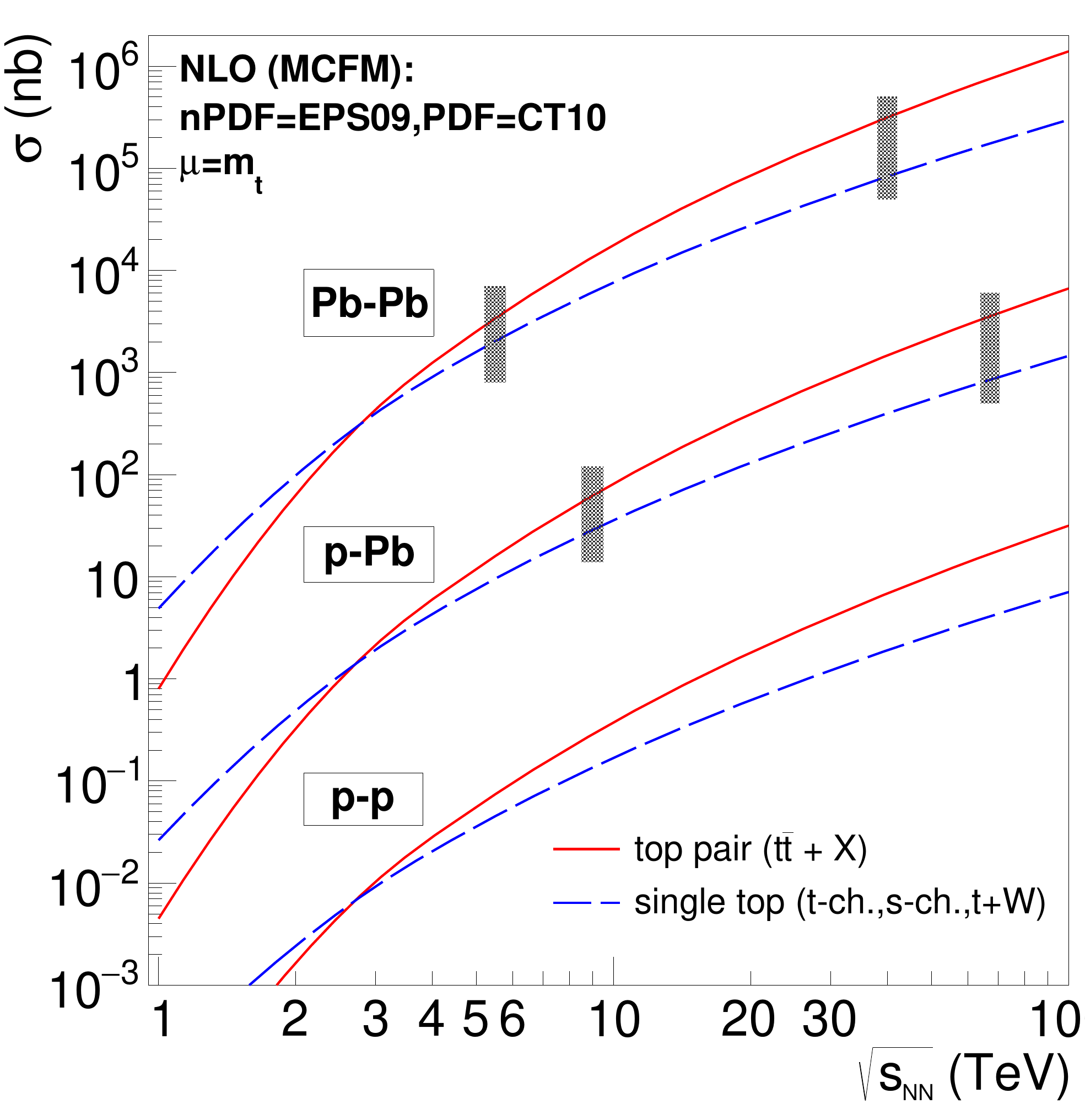}
\includegraphics[width=0.49\columnwidth,height=7.cm]{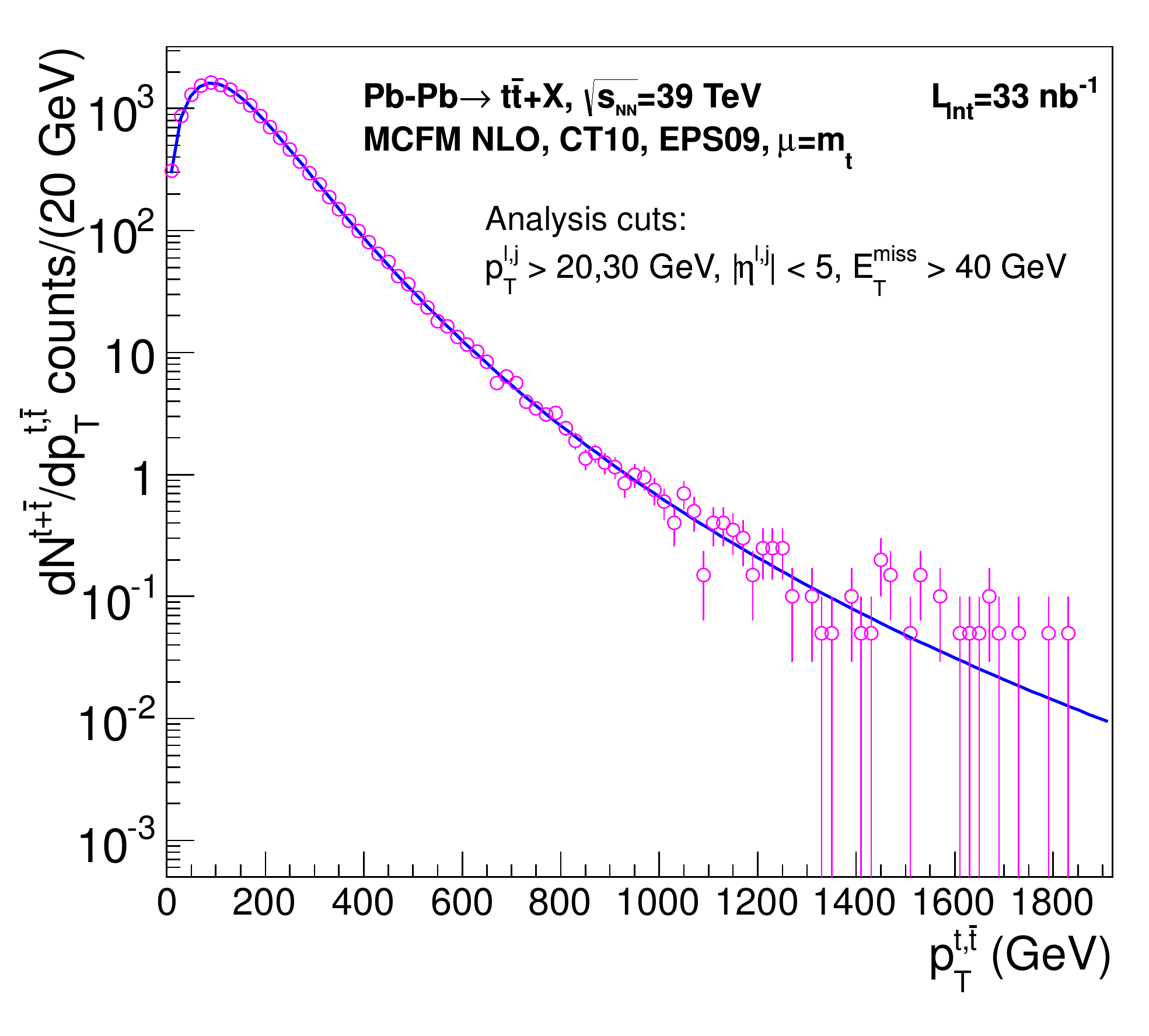}
\caption{Left: Computed total cross sections for top-pair and single-top (sum of $t$-,$s$-, and
  $t\,W$-channels) production in pp, p--Pb and Pb--Pb collisions as a function of $\sqrtsNN$
  (the boxes indicate the LHC and FCC energies). 
  Right: Expected top-quark $\pt$ distributions ${\rm d}N/{\rm d}p_{\rm T}^{t,\overline{t}}$ in \PbPb\ in the
  fully-leptonic decay modes at $\sqrtsnn=39$~TeV after acceptance and efficiency cuts. The markers
  correspond to a set of pseudodata with the statistical uncertainties
  expected for $L_{\rm int}=33$~nb$^{-1}$. 
The figures are adapted from Ref.~\cite{d'Enterria:2015jna}.}
\label{fig:txsec}
\end{figure}

\begin{table}[t]
\caption{The expected number  per run of top and antitop quarks in fully-leptonic final states, after typical acceptance cuts and efficiency
  losses (see text), for $\ttbar$ and $t\,W$ production in \pPb\ and \PbPb\ collisions at FCC energies~\cite{d'Enterria:2015jna}.  
\label{tab:tyields}}
\begin{center}
\begin{tabular}{lcccc}
\hline
\hspace{0mm} System \hspace{0mm} & \hspace{0mm} $\sqrtsNN$ \hspace{0mm} & \hspace{0mm} $\LumiInt$ \hspace{0mm} 
& \hspace{0mm} $\ttbar\to \bbbar\,\ell\ell\,\nu\nu$ \hspace{0mm} & \hspace{0mm} $t\,W \to b\,\ell\ell\,\nu\nu$  \hspace{0mm} \\ \hline
\hspace{0mm} \PbPb & 39~TeV \hspace{0mm} & \hspace{0mm} 33 nb$^{-1}$
\hspace{0mm} & \hspace{0mm} $3.1\times 10^5$\hspace{0mm} &
\hspace{0mm} $8.6\times 10^3$ \hspace{0mm} \\
\hspace{0mm} \pPb  & 63~TeV \hspace{0mm} & \hspace{0mm} 8 pb$^{-1}$
\hspace{0mm} & \hspace{0mm}  $8\times 10^5$\hspace{0mm} & \hspace{0mm}
$2.1\times 10^4$ \hspace{0mm} \\
\hline
\end{tabular}
\end{center}
\end{table}

As mentioned above, the $p_{\rm T}$ reach of top quarks in Pb--Pb collisions is of special importance for QGP
studies. To this end, Figure~\ref{fig:txsec} (right) shows the estimated $p_{\rm T}$ spectrum
of the top+antitop yields (per year) in Pb--Pb collisions for top-quark pair production, which is the most
promising channel due to the higher yields, as shown in Table~\ref{tab:tyields}. The figure indicates that one
could measure top quarks approximately up to $p_{\rm T} \approx 1.8~{\rm TeV}/c$. At mid-rapidity, $p_{\rm T}$ as large
as this would correspond approximately to a factor of 10 time dilation
in the top decay (see Section~\ref{sec:HI_boostedtops}).

\subsubsection{Boosted tops and the space-time picture of the QGP}
\label{sec:HI_boostedtops}

The large centre-of-mass energy of the FCC will provide high rates of  highly-boosted heavy particles, such as tops, $Z$ and $W$ bosons. It is expected that when these particles decay the density profile of the QGP has already evolved. By using this time delay, and by comparing the reconstructed energy to the one expected from usual energy loss processes, it should be possible to get unique insight into the time structure of the jet--QGP interaction. 

A key feature that becomes accessible at FCC energies is the role of colour coherence effects in the parton cascade in the presence of a QCD medium, as proposed in~\cite{CasalderreySolana:2012ef}. The physics is rather simple: in a given time interval $t$, fast coloured objects, either fundamental ($q$ or $g$) or composite (e.g.\,$\qqbar$, $gg$ or $qg$), probe the medium with a typical spatial resolution  $r_\perp\sim 1/Q$, where $Q$ is the transverse energy scale of the object.
For example, for a gluon that is produced in the fragmentation of a jet (hard parton) we have $1/Q\sim \theta\,t$, where $\theta$ is the angle between the gluon and the hard parton; for a $\qqbar$ pair produced in a $W$ or $Z$ decay, $\theta$ is the angle between the $q$ and the $\overline q$.
The spatial resolution $r_\perp$ has to be compared with the typical colour correlation length in the medium $L_{\rm corr}\sim 1/\sqrt{\hat{q}\,t}$. Here, $\hat{q}$ is the transport coefficient of the medium, that translates the average transverse momentum squared that particles exchange with the medium by mean-free path. When the colour correlation length of the medium is smaller than the typical transverse size of the probe, $L_{\rm corr}<r_\perp$, the different components of the coloured object (jet or $\qqbar$ pair) undergo independent colour rotations (as they are separated by more than the typical correlation length) losing coherence. This happens parametrically at the time
\begin{equation}
	t_{\rm singlet} \sim \left[ \frac{12}{\hat{q}\, \theta^2} \right]^{1/3} \, ,
	\label{eq:time}
\end{equation}
where 12 is a numerical factor depending on the actual model for the medium and the definition of the variables. For a highly boosted quark-antiquark pair this time increases, which gives the possibility to investigate further the coherence/decoherence jet quenching phenomena. Of particular interest are the colour singlet probes, as the $\qqbar$ decay products of a highly-boosted $W$ or $Z$ boson, as, in this case, the  effect of the medium will be absent.

Top--antitop events provide a unique topology to study the space-time picture of the hot QCD medium interacting with the jets. We are mainly interested here in the channel in which one of the $W$ decays leptonically while the other decays hadronically. The time dilation of the decay $W\to q\bar q$ and the color coherence discussed above ensure that also in the direction of the resulting jets, the system was traveling through the medium in a color singlet state for some time. This time increases with the boost due to both time dilation and the smaller angle of the system which determines the degree of coherence in (\ref{eq:time}).  
As shown in Fig.~\ref{fig:decay}, we obtain times in the range $\sim 0.3$--$3$~fm/$c$ when adding the time delay from Lorentz boosts and the 
time in which a singlet antenna remains in a colour coherent state (estimated for $\hat q =4~{\rm GeV^2/fm}$).

A detailed discussion about all possibilities to be exploited with this and other related observables is out of the scope of this note and will be presented elsewhere~\cite{AMSS-tops}. We study here the effect on the reconstructed masses of the top and W with different energy loss scenarios as a proof of concept of the potential of these observables to access completely novel quantities in heavy-ion collisions. Exploiting all this potential will require a good control over the energy loss of, e.g. the $b$ quarks, something which will be studied at the LHC in the coming years and for which further information could be obtained by considering together with the discussed $\ttbar\to \bbbar\,W^+W^- \to \bbbar\,\qqbar\, \ell^-\,\nu$ channel, the leptonic decay channel $\ttbar\to \bbbar\,W^+W^- \to \bbbar\,\ell^+\ell^-\,\nu\overline{\nu}$ sensitive to the energy loss of the $b$ quark.

\begin{figure}
    \centering
        \includegraphics[width=0.9\textwidth]{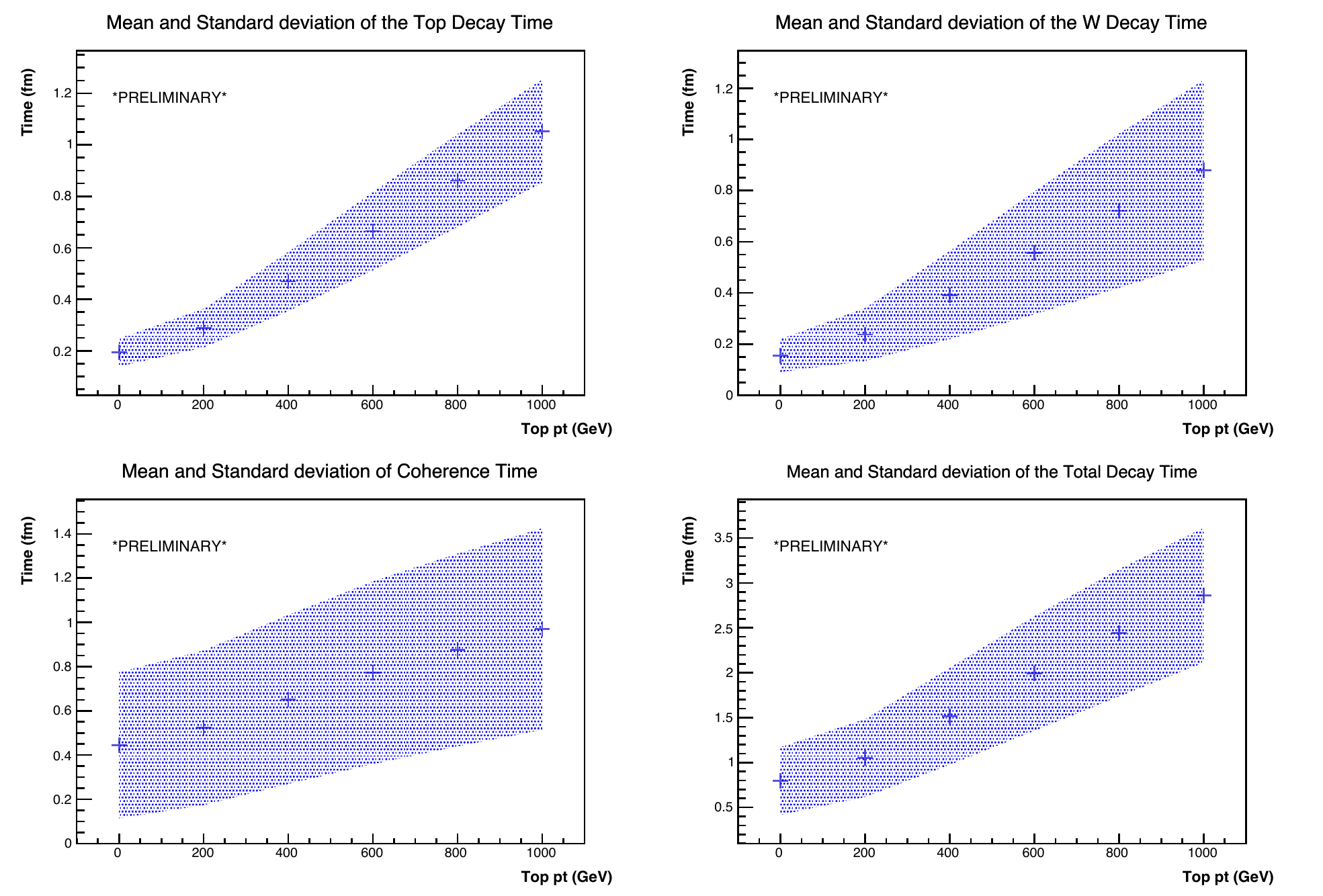}
    \caption{Average decay times of top (upper, left) and $W$ boson (upper, right), in the laboratory frame, as a function of top $\pt$. On the bottom-left, the coherence time of the $\qqbar$ pair from the $W$ decay, as given by Eq.~(\ref{eq:time}), and on bottom-right, the sum of the three components (top decay, $W$ decay and coherence time). The dots correspond to the average decay times and the shaded region to the standard deviation.}
    \label{fig:decay}
\end{figure}

Events with $\ttbar$ pairs were generated using the PYTHIA\,8 generator. 
The events were selected with the following criteria: (i) having at least 4 hadronic anti-$k_{\rm T}$ jets (2 $b$-tagged jets + 2 non-$b$-tagged jets) with $R = 0.3$, $\pt > 30$~GeV/$c$ and $|\eta| < 2.5$ (ii) having a single muon with $\pt > 25$~GeV/$c$ and $|\eta| <  2.5$.
A $b$-tagging efficiency of  $70\%$ was assumed. 
The reconstructed top and $W$ jet mass as a function of top transverse momentum at $\sqrtsNN=5.5$~TeV are shown in Fig.~\ref{fig:lhc_stat}. The same but at $\sqrtsNN=39$~TeV are shown in Figs.~\ref{fig:fcc_stat} and \ref{fig:fcc_stat2}. 
The shaded region corresponds to the statistical uncertainty estimated for $L_{\rm int}=10$~nb$^{-1}$ at $\sqrtsNN=5.5$~TeV and for $L_{\rm int}=33$~nb$^{-1}$ (a one-month run) and $L_{\rm int}=100$~nb$^{-1}$ (three runs) at $\sqrtsNN=39$~TeV.
 The number of events was determined considering the 0--10\% centrality class. Experimental effects will somewhat increase the errors shown in Figs.~\ref{fig:lhc_stat}--\ref{fig:fcc_stat2} due to the mass resolution: this remains to be studied.

\begin{figure}[!t]
    \centering
    \begin{subfigure}[b]{0.45\textwidth}
        \includegraphics[width=\textwidth]{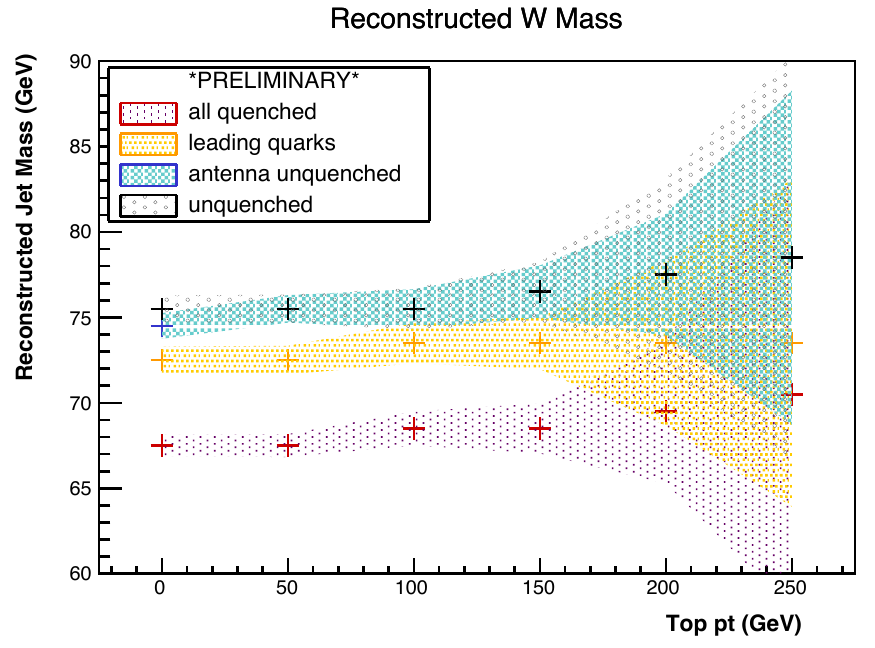}
        \caption{Reconstructed $W$ boson mass}
        \label{fig:lhc_w_stat}
    \end{subfigure}
    \begin{subfigure}[b]{0.45\textwidth}
        \includegraphics[width=\textwidth]{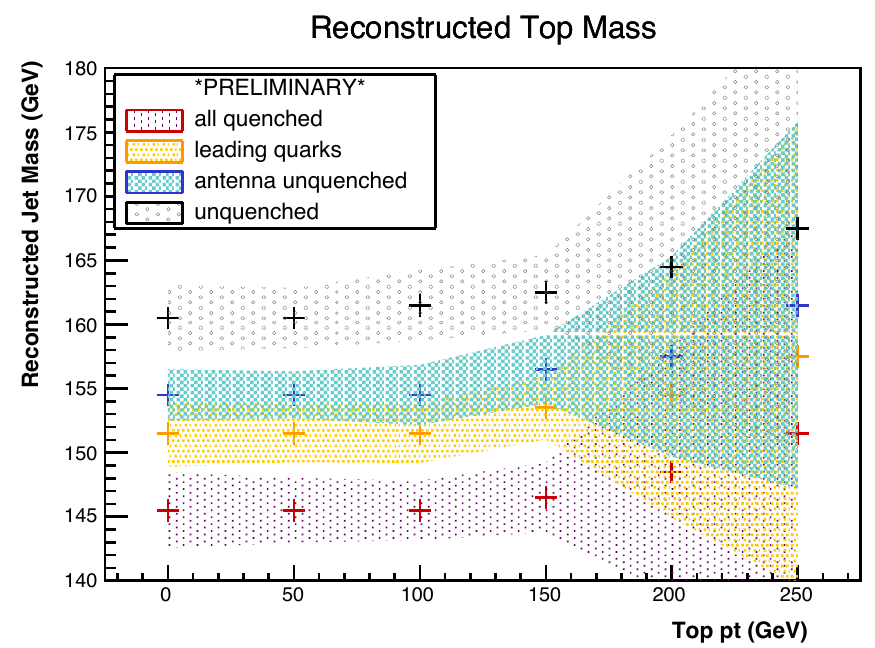}
        \caption{Reconstructed top mass}
        \label{fig:lhc_top_stat}
    \end{subfigure}
    \caption{Reconstructed masses of top and $W$ boson at the LHC energies, $\sqrtsNN = 5.5$~TeV. The shaded region corresponds to the statistical error for $L_{\rm int}=10~$nb$^{-1}$.}
    \label{fig:lhc_stat}
\end{figure}

\begin{figure}[!t]
    \centering
    \begin{subfigure}[b]{0.45\textwidth}
        \includegraphics[width=\textwidth]{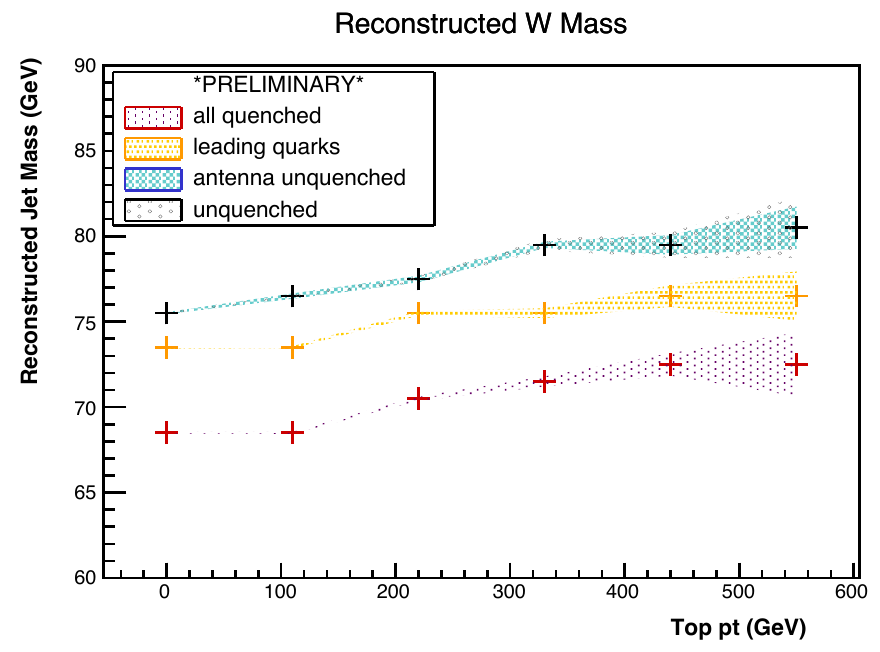}
        \caption{Reconstructed $W$ boson mass}
        \label{fig:fcc_w}
    \end{subfigure}
    \begin{subfigure}[b]{0.45\textwidth}
        \includegraphics[width=\textwidth]{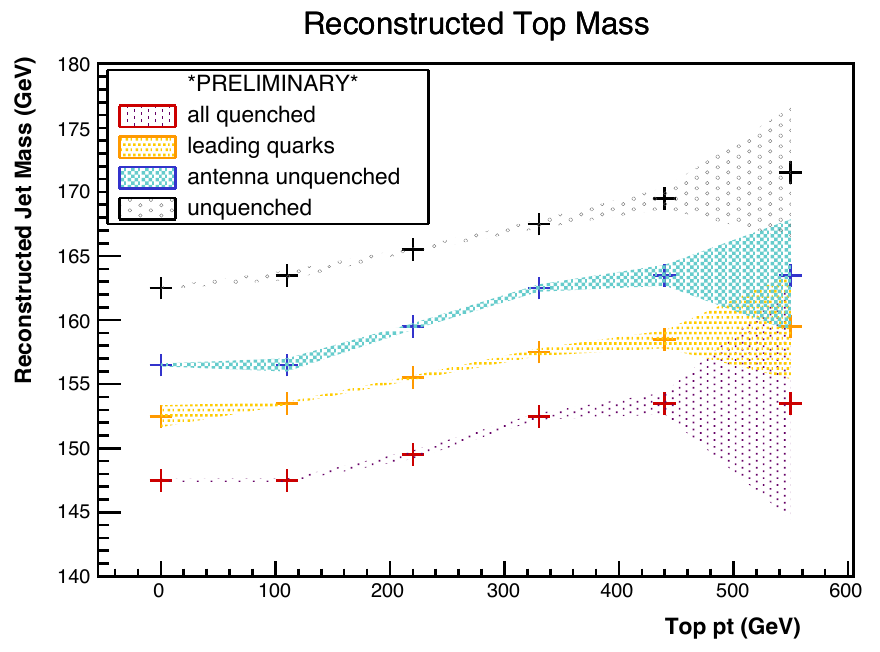}
        \caption{Reconstructed top mass}
        \label{fig:fcc_top}
    \end{subfigure}
    \caption{Reconstructed masses of top and $W$ boson at the FCC energies, $\sqrtsNN = 39$~TeV. The shaded region corresponds to the statistical error for $L_{\rm int}=33~$nb$^{-1}$.}
    \label{fig:fcc_stat}
\end{figure}

\begin{figure}[!t]
    \centering
    \begin{subfigure}[b]{0.45\textwidth}
        \includegraphics[width=\textwidth]{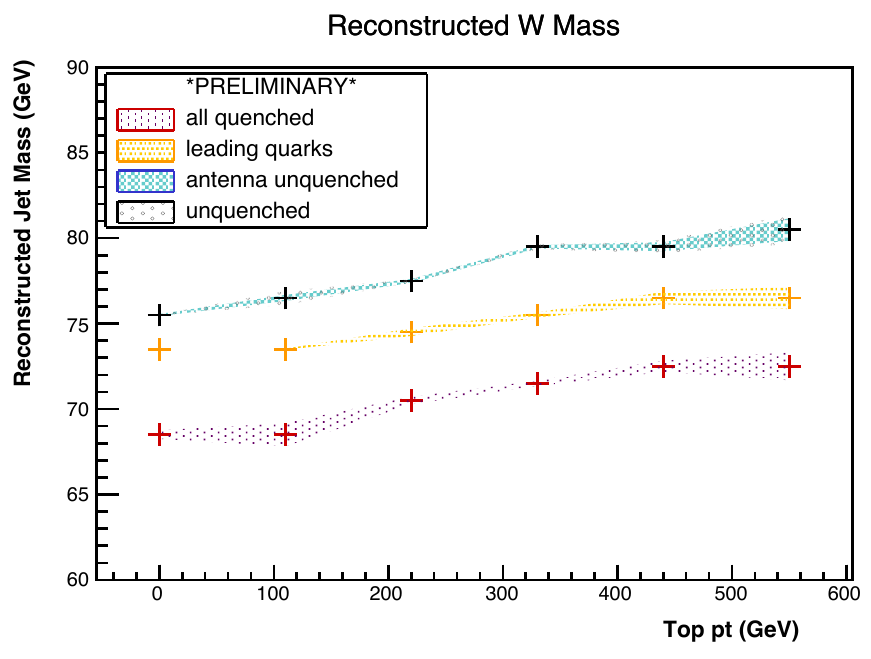}
        \caption{Reconstructed $W$ boson mass}
        \label{fig:fcc_w2}
    \end{subfigure}
    \begin{subfigure}[b]{0.45\textwidth}
        \includegraphics[width=\textwidth]{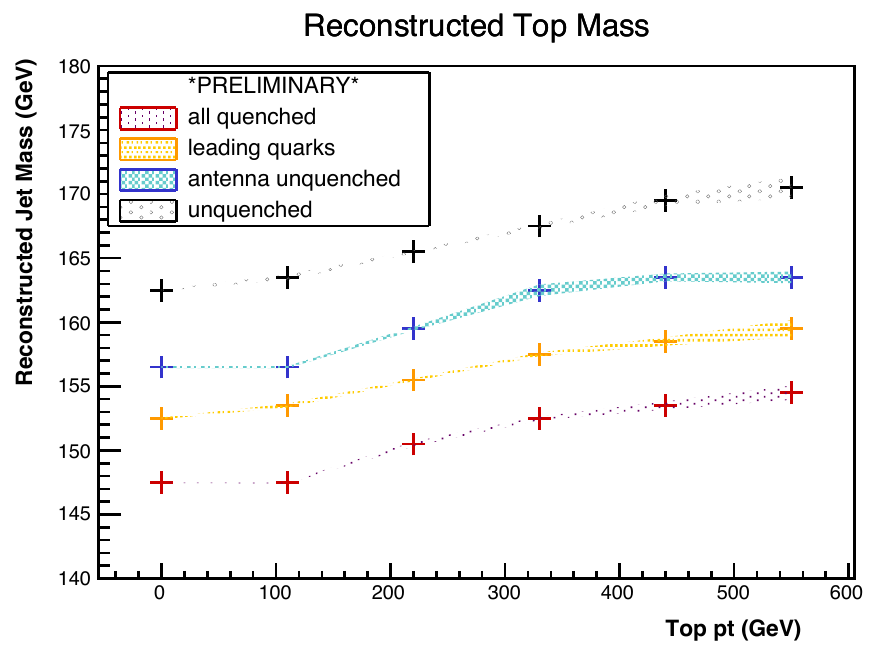}
        \caption{Reconstructed top mass}
        \label{fig:fcc_top2}
    \end{subfigure}
    \caption{Reconstructed masses of top and $W$ boson at the FCC energies, $\sqrtsNN = 39$~TeV. The shaded region corresponds to the statistical error for $L_{\rm int}=100~$nb$^{-1}$.}
    \label{fig:fcc_stat2}
\end{figure}

Energy loss was simulated considering that all quarks lose about $10\%$ of their initial four-momenta. As a simple toy model to check the effects of coherence in the reconstructed mass, this energy loss is applied equally to all quarks (\textit{all quenched}), to all quarks but the $W$ boson decay products (\textit{antenna unquenched}) or to all quarks but the leading quark-antiquark from the $W$ boson decay (\textit{leading quarks}). 
Energy loss causes a reduction of the reconstructed $W$ and top masses with respect to their true values of about 80 and 175~GeV.
The results shown in Figs.~\ref{fig:lhc_stat}, \ref{fig:fcc_stat} and \ref{fig:fcc_stat2} reveal a clear separation between what one would expect if coherence effects were completely suppressed in the presence of a medium (\textit{all quenched}) or still present within two limits (\textit{leading quarks} and \textit{antenna unquenched}). In the \textit{all quenched} case, the reconstructed masses of the $W$ and top would drop below 70 and 150~GeV, respectively. Instead, in the case of coherence they would be of the order of 75 and 155~GeV, respectively. The difference between the two limits \textit{leading quarks} and \textit{antenna unquenched} is expected to be small (a few GeV), but larger than the projected statistical uncertainties.

In summary, the reconstructed values of top and $W$ masses provide robust sensitivity to the degree of quenching. By correlating them with the expected decay and decoherence times of the top and $W$, as deduced from their $\pt$, one has an opportunity to gain unique insight into the time dependence of medium properties.

\subsection{Open and closed charm and bottom production}
\label{sec:HI_hf}


Heavy quarks (charm and bottom) are among the hard probes that have
provided important insights on the formation and the characterics of
the QGP in the heavy-ion programmes at SPS, RHIC and
LHC. A recent review of the theoretical and experimental aspects of 
heavy-flavour probes can be found in Ref.~\cite{Andronic:2015wma}.

Briefly, on the one hand, quarkonium states are sensitive to the formation and to the
temperature of a deconfined plasma via the mechanism of colour-charge
screening, which is thought to be to some extent balanced by the
recombination of heavy quarks and antiquarks from the plasma.
On the other hand, the production of hadrons with open heavy flavour
is sensitive to the QGP-induced modification of the momentum value and
direction of heavy quarks, that are created in initial hard
collisions before the formation of the QGP. In particular, it provides
information on the interaction mechanisms of heavy quarks
with the constituents of the QGP (energy loss, and gain) and on its
transport properties.

In this section, we focus on a few selected aspects that could
represent novel or particularly remarkable observations at FCC energy,
namely:
\begin{itemize}
\item large production of so-called thermal, or secondary, charm from
  interactions of light quarks and gluons within the QGP;
\item observation of an enhancement of charmonium production with
  respect to the binary scaling of the production in pp collisions, as
  consequence of (re)generation;
\item observation of a colour screening and (re)generation for the most tightly-bound
  quarkonium state, the $\Upsilon$(1S).
\end{itemize}

\subsubsection{Thermal (or secondary) charm production}

Interactions between gluons or light quarks of the QGP can lead to the
production of $c\overline c$ pairs if the energy in the centre of mass 
of the interaction is of the order of twice the charm quark mass
$\sqrt{\hat s}\sim 2\,m_c\sim 3$~GeV. This requires the energies of
the two interacting gluons (or quarks) to be of the order of $E\sim
m_c\sim 1.5$~GeV. If the gluons (or quarks) are thermalised in a
medium, their energy is of the order of the temperature $T$, with a
thermal-like exponential distribution. Therefore, for a QGP with $T$
of several hundreds of MeV (say larger than 500~MeV), there is a
significant probability that $c\overline c$ pairs are produced in
these in-medium interactions. This production is indicated as thermal, 
or secondary, in contrast with the primary production that occurs in
initial hard-scattering processes between partons of the two incident nuclei.


In Section~\ref{sec:HI_global} we have estimated the parametric dependence
of $T$ on time and on the measured transverse energy density
$T(t)\sim [({\rm d}E_{\rm T}/{\rm d}\eta)/t]^{1/4}$.
From this simplified estimate, we observe that the temperature at a
given time increases slowly with the $E_{\rm T}$ density (e.g. about
20--30\% from LHC to FCC with the $E_{\rm T}$ density estimated in
Section~\ref{sec:HI_global}). However, the thermalization time $\tau_0$ of the QGP is 
expected to decrease substantially when $\sqrtsNN$ increases, 
typically by a factor of 2 from LHC to FCC (see e.g.~\cite{Zhou:2016wbo}).
Therefore, the overall increase of the initial temperature $T_0$ of the QGP from LHC to
FCC can be about 50\%. 
In Ref.~\cite{Zhou:2016wbo} a detailed hydrodynamical calculation gives 
$T_0=580~$MeV at initial time $\tau_0=0.6$~fm/$c$ for LHC ($\sqrtsNN=5.5~\tev$) and
$T_0=840~$MeV at $\tau_0=0.3$~fm/$c$ for FCC.
With these QGP temperatures a sizeable fraction of the gluons and
light quarks have energies larger than the charm quark mass 
and $c\overline c$ pairs can be produced in their interactions. This
production is concentrated in the initial  $\sim 1$~fm/$c$ of the QGP evolution.

Predictions for the production of thermal charm at LHC and FCC energies
were reported by three groups: BAMPS~\cite{Uphoff:2010sh,bampsFCCws}, 
Ko et al.~\cite{Zhang:2007dm,Liu:2016zle} and Zhou el al.~\cite{Zhou:2016wbo}.
In the BAMPS transport model, which is based on the Boltzmann
equation, 
secondary charm production is calculated
using leading order (LO) gluon fusion and $q\overline q$ annihilation
processes. The calculations by Ko et al.\,and Zhou et al.\,use dynamical
kinetic equations where the charm quark
density in the medium is evolved in time using a gain term 
(secondary $c\overline c$ production) and a loss term ($c\overline c$
annihilation).
The gain and loss terms are calculated at next-to-leading order (NLO)
considering also $2\to 3$ and $3\to 2$ processes, respectively.
It has been shown~\cite{koFCCws,zhouFCCws} that the NLO cross
sections are significantly larger than the LO ones. Therefore, we
mainly focus on these predictions.

Figure~\ref{fig:thermalcharm} shows the results by Zhou el al. (left)
and Ko et al.\,(right) for the time-dependence of the $c\overline c$
rapidity density at mid-rapidity. The value at the initial time
$\tau_0$ corresponds to the initial hard-scattering cross section, 
which is taken from FONLL calculations~\cite{Cacciari:2012ny} with a correction
for PDF nuclear shadowing based on EKS98~\cite{Eskola:1998df} by Zhou et al.\,and
from the PYTHIA\,6 event generator~\cite{Sjostrand:2001yu} by Ko et al., with
slightly different values. Both calculations show a rapid increase
after $\tau_0$ with a final value that is larger by up to 80\% than
the hard-scattering value. The increase obtained for top LHC energy 
is of about 15\%.
The predictions by BAMPS, with LO secondary
charm production cross sections, show smaller
increases of about 30\% and 10\% at FCC and LHC energies, respectively~\cite{bampsFCCws}.

The thermal charm production would result in an enhancement of charmed hadron 
production at very low $p_{\rm T}$, with respect to the expectation
from binary scaling of the production in pp collisions, after
correction for the nuclear initial-state effects (PDF modification),
that should be measured using proton--nucleus collisions. 
This enhancement provides a handle on the
temperature of the QGP. 

The abundance of charm quarks also has an effect on the QGP equation of the state, which includes a dependence on the number of degrees of freedom.
Inclusion of the charm quark in the lattice QCD calculations results in a sizeable 
increase of $P/T^4 \propto n_{\rm d.o.f.}$ for temperatures larger than about 400~MeV. This
was discussed in more detail in Section~\ref{sec:HI_charmlattice}.

\begin{figure}[!t]
\begin{center}
\includegraphics[width=0.45\textwidth]{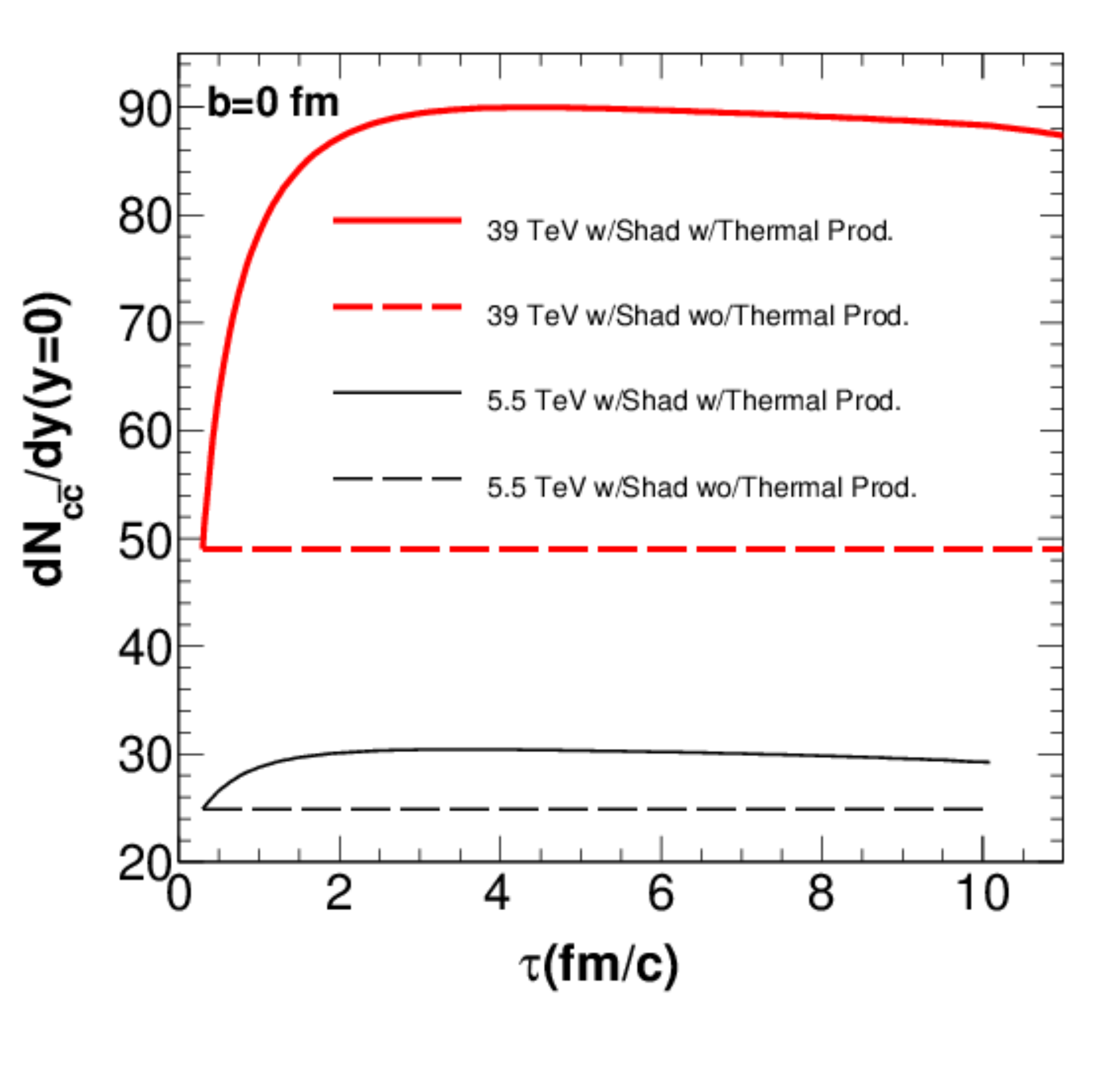}
\hfill
\includegraphics[width=0.54\textwidth]{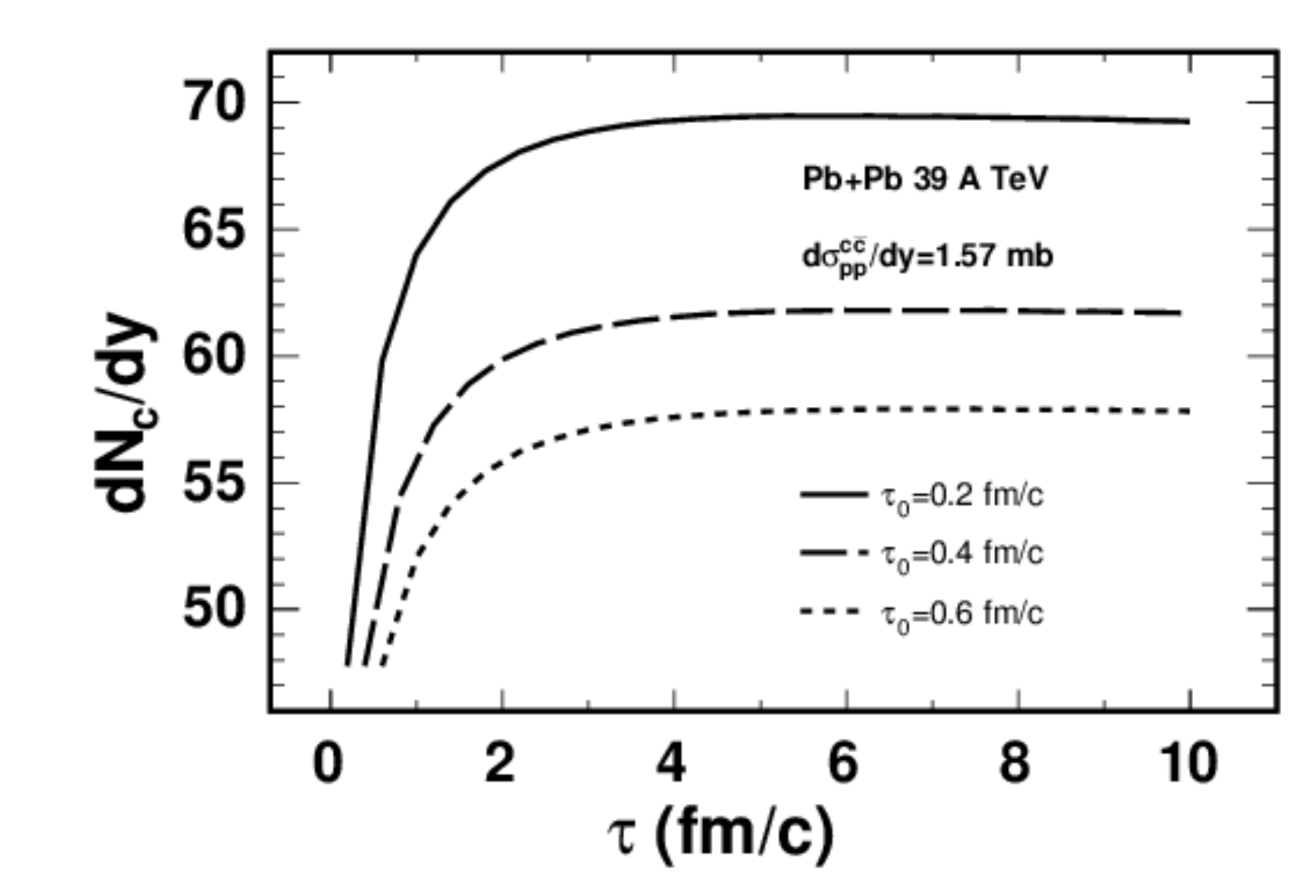}
\caption{Time-evolution of the charm and anti-charm quark pair yield
  (per unit of rapidity at midrapidity) for central Pb--Pb collisions
  at $\sqrtsNN=39$~TeV: results by Zhou et al.~\cite{Zhou:2016wbo}
  (also shown for $\sqrtsNN=5.5$~TeV) and by
  Ko et al.~\cite{Liu:2016zle} (the results by Ko et al. at
  5.5~TeV are similar to those by Zhou et al.).}
\label{fig:thermalcharm}
\end{center}
\end{figure}

\subsubsection{J/$\psi$ enhancement from (re)generation}

The measurements of the nuclear modification factor of J$/\psi$ at the LHC~\cite{Adam:2015isa,Adam:2015rba,Abelev:2013ila,Chatrchyan:2012np} 
are described by models that include dissociation caused by
colour-charge screening and a contribution of recombination
(usually denoted (re)generation) from deconfined $c$ and $\overline c$
quarks in the QGP. 
In particular, this contribution describes the larger nuclear
modification factor $\RAA$ (smaller suppression) at the LHC with respect
to RHIC, the larger $\RAA$ at low $\pt$ than at high $\pt$ 
and the fact that $\RAA(\pt>0)$ at the LHC is almost constant
from semi-peripheral to central collisions.
There are essentially two classes of such models: 
the kinetic transport models~\cite{Liu:2009nb,Zhao:2011cv} calculate the
time-evolution of the J/$\psi$ yield using loss (dissociation) and
gain (recombination) terms; the Statistical Hadronization
Model~\cite{Andronic:2011yq} assumes complete dissociation of the
initially-produced J/$\psi$ mesons and computes the J/$\psi$ yield
considering statistical hadronization at the chemical freeze-out
temperature.

In both approaches the (re)generation contribution is proportional to
the rapidity density of $c\overline c$ pairs in the QGP. It is,
therefore, clear that this contribution is predicted to be much larger
at FCC than LHC energies, as a consequence of a) the larger hard-scattering
production cross section of $c\overline c$ pairs and b) the possible 
sizeable thermal production, that we discussed in the previous
Section.
In particular, the hard-scattering production cross section is
expected to increase by a factor about 2--2.5, depending on the considered
nuclear modification of the PDFs (e.g. with FONLL calculations and 
EKS98 nuclear PDFs the factor is about 2, as shown in Fig.~\ref{fig:thermalcharm}-left).
Thermal production could lead to an additional increase by a factor 
about 1.5, as shown in the previous Section.
This could lead to the observation of an enhancement of J$/\psi$
production with respect to binary scaling of the yield in pp
collisions, i.e. $R_{\rm AA}>1$, which would be a striking evidence of 
$c\overline c$ recombination from a deconfined QGP.

Figure~\ref{fig:JpsiRAA} shows the predicted J$/\psi$ $\RAA$ at FCC
energy, as obtained with the Statistical Hadronization Model
(left)~\cite{jpsiFCCshm} and with the kinetic transport model by Zhou
et al.\,(right)~\cite{Zhou:2016wbo}. The Statistical Hadronization Model
uses only the initial hard-scattering $c\overline c$ cross section,
without a thermal contribution. Even in this conservative scenario,
the model predicts $\RAA(\pt>0)>1$ in central collisions and an
increase of about 40\% with respect to top LHC energy.
The kinetic model by Zhou et al.\,uses also the thermal $c\overline c$
contribution as shown in Fig.~\ref{fig:thermalcharm} (left).  This
model predicts an enhancement at low $\pt$ with $\RAA$ values between 
1.1 and 1.6, depending on the assumed nuclear modification of the PDFs.

\begin{figure}[!t]
\begin{center}
\includegraphics[width=0.45\textwidth]{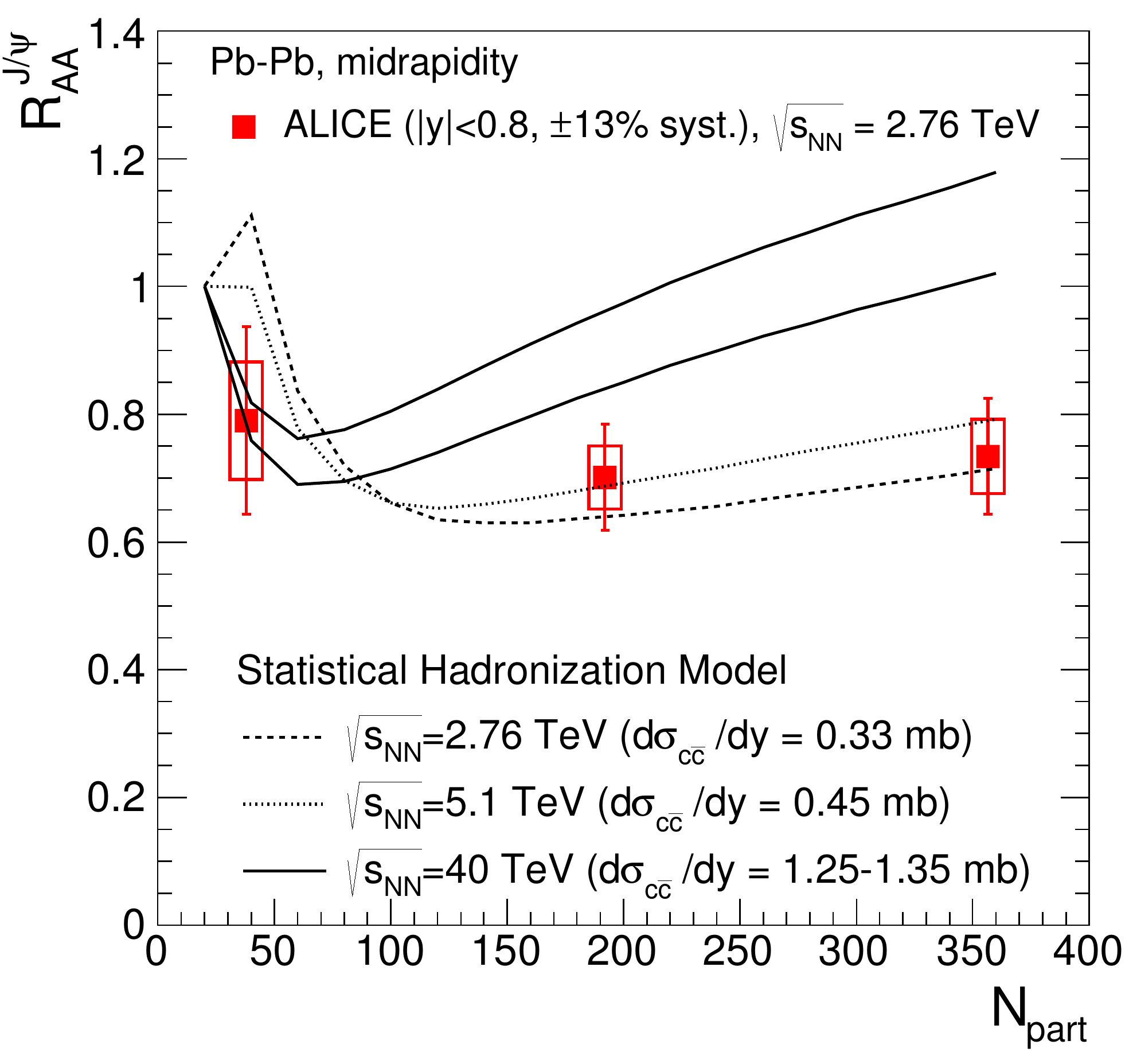}
\hfill
\includegraphics[width=0.54\textwidth]{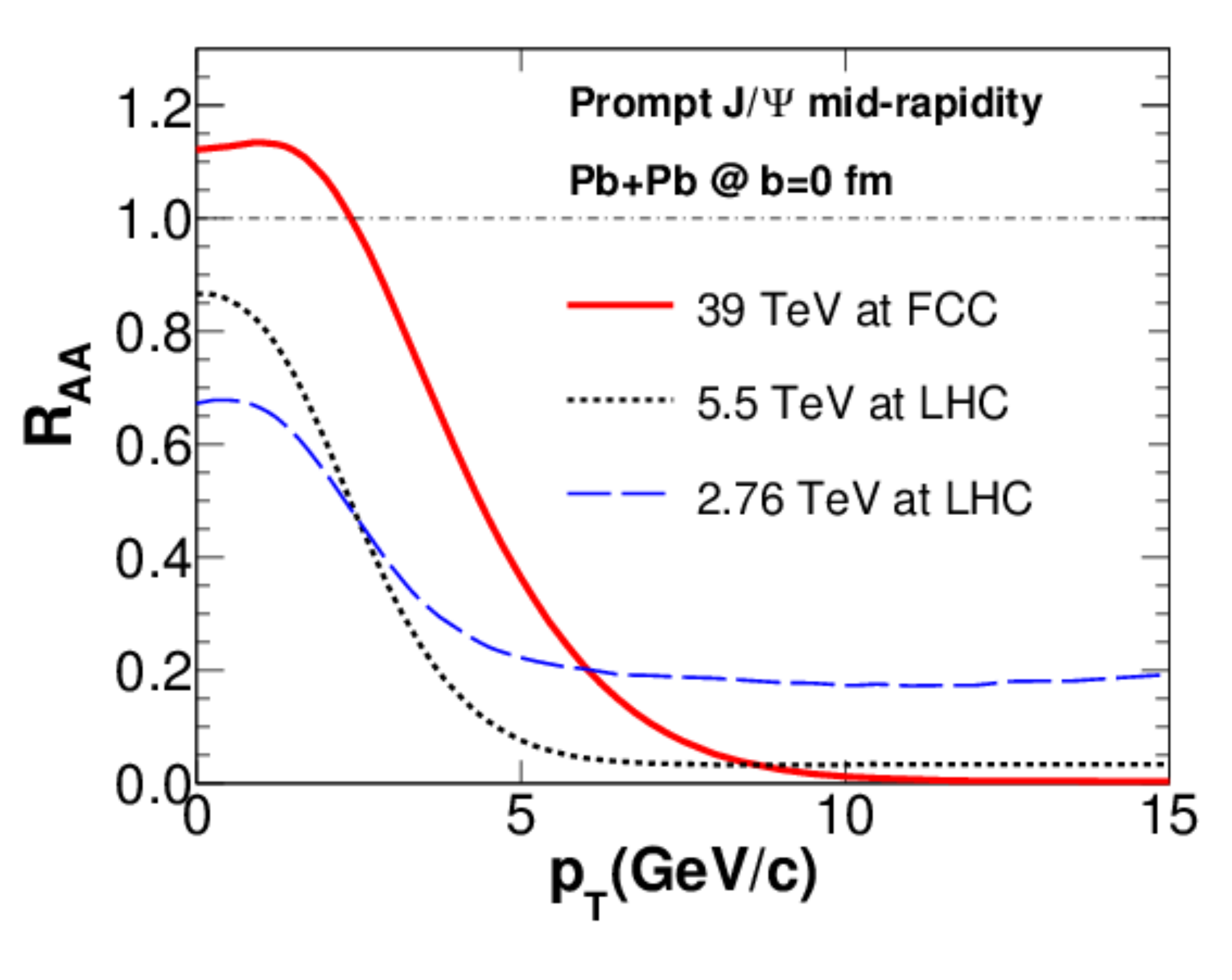}
\caption{Nuclear modification factor $\RAA$ of J$/\psi$ mesons 
at LHC and FCC energies. Left: $\RAA(\pt>0)$ as a function of centrality (number
of nucleons participating in the collision) from the
Statistical Hadronization Model~\cite{Andronic:2011yq,jpsiFCCshm}, without 
considering thermal charm production. Right: $\RAA(\pt)$ in 
central collisions from the kinetic model by Zhou et
al.~\cite{Zhou:2016wbo}.}
\label{fig:JpsiRAA}
\end{center}
\end{figure}

\subsubsection{Colour screening and (re)generation for $\Upsilon$(1S)?}

The measurement of $\Upsilon$ production would be particularly interesting
at the high
energies and temperatures reached at the FCC.
The LHC data are consistent with a scenario in which the excited
states 2S and 3S are partially or totally suppressed by colour
screening, while the 1S, which is the most tightly bound state, has no
or little direct melting. Its suppression by about 50\% can be
attributed to the lack of feed-down from the (melted) higher states
(see e.g. Ref.~\cite{Andronic:2015wma} for a recent review).
At FCC energies, on the one hand, the temperature could be large
enough to determine a full melting even of the tightly-bound 1S state,
on the other hand the large abundance of $b\overline b$ pairs in the
QGP could induce substantial $\Upsilon$ (re)generation.

Results from a recent lattice-QCD calculation of spectral functions
of the bottomonium states are shown in the left panel of 
Fig~\ref{fig:Raa_ups}: they have been obtained for different,
increasing temperatures, indicated in units of the critical temperature
$T_c$ for QGP formation ($T_c\approx 155$~MeV).
The suppression of the excited states is
quite evident, as well as the persistence of the fundamental $\Upsilon$
state, up to about twice the critical temperature. However, should
this trend persist, one could anticipate a strong suppression 
of the fundamental
state $\Upsilon$(1S) at FCC energies.  

\begin{figure}[!t]
\includegraphics[width=10.5cm]{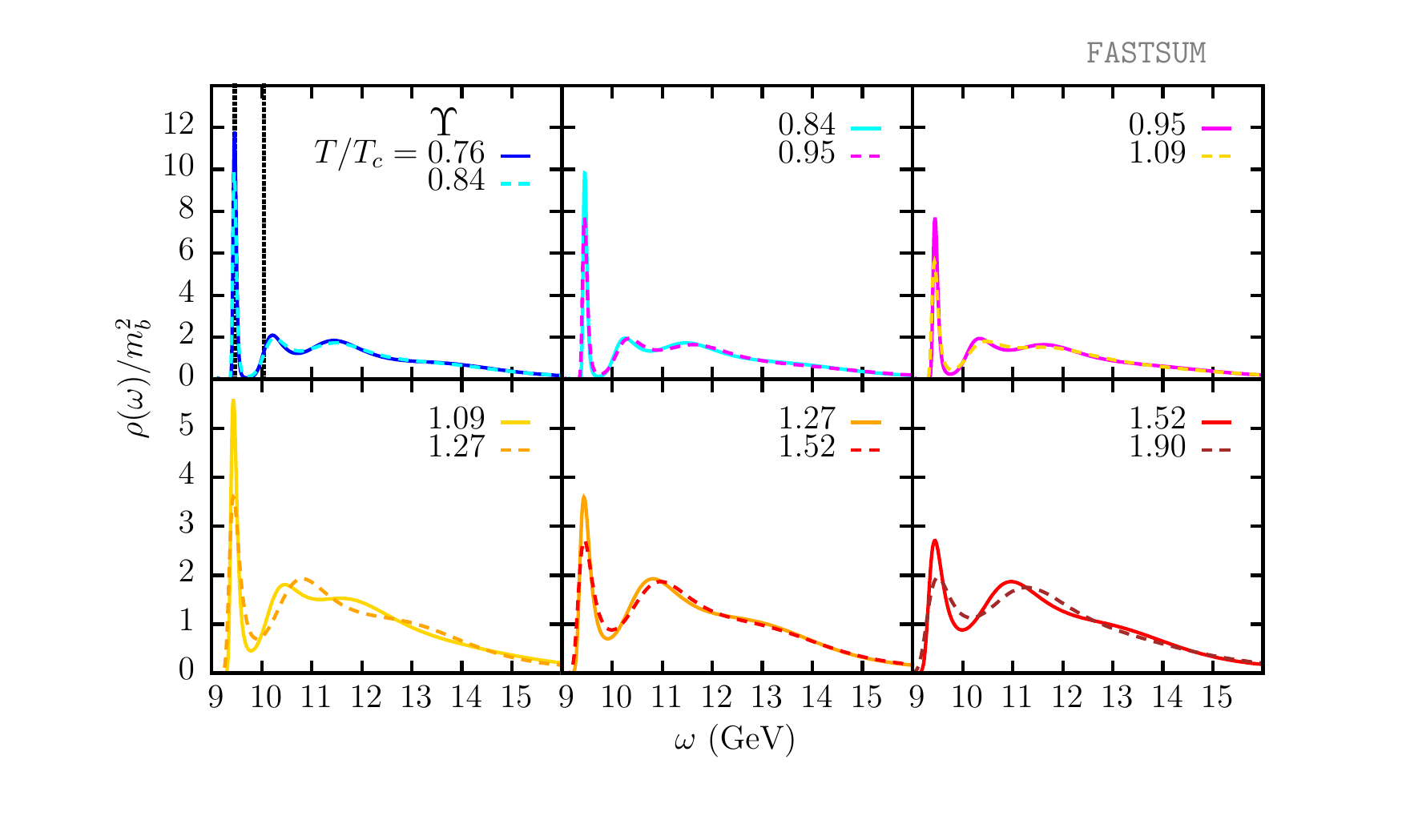}
\includegraphics[width=5.5cm]{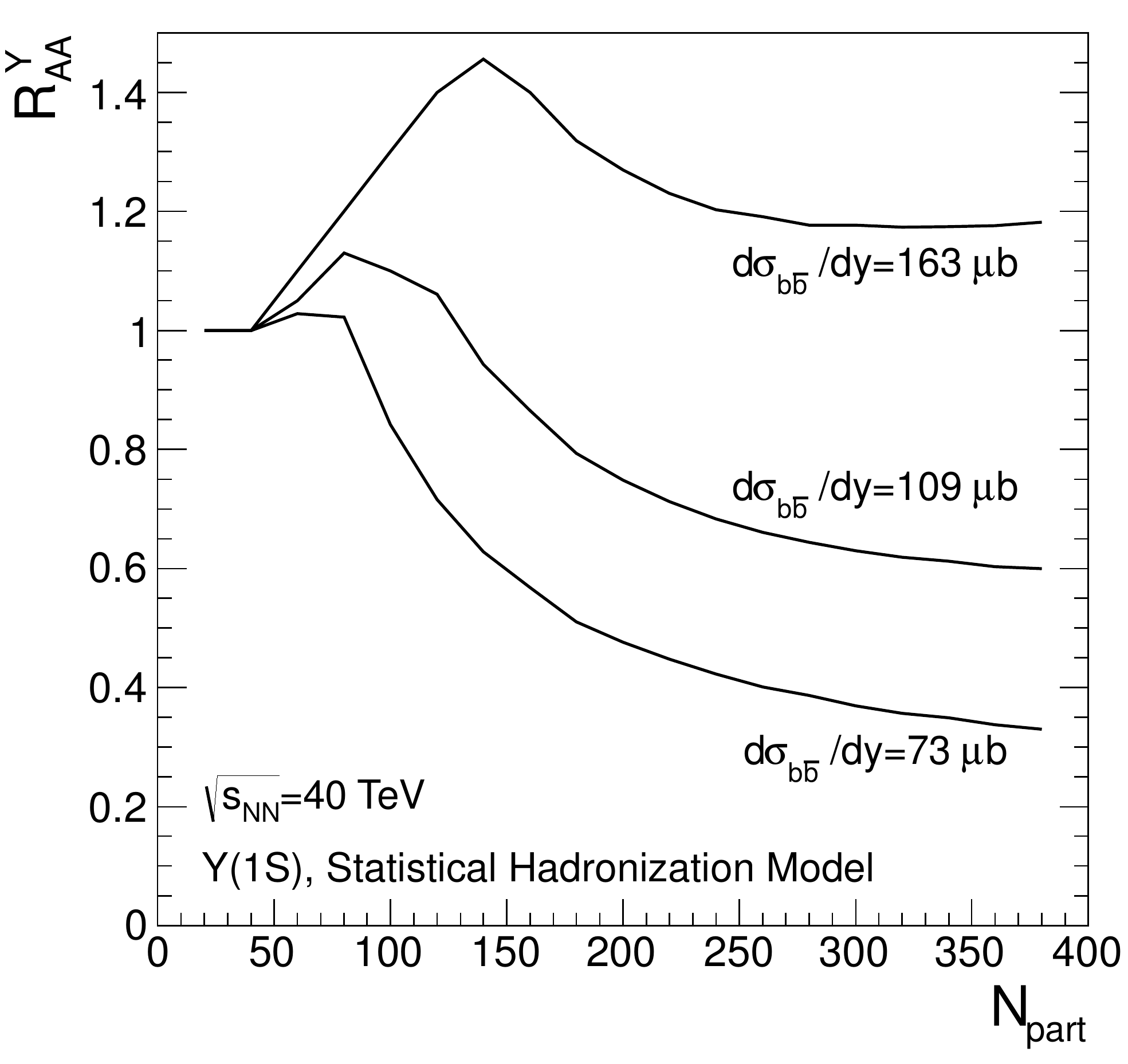}
\caption{Left: the spectral functions for the $\Upsilon$ at different
  temperatures, obtained using the maximum entropy method, from
  Ref.~\cite{Aarts:2014cda}; the (leftmost) 1S state is progressively
  disappearing
with increasing temperature. Right: $R_{\rm AA}$ for $\Upsilon$(1S) in Pb--Pb collisions at
  $\sqrtsNN=40$~TeV as predicted by the Statistical Hadronization Model~\cite{Andronic:2011yq,upsilonFCCshm}.}
\label{fig:Raa_ups}    
\end{figure}

Accurate estimates of the bottomonium spectral functions up to higher 
temperature of 4--5 $T_c$ (i.e. 700--800~MeV)
are needed to
refine these expectations. 
For what concerns accuracy, while general, qualitative features of the 
results are robust, and all consistent with the sequential suppression 
scenario,
quantitative results are very difficult to obtain. Systematic comparisons
of different methods are now starting and will produce robust
results soon~\cite{Kim:2015rdi,Burnier:2015tda,Shu:2015tva}. This is 
especially likely for bottomonium, whose analysis
can be done within the NRQCD (non-relativistic QCD) 
formalism, which remains valid in the range
of temperatures explored at the FCC.
A further challenge is
posed by the high temperature reached at the FCC: 
on the lattice the temperature is
realised as the inverse of the temporal extent of the box. Since many points
are needed in the same temporal direction, high temperatures imply very fine
lattices, hence a very large number of spatial points, needed to approximate
an infinite spatial volume. To some extent this issue can be dealt with by
use of asymmetric lattices, but such simulations remain
computationally very demanding, and so far the temperatures have been limited
to $T < 2\, T_c$. A very fascinating possibility is to be able to make
contact with high temperature perturbation theory, matching lattice
spectral functions at FCC temperatures with perturbative ones. 
In summary, an accurate calculation of bottomonium spectral functions
at FCC temperatures requires advances in methodology, which are
underway, to keep systematic errors under control, 
and a consistent investment in computer time to reach the required high temperatures.

Another important question is whether the $\Upsilon$ states reach
equilibrium with the surrounding QGP constituents. This is
prerequisite for colour-screening to apply and it is implicitly
assumed in the lattice QCD calculations of the spectral functions.
Measurements of $\pt$ distributions and elliptic flow of bottomonium
states of $B$ mesons would shed light on this aspect.

The possibly dramatic effect of (re)generation of bottomonia from $b$
and $\overline b$ quarks 
is illustrated by the prediction of the Statistical
Hadronization Model~\cite{Andronic:2011yq,upsilonFCCshm} for the $\RAA$ of $\Upsilon$(1S) as a function of centrality, shown in the right panel of  
Fig.~\ref{fig:Raa_ups}. Like for charmonium, this model assumes full
melting of the initially-produced bottomonia and generation at the
phase boundary. The predictions are calculated for values of
$d\sigma_{b\overline b}/dy$ in nucleon--nucleon collisions at $\sqrt
s=40$~TeV ranging from 73 to 163~$\mu$b, as obtained from the MNR NLO
calculation~\cite{Mangano:1991jk}
with usual parameter variations and without nuclear modification of
the PDFs (nuclear shadowing of small-$x$ PDFs is expected to decrease
the cross section by about 60--90\%). 
These cross sections result in a total number of $b\overline b$ pairs
ranging from 15 to 40 in central Pb--Pb collisions.
Depending on the value of the bottom cross section, 
the $\Upsilon$(1S) $\RAA$ in central Pb--Pb collisions is predicted
to range between 0.3 and 1.2.

The role of the two effects ---degree of survival of initial bottomonia and contribution of
(re)generation--- could be separated by means of precise measurements of
the $b\overline b$ cross section, an essential ingredient for
(re)generation calculations, 
and of the $B$ meson and $\Upsilon$ $\RAA$ and elliptic flow $v_2$.
The elliptic flow measurements would be particularly important because 
the regenerated $\Upsilon$ states could exhibit a $v_2$ such that $0<v_2^{\Upsilon}<v_2^B$.

\clearpage 
\section{Small-$x$ and nPDF studies \footnote{Editors: N.~Armesto, D.~d'Enterria, M.~van~Leeuwen}}
\label{sec:HI_smallx}

\newcommand{\pizero}{\ensuremath{\pi^{0}}}
\newcommand{\ptjet}{\ensuremath{p_\mathrm{T,jet}}}

\subsection{Small-$x$ and nPDF studies in hadronic p--A and A--A collisions}
\label{sec:HI_smallxhadro}

\subsubsection{Introduction: small $x$ and factorisation}
\label{sec:HI_smallxintro}

More than 30 years ago, the idea of parton saturation was proposed~\cite{Gribov:1984tu,Mueller:1985wy}: with
BFKL~\cite{Kuraev:1977fs,Balitsky:1978ic} linear evolution, the multiplication of partons with small values of momentum fraction $x$ leads to parton densities so high that non-linear dynamics (gluon recombination, multiple scattering, $\dots$) becomes important. Such non-linear effects would tame the growth of parton densities from power-like to logarithmic, a phenomenon known as ``saturation''.

In the case of proton--nucleus and nucleus--nucleus collisions, 
where nuclei with large mass number $A$ 
are involved, the non-linear effects are enhanced by the larger density of 
gluons per unit transverse area of the colliding nuclei.
The high density of gluons at 
small $x$ and small $Q^2$ induces a suppression of the observed hard scattering
yields 
with respect to expectations based on a scaling with the
 number of binary nucleon--nucleon collisions. 
This reduction affects the kinematic region dominated by small-$x$ gluons: 
low transverse momentum $\pt$ and forward rapidity $y$, since, at leading 
order, we have $x\approx \pt\exp(-y)/\sqrtsNN$.

Data from Deeply Inelastic Scattering (DIS) experiments on nuclear targets were analysed in terms of nuclear Parton Density Functions (nPDFs) within the linear-evolution DGLAP framework. The results are normally reported as 
a modification ratio $R_i$ of the parton distribution functions of the nucleon
in the nucleus, $f_i^{\rm A}(x,Q^2)$, with respect to those of
the free nucleon, $f_i^{\rm N}(x,Q^2)$,
\begin{equation}
  \label{eq:shad}
  R_i^{\rm A}(x,Q^2) = \frac{f_i^{\rm A}(x,Q^2)}{f_i^{\rm N}(x,Q^2)}\, ,
\end{equation}
where $i = q_{\rm v},\,q_{\rm sea},\,g$ for valence quarks, 
sea quarks, and gluons. We have shadowing, $R_{g}^{\rm A}<1$, 
for $x\, \lsim\,  5\times 10^{-2}$. Significant differences between nPDFs of nuclei and the PDFs of free protons were found, both at high $x$ (the `EMC effect' \cite{Ashman:1992kv}) and at low $x$, where a depletion is seen which is referred to as `nuclear shadowing'. 

The usage of nPDFs allows some of 
the high-density effects at small $x$ to be absorbed in the non-perturbative description of the PDFs within the framework of 
perturbative QCD collinear factorization. However, factorization 
is expected to break down when the gluon phase-space becomes saturated.
In these conditions, in the collision with an incoming projectile parton, 
the partons in the target nuclear wave function at small $x$ would 
act coherently, not independently as assumed with factorization.
In the limit, they may form 
a Colour Glass Condensate (CGC, see e.g. Ref.~\cite{Gelis:2010nm} for a recent review): 
a system, that can be described in analogy to a 
spin glass, where gluons (colour charges) have a large occupation number, as in a condensate.
The CGC theory relies on the resummation of powers of parton density. 

The onset of saturation is usually discussed in terms of the 
so-called saturation scale $Q_{\rm S}^2$, defined as the 
scale at which the transverse area of the nucleus is completely saturated and gluons start to overlap.
This happens when the number of gluons, $\sim A\,xg(x,Q^2_{\rm S})$, multiplied by the typical gluon size, $\sim 1/Q^2_{\rm S}$, is equal to the transverse area, $\sim\pi R^2_{\rm A}$. Thus:
\begin{equation}
\label{eq:qs}
Q^2_{\rm S}\sim\frac{ A\,xg(x,Q^2_{\rm S})}{\pi R^2_{\rm A}}\sim\frac{A\,xg(x,Q^2_{\rm S})}{A^{2/3}}\sim {A}^{1/3}x^{-\lambda}\sim{A}^{1/3} \big(\sqrtsNN\big)^\lambda e^{\lambda y}\,,~~~~{\rm with~}\lambda\approx0.3. 
\end{equation}
$Q^2_{\rm S}$ grows at forward rapidity, at high c.m.s. energy, 
and it is enhanced by a factor about $6 \approx 200^{1/3}$ in the Au or Pb nucleus, 
with respect to the proton.
Saturation affects the processes in the region $Q^2\,\lsim\, Q^2_{\rm S}$,
where gluon recombination dominates and factorization may start to become
invalid. Figure \ref{fig:saturation} illustrates how saturation comes about in the high density regime, which can be achieved by decreasing the value of $x$ (left panel) and/or increasing the mass number $A$ of the colliding objects (i.e. using nuclei instead of nucleons; right panel).

\begin{figure}
\includegraphics[width=0.48\textwidth]{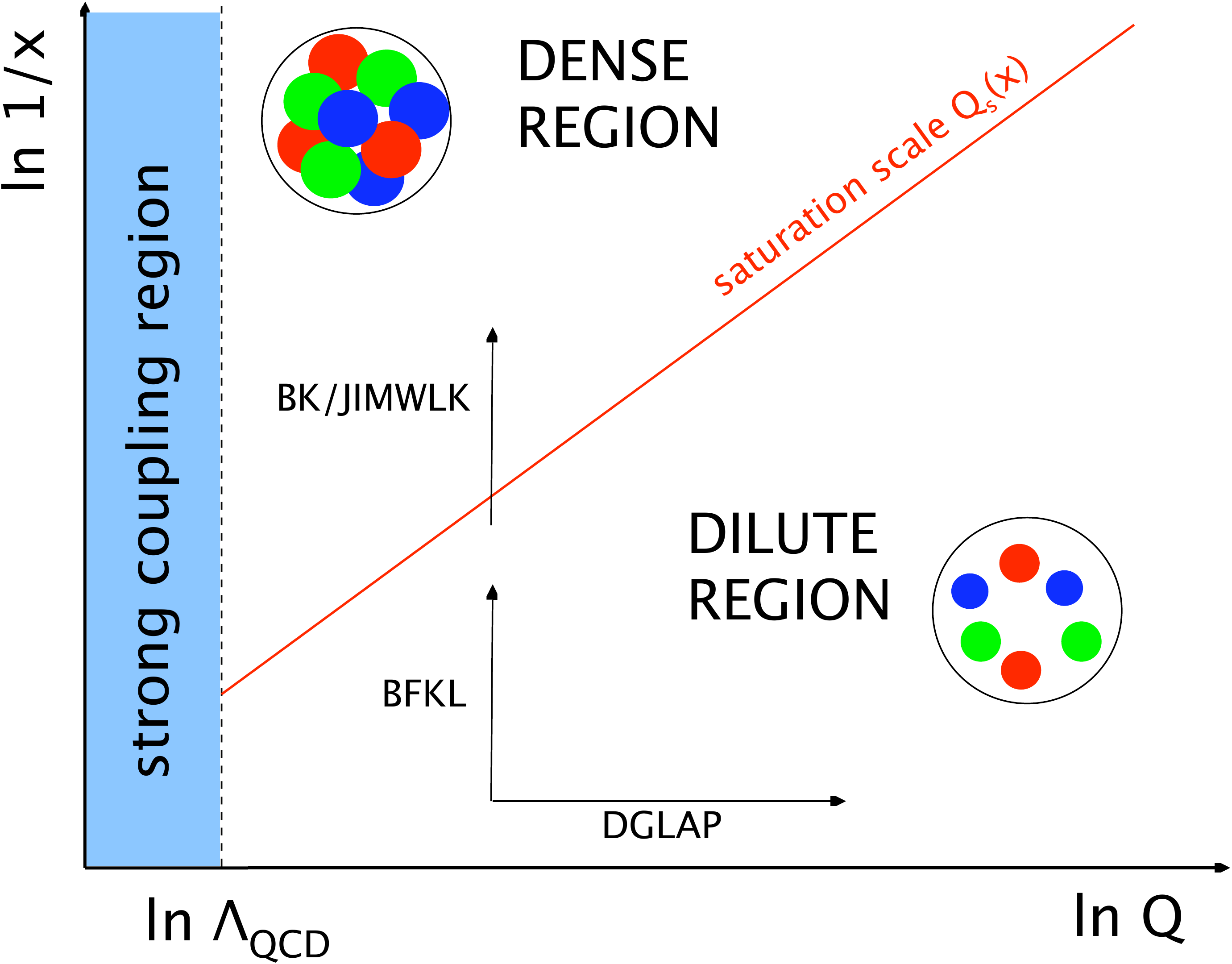} \hfill\includegraphics[width=0.48\textwidth]{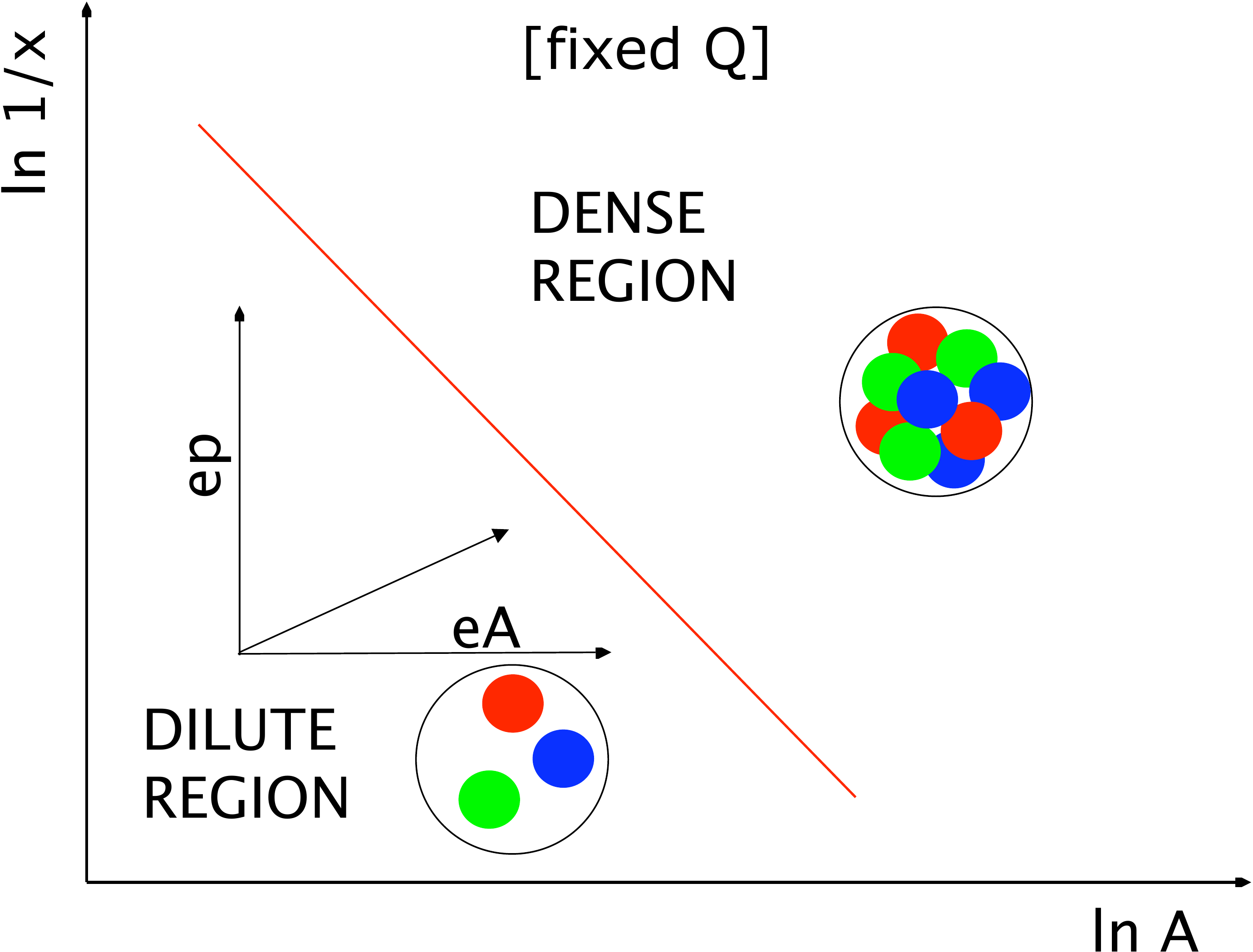}
\caption{\label{fig:saturation}Left: illustration of linear (dilute) and non-linear (dense) domains in the kinematic $x$--$Q^2$ plane, $Q^2$ being the inverse squared transverse resolution, and the different evolution equations. Right: the same but in the $x$--$A$ plane. The red line indicates the separation between the dilute and the dense regimes, given by the saturation scale $Q_{\rm S}(x)$.}
\end{figure}

There is an intrinsic relation between the parton density evolution that can be used in the dilute and dense regimes and the kind of factorisation that, if existing, should be employed to compute particle production in collisions involving hadrons and nuclei. In the dilute regime, collinear~\cite{Collins:1989gx} or $k_{\rm T}$~\cite{Collins:1991ty,Levin:1991ry,Catani:1990xk,Catani:1990eg} factorisations can be applied in the hard and semihard regions, respectively. The corresponding evolution equations are DGLAP ~\cite{Gribov:1972rt,Gribov:1972ri,Altarelli:1977zs,Dokshitzer:1977sg} and BFKL.
In the dense regime, the situation is not yet clear. For scattering of a dilute projectile on a dense target, single inclusive hadron production has been computed at NLO~\cite{Chirilli:2012jd,Altinoluk:2014eka} and the corresponding JIMWLK evolution equation, an infinite hierarchy of coupled evolution equations for traces of Wilson lines, is also known at NLO~\cite{Balitsky:2008zza,Kovner:2013ona,Balitsky:2013fea}\footnote{The pathologies of this equation, related to those of NLO BFKL, are understood and several solutions involving resummation have been proposed~\cite{Beuf:2014uia,Iancu:2015vea,Lappi:2016fmu}.}.

At present, no conclusive evidence has been provided for the existence of saturation, although a number of observations are consistent with expectations from gluon saturation, as discussed in the following. 
\begin{enumerate}
\item[i.]
The degrading quality of DGLAP fits when HERA data at moderate $Q^2$ (which by DIS kinematics is linked to small $x$) are included in the fit~\cite{Abramowicz:2015mha}, while fits done within saturation do not show such degradation~\cite{Albacete:2012rx}; the degradation, however, is quantitatively small and the freedom in the choice of initial conditions and of the scheme for treating heavy flavours still sizeable. 
\item[ii.]
The success in describing the factorisation between the energy and centrality dependencies of charged particle multiplicities at mid-rapidity~\cite{Adam:2015ptt}; this observable is likely to be affected by final-state collective and non-perturbative effects that are not included in saturation models. 
\item[iii.]
The ridge structure (two-particle correlations that are independent of rapidity and strongly collimated in azimuth at 0 and 180 degrees) observed in pp and p--Pb collisions at the LHC that can be explained in CGC-type models~\cite{Schlichting:2016xmj}; yet, the explanation is not unique and the possibility of final-state collective effects in such small systems is currently under debate. 
\item[iv.]
The suppression of inclusive particle production \cite{Arsene:2004ux,Adams:2006uz} and the reduction of back-to-back correlated yield at forward rapidities in d--Au collisions at RHIC~\cite{Braidot:2010ig,Adare:2011sc,Albacete:2010pg,Stasto:2011ru}; still, the experimental data lie so close to the kinematical limit that tiny additional effects may affect their interpretation, and at RHIC the probed values of $x$ are not that small. 
\end{enumerate}
Note that points (i) and (ii) are related to the fact that the geometric scaling that experimental data show~\cite{Stasto:2000er,Armesto:2004ud,McLerran:2010ex} finds a natural explanation in saturation models, but it is also shown by e.g. DGLAP dynamics~\cite{Caola:2008xr}.

In order to firmly establish the existence of this new high-energy regime of QCD and clarify the validity of the different approaches to factorisation and evolution, new kinematic regions must be explored using higher collision energies in order to have a large lever arm in $Q^2$ in a region that, while perturbative, lies inside the saturation domain. The FCC offers such energies and the possibility of combining proton and nuclear beams, as required for a detailed understanding of the mechanism underlying saturation. 
In Fig.~\ref{fig:kinplane} the coverage of the $x$--$Q^2$ plane of present facilities (nuclear DIS and Drell-Yan experiments, proton--nucleus colliders RHIC and the LHC
in the left panel, and ultra-peripheral nucleus--nucleus collisions at RHIC and
LHC using exclusive quarkonium production in the right panel) is compared with that of p--Pb collisions at FCC energies (proton momentum of 50~TeV and Pb momentum per nucleon of 19.7~TeV) and photon--Pb collisions at FCC energies (the latter will be discussed in Section~\ref{sec:HI_upc}). 
In the left panel, the coverage is calculated for an experimental acceptance reaching rapidity in the laboratory frame $y_{\rm lab}\approx\eta=6.6$. The diagonal lines represent constant $y_{\rm lab}$ values 0, 1, ..., 6. The estimate of the saturation scale $Q^2_{\rm S}(x)$ for Pb is also shown. The figure shows that coverage below the saturation scale is much larger at FCC than at LHC, reaching for example down to $x$ values well below $10^{-6}$ and $Q^2$ values of 10--20~GeV$^2$ at a rapidity of 5, which could be well within reach of an experiment at the FCC.

\begin{figure}
\begin{center}
\includegraphics[width=0.49\textwidth]{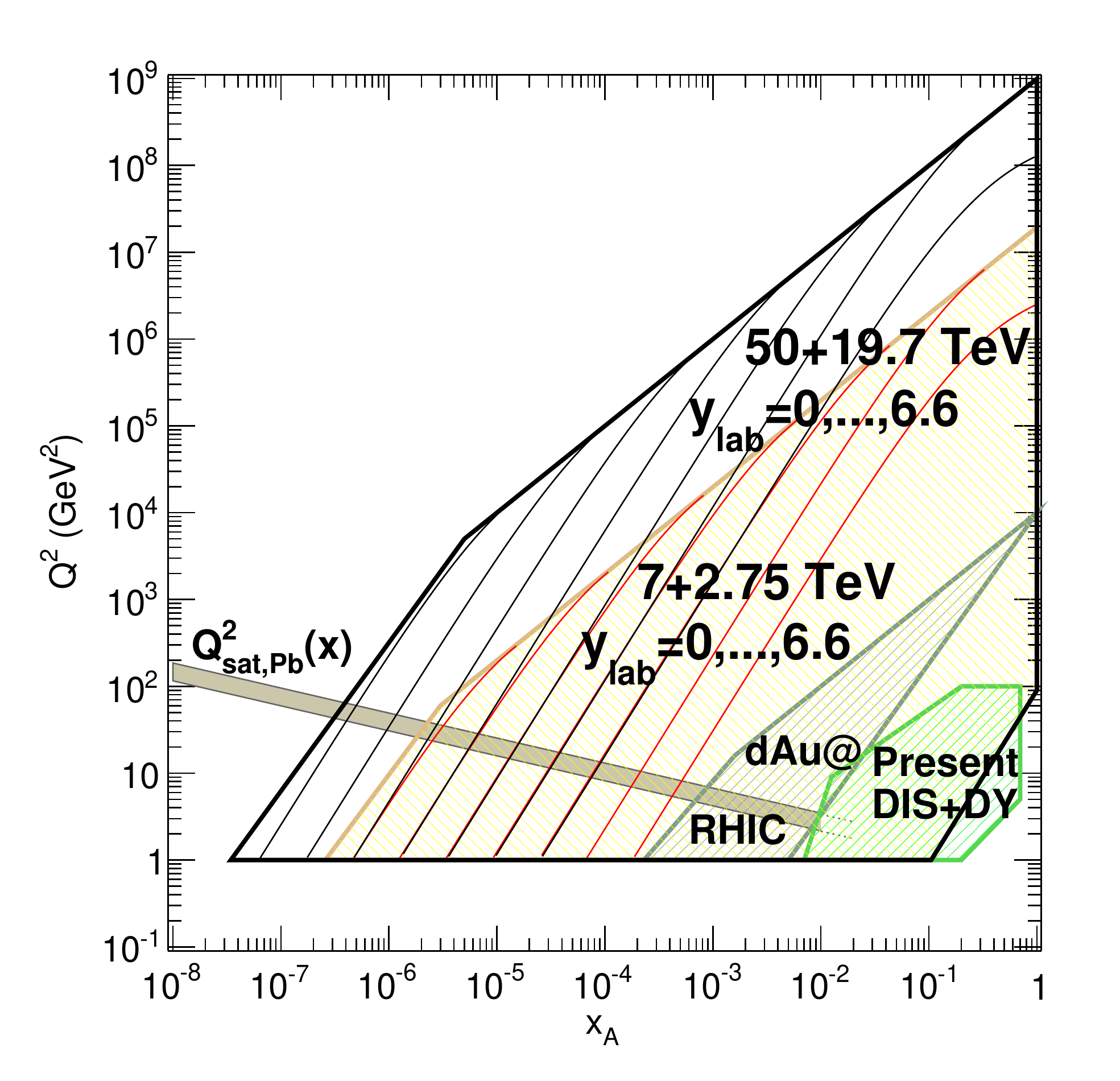}
\includegraphics[width=0.49\textwidth]{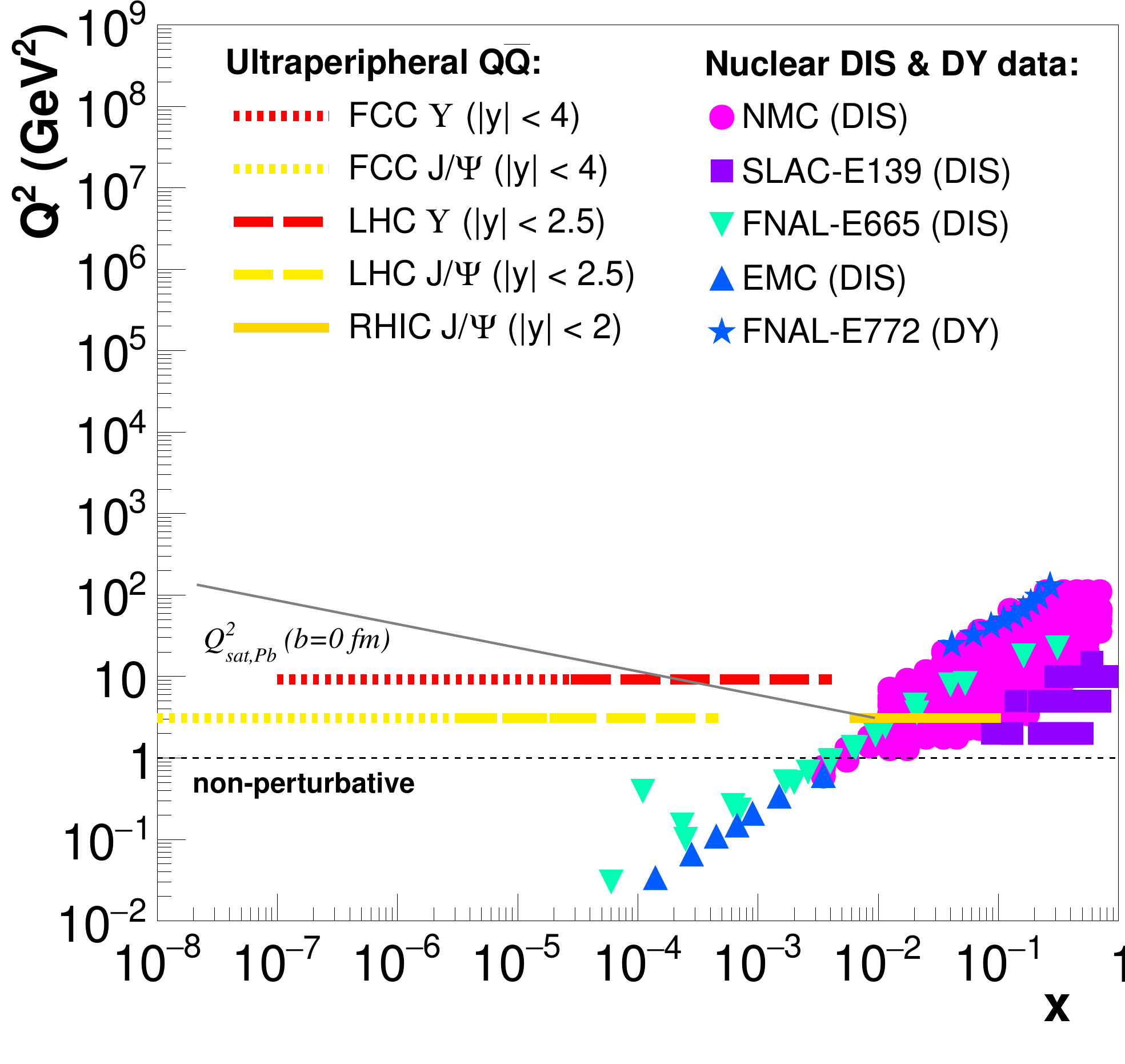}
\end{center}
\caption{Regions of the $x$--$Q^2$ plane covered 
with nuclear DIS and Drell-Yan data (left) and exclusive $J/\psi$ and $\Upsilon$ photoproduction in ultraperipheral ion collisions (right). The left
panel shows the range covered in nuclear DIS and DY experiments (light green), and in collisions
involving nuclei at RHIC (dark green) and at the LHC (yellow), and accessible at the FCC (black). For
p--Pb collisions at the LHC and the FCC, thin lines correspond to different rapidities in the
laboratory frame $y_{\rm lab} = 0$,1, 2, 3, 4, 5, 6 from right to left, with the left edge defined by $y_{\rm lab} = 6.6$. Values of the saturation scale $Q^2_{\rm S}(x)$ for $A = 208$ are shown for illustration.}
\label{fig:kinplane}
\end{figure}

Establishing the dynamics at small $x$ and the type of factorisation to be used to compute particle production will constrain the initial conditions for the collective behaviour of the medium produced in high-energy nucleus-nucleus collisions
\cite{Albacete:2014fwa}. It will also clarify the dynamics leading the approximate thermalisation or isotropisation~\cite{Kurkela:2016vts} that allows the use of relativistic hydrodynamics to describe such collectivity. Note that the extraction of properties of the QGP is significantly affected by both the initial conditions and the pre-equilibrium dynamics, see e.g.~\cite{Liu:2015nwa}. The need for more precise information about the partonic structure of nuclei also holds for hard processes, where the lack of knowledge of nPDFs, illustrated in Fig. \ref{fig:comparnpdfs}, compromises the precision for extraction of transport properties of the QGP from the comparison of theoretical calculations to data on hard probes. The information that is coming from p--Pb collisions at the LHC has not yet provided large constrains, see~\cite{Armesto:2015lrg}.

\begin{figure}
\begin{center}
\includegraphics[width=0.5\textwidth]{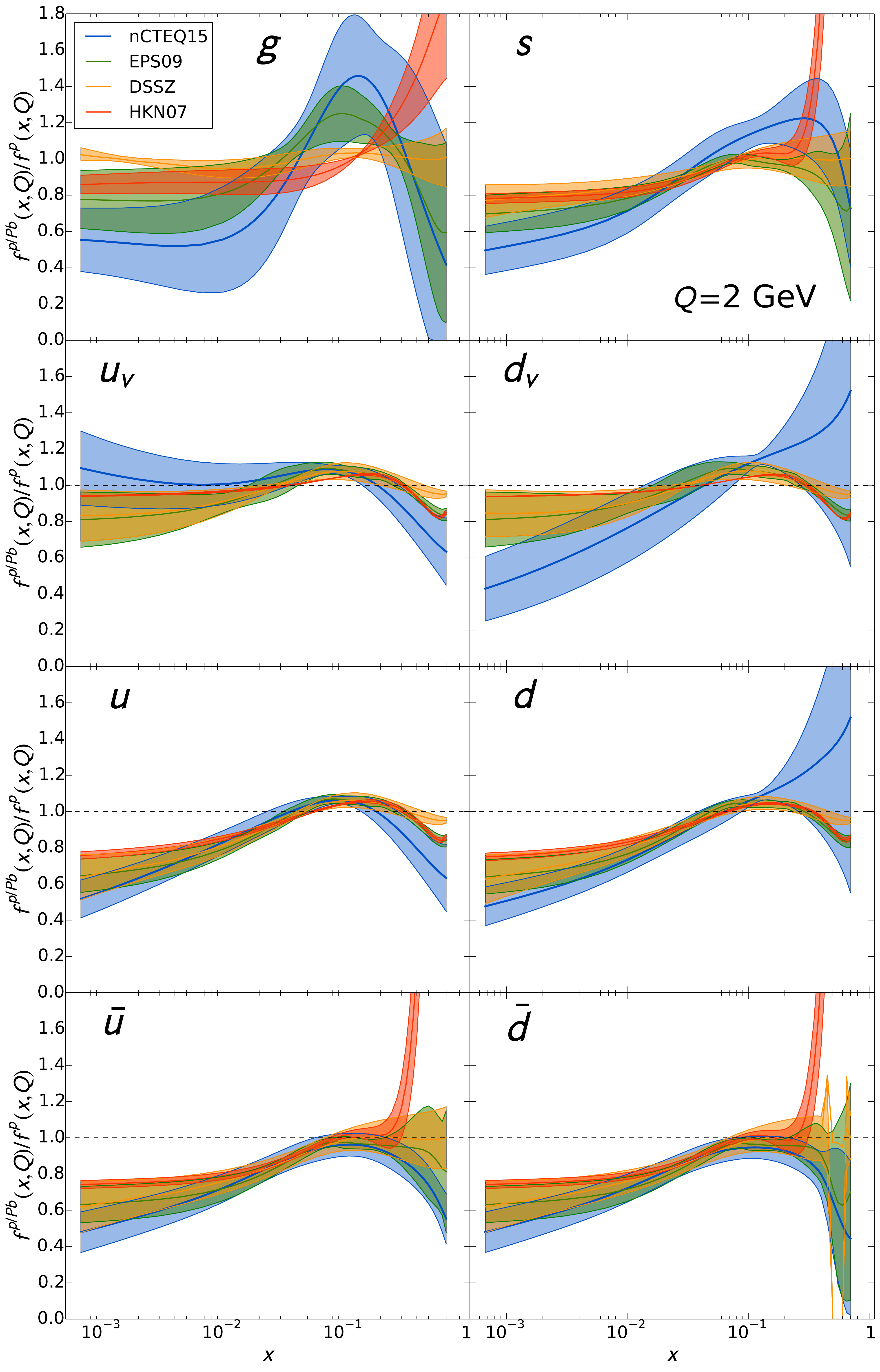}
\end{center}
\caption{\label{fig:comparnpdfs}Comparison of the nuclear modification factor (defined as the ratio of a given parton density in a bound proton over that in a free proton) for different parton species at $Q^2=4$ GeV$^2$ for a Pb nucleus. Lines correspond to DGLAP analyses at NLO: EPS09~\cite{Eskola:2009uj}, HKN07~\cite{Hirai:2007sx}, DSSZ~\cite{deFlorian:2011fp} and nCTEQ15~\cite{Kovarik:2015cma}. Bands correspond to the uncertainty in the respective Hessian analyses. Taken from~\cite{Kovarik:2015cma}.} 
\end{figure}

As a final remark, we should indicate that although important progress has been achieved lately in advancing calculations of different observables in the CGC, several aspects are still missing. For example, the resummed evolution equations discussed previously have not been fully used although the NLO impact factor for DIS is available~\cite{Balitsky:2012bs,Beuf:2011xd}; the calculation of many single-inclusive observables apart from hadrons is still missing, see~\cite{Albacete:2014fwa}; much progress is undergoing on two-particle correlations to understand the origin of azimuthal asymmetries and the ridge in pp and p--A collisions~\cite{Lappi:2015jka}.

\subsubsection{Possible signatures of the CGC using forward-rapidity hadrons and photons in p--Pb collisions}

The simplest way to probe the gluon density in protons and nuclei is by studying inclusive particle production. In particular, nuclear modification ratios \rppb{} are used to explore the nuclear modification of parton densities and saturation effects due to the larger density of gluons in nuclei than in protons included in the CGC. The nuclear modification factor \rppb{} is defined as:
\begin{equation}
\rppb = \frac{\left. \dd^2\sigma/\dd\pt \dd y \right|_{\rm pPb}}{A \left. \dd^2\sigma/\dd\pt \dd y \right|_{\pp}}\,, 
\end{equation}
i.e.\,by taking the $\pt$-differential cross section measured in p--Pb collisions and dividing by that in pp collisions at the same energy, multiplied by the Pb mass number $A$. If the production in p--Pb follows a scaling with the number of binary nucleon--nucleon collisions, then $\rppb=1$.


\begin{figure}
\begin{center}
\includegraphics[width=0.6\textwidth]{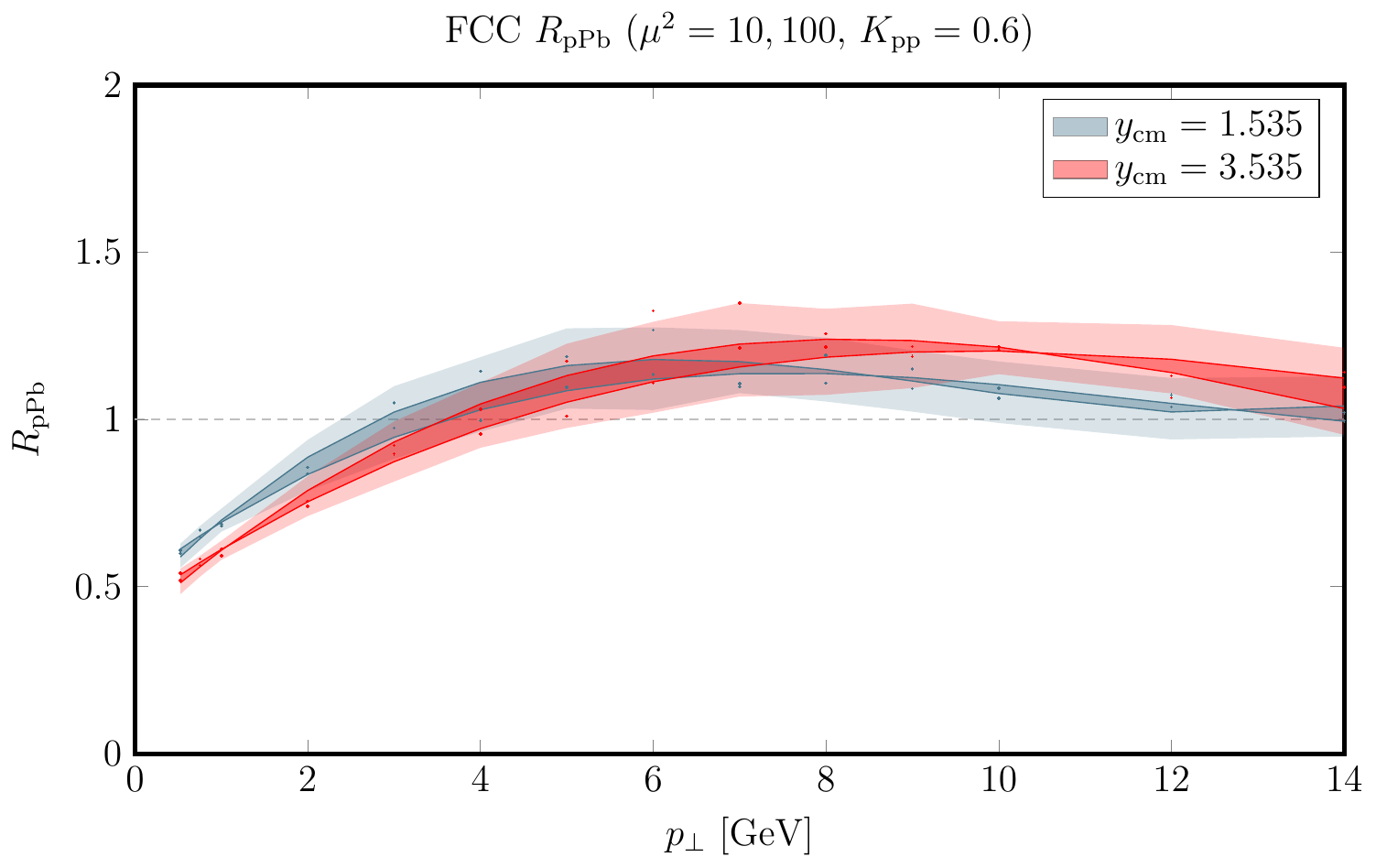}
\end{center}
\caption{\rppb as a function of $\pt$ for charged hadrons at the FCC for two different rapidities \cite{Watanabe:2015tja}.}
\label{fig:davidbowen}
\end{figure}


\begin{figure}
\begin{center}
\includegraphics[width=0.49\textwidth]{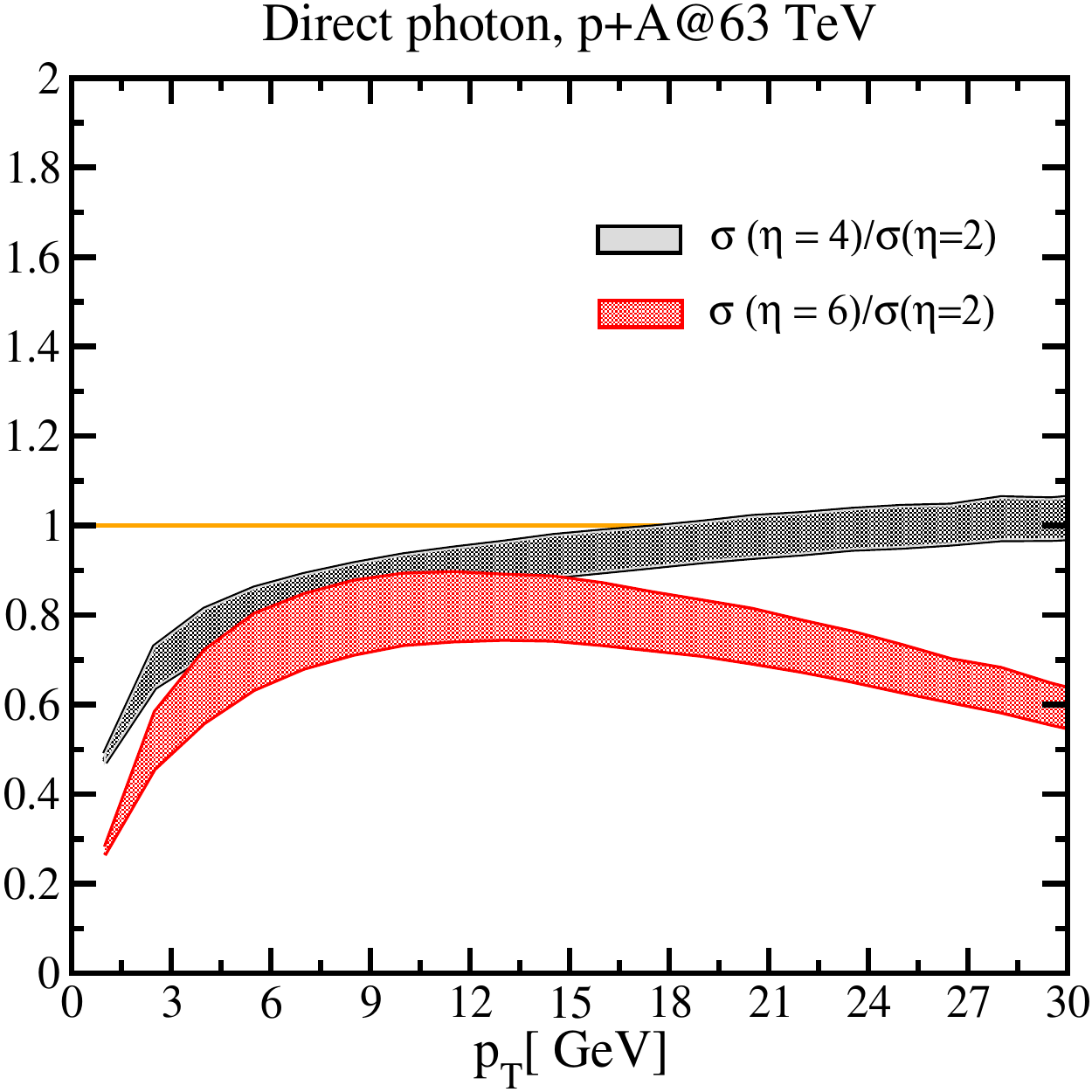}
\includegraphics[width=0.49\textwidth]{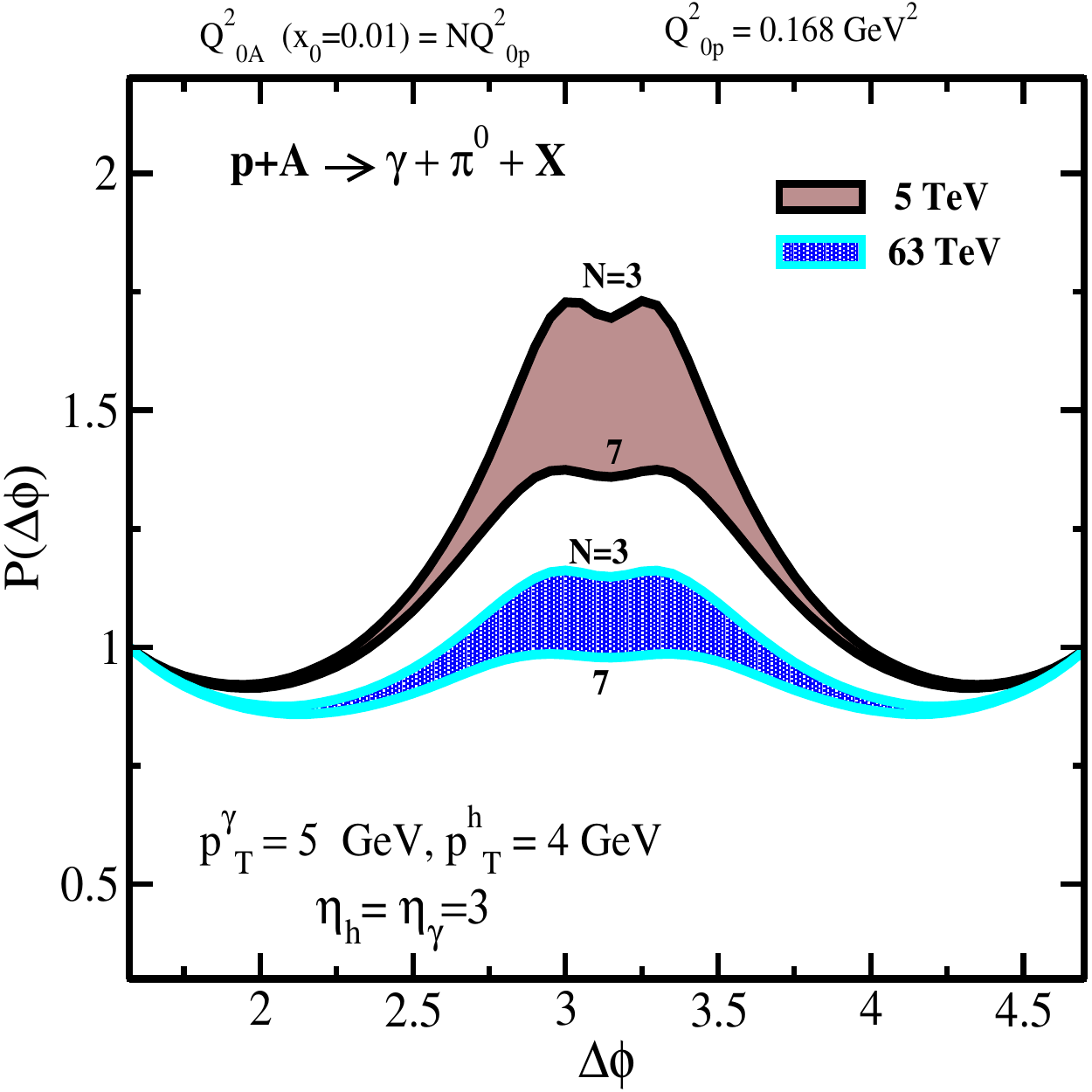}
\end{center}
\caption{Left: Ratio of direct photon spectra at forward ($\eta=4, 6$) over central ($\eta=2$) rapidities obtained in the CGC formalism in minimum-bias p+A collisions at the FCC~\cite{Rezaeian:2012wa,Rezaeian:2012ye}.
Right: Normalized azimuthal correlation of prompt photon and \pizero{} as a function of the angle between the produced prompt photon and $\pi^0$ at forward rapidity obtained in the CGC formalism at the LHC and the FCC in minimum-bias p+A collisions~\cite{Rezaeian:2012wa,Rezaeian:2012ye}. The bands show the uncertainty related to the scaling of the initial saturation scale in nuclei.}
\label{fig:rezaeian}
\end{figure}

It can be seen from Fig.~\ref{fig:kinplane} (left) that even measurements at mid-rapidity and $\pt < 10 \GeVc$ cover the saturation region with $Q\approx \pt$ and $x$ in the range $10^{-5}$--$10^{-4}$, which is at much lower $x$ and therefore larger gluon density than measurements at the LHC. A forward measurement, for example at $\eta \approx 4$, would be even more interesting, as it covers $x \approx 10^{-6}$.

To illustrate how future measurements at the FCC are sensitive to gluon saturation, we present a few selected calculations. It is, however, important to realise that there are significant uncertainties in these calculations, since the currently available measurements do not provide precise constraints for the saturation scale and some of the model details, such as the dipole cross section parametrisation. 
This uncertainty is closely related to our earlier statement that the CGC or saturation effects have not been unambiguously identified yet. In addition to this, the calculations for FCC energies require a large extrapolation of available constraints from data at $x\approx 10^{-4}$--$10^{-3}$ to lower $x$. Measurements at the larger energies available at the FCC are the only possibility to find out how the gluon density at small $x$ behaves in Nature.

Figure~\ref{fig:davidbowen} shows the expected nuclear modification factor for charged particles at two different rapidities, computed using the state-of-the-art NLO calculations available \cite{Chirilli:2012jd,Altinoluk:2014eka,Watanabe:2015tja}. 
A modest increase of the suppression at low $\pt \lesssim 4\;\GeVc$  compared to the expectation for LHC is visible\footnote{The magnitude of the effect is very sensitive to the specific form employed for evolution and to whether the calculation is done at LO or at NLO. Furthermore, the formalism employed in these calculations is expected to work better in p--A than in pp. This fact would amount to a sizeable uncertainty in the ratio.}.
Figure~\ref{fig:rezaeian} (left) shows an example of the expected direct photon production in the CGC framework at LO approximation in the hybrid approach using the rcBK formalism. 
The ratios of particle production at different pseudorapidities explicitly probe the evolution of the gluon density in $x$. A clear suppression of direct photon production at moderate $\pt \lesssim 6 \GeVc$ is visible in the figure. At higher $\pt$, a modest suppression by about 10-20\% is still visible. 
When comparing Fig.~\ref{fig:davidbowen} and the left panel of Fig.~\ref{fig:rezaeian}, a larger effect of gluon saturation is seen for direct photon production. Such a difference between charged hadrons and photons could be qualitatively expected\footnote{Nevertheless, it must be kept in mind the existence of sizeable uncertainties due to differences in the order in perturbation theory and in phenomenological details between both calculations.} since hadron production is sensitive to both the quark and gluon densities, while the dominant production mechanism for direct photons is quark--gluon Compton scattering, which is directly sensitive to the gluon density~\cite{Helenius:2014qla}.



Measurements of dijets or back-to-back two-particle correlations offer more potential to experimentally constrain the probed $x$ region, in particular at low \pt where multiple-interaction effects may also play a role. 
Color Glass Condensate models make a specific prediction that the recoil jet is suppressed, because (mini-)jets can be produced by scattering a parton off the color field in the nucleus where the recoil momentum is carried by multiple gluons, unlike in a standard (semi-)hard 2-to-2 scattering where all the recoil momentum is carried by a single jet~\cite{Kharzeev:2004bw,Albacete:2010pg,Rezaeian:2012wa}.
A suppression of the recoil yield has been observed at lower energies at RHIC~\cite{Adare:2011sc}, close to the kinematic limit, where suppression of the yield by multiple scattering or energy loss in the initial state may also be important~\cite{Kang:2012kc,Arleo:2012hn}.
At the FCC, such measurements can be performed over a broad kinematic range in both $x$ and $Q^2$, which will allow to disentangle different effects.

The potential of recoil measurements at FCC is illustrated in Fig.~\ref{fig:rezaeian} (right), which shows the azimuthal distribution of $\pi^0$ recoiling from a prompt photon trigger particle, both at forward $\eta=3$.
The double-peak structure on the away side is a characteristic prediction of CGC calculations~\cite{Rezaeian:2012wa,Stasto:2012ru}.
The recoil signal at FCC energies is smaller than at the LHC due to the larger gluon density at lower $x$.

%
%
\begin{figure}
\begin{center}
\includegraphics[width=0.325\textwidth]{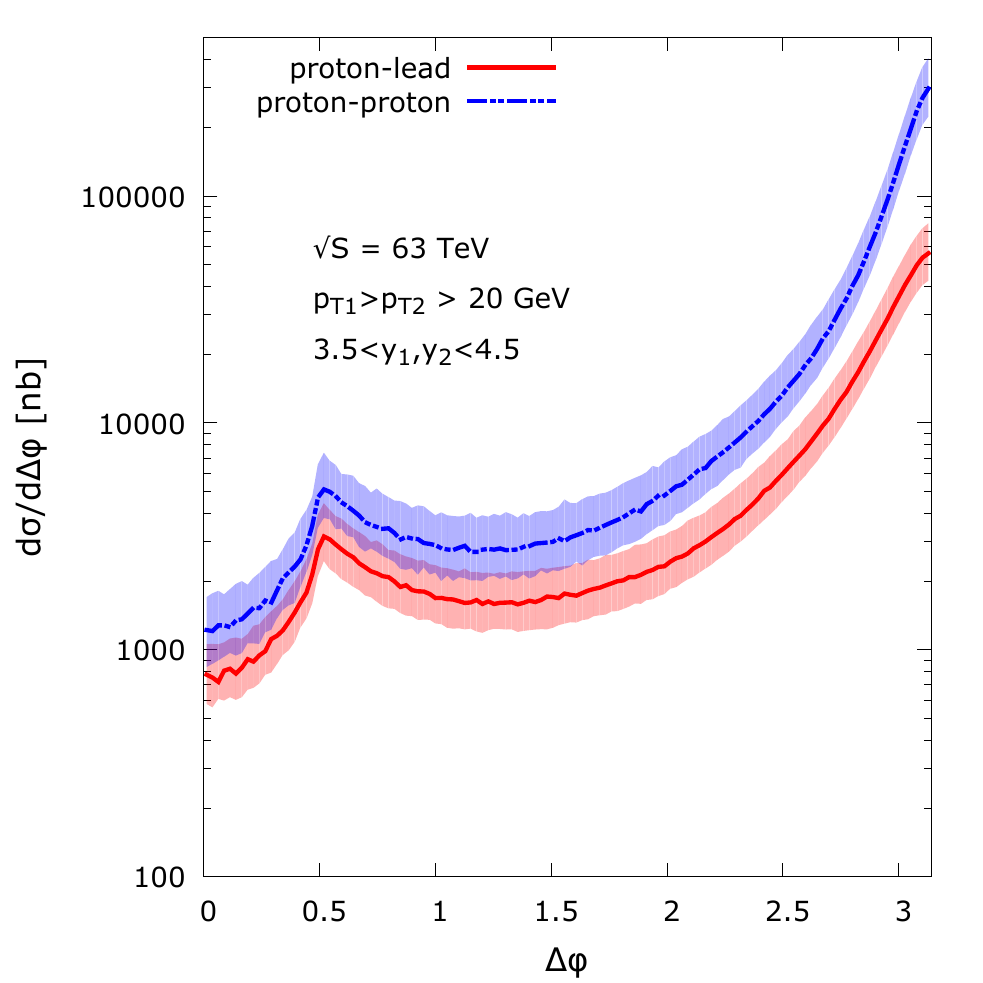}
\includegraphics[width=0.325\textwidth]{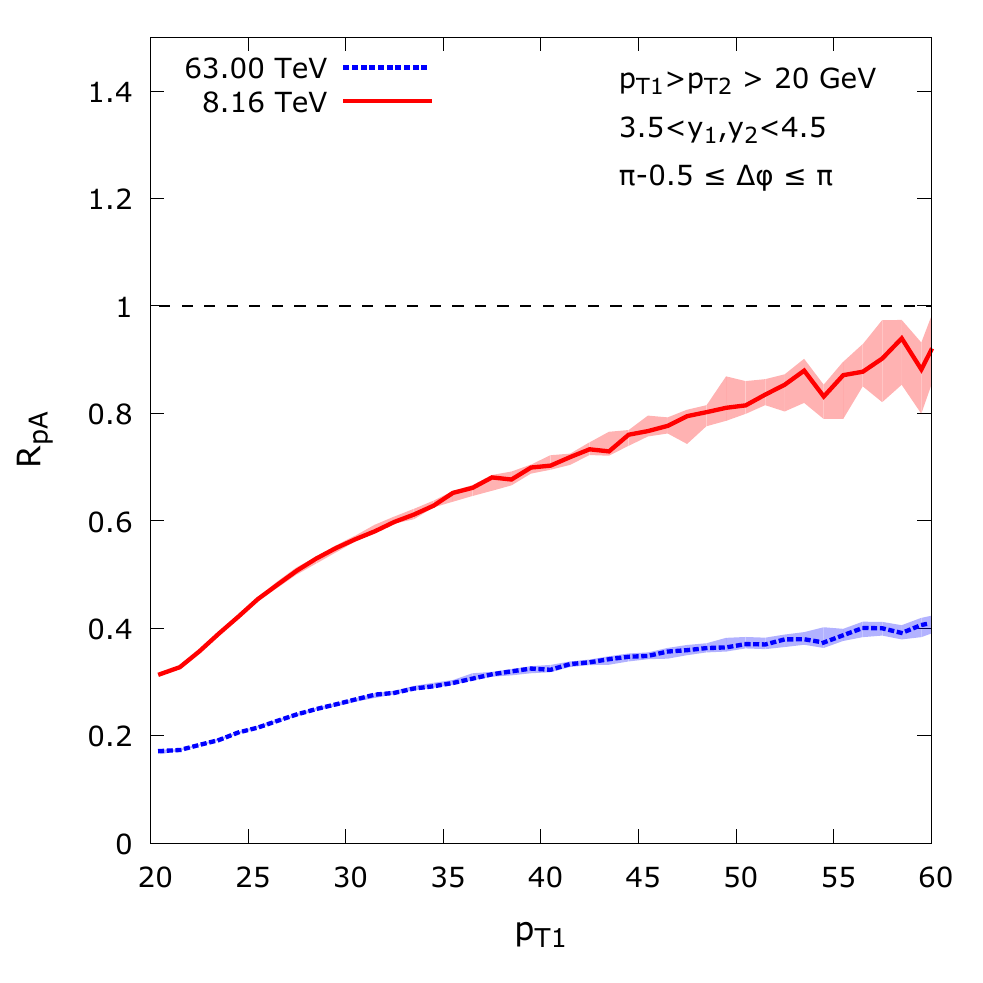}
\includegraphics[width=0.325\textwidth]{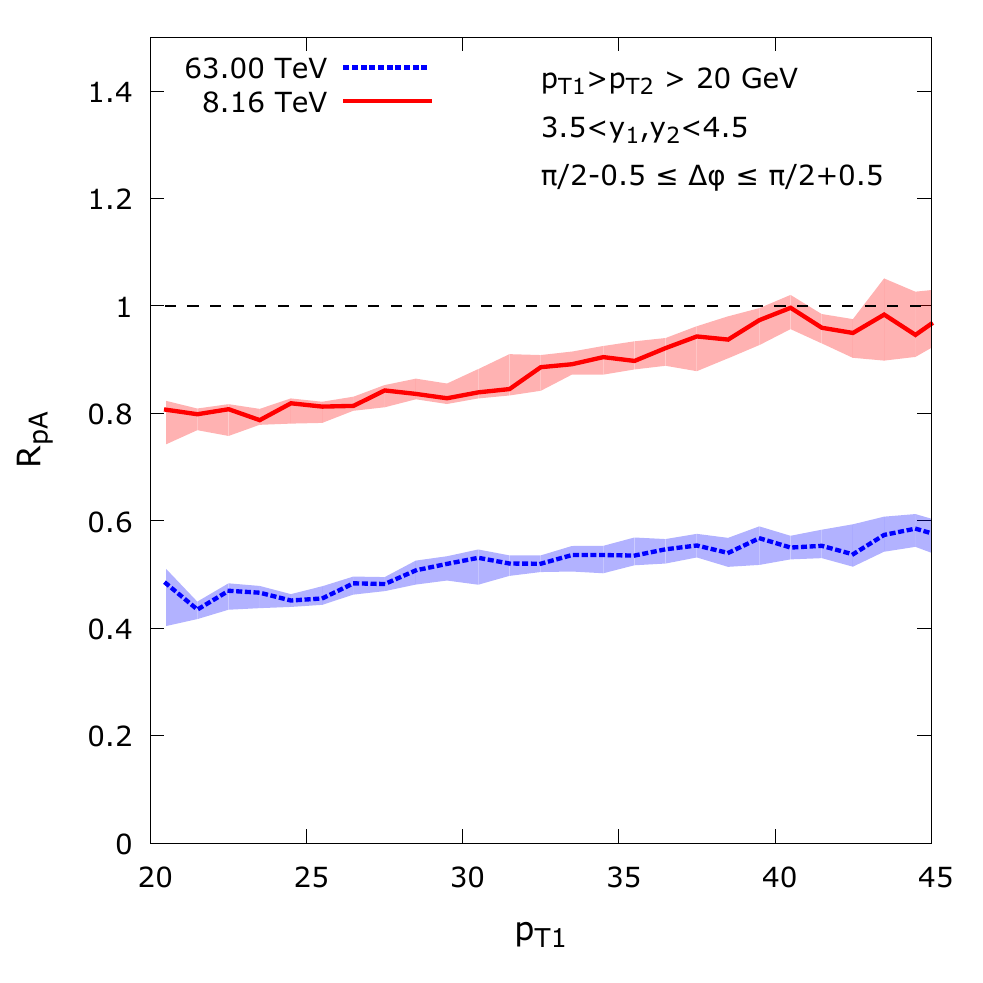}
\end{center}
\caption{Dijet yield suppression \cite{vanHameren:2014lna}. Left: azimuthal angle distribution in \pp{} and \pPb{} collisions (\pPb{} curve divided by $A$). Middle and right: ratio of dijet yield in \pPb{} and \pp{} collisions as a function of $\ptjet$ for jet pairs with $\ptjet>20$~\GeVc{} and $3.5 < y_\mathrm{jet} < 4.5$ at LHC and FCC energies, for back-to-back pairs (right panel) and pairs in the underlying even region $\Delta\phi \approx \pi$ (middle panel). }
\label{fig:marquet}
\end{figure}

The recoil suppression can be also be explored using dijets at forward rapidity~\cite{vanHameren:2014lna,Kotko:2015ura}. 
Figure~\ref{fig:marquet} shows the expected broadening of the $\Delta\phi$ distribution in p--Pb versus pp collisions at the FCC (left panel, the p--Pb curve as been divided by $A$), as well as the expected nuclear modification factors for dijets as a function of the transverse momentum of the leading jet \ptjet{}, in the recoil region (middle panel, for $\Delta\phi \approx \pi$) and in the ``underlying event'' region (right panel, for $\Delta\phi$ away from $\pi$). Both jets have rapidity $3.5 < y_\mathrm{jet} < 4.5$.
 A clear suppression is visible --strongest in the recoil region-- which persists to much larger $\ptjet > 100$~\GeVc{} than at LHC, where the suppression is small at $\ptjet \approx 50$~\GeVc. These calculations clearly show that the effects of saturation are expected to persist to high \pt, much larger than the saturation scale, as long as the transverse momentum imbalance of the dijet system does not exceed a few times $Q_{\rm S}$. We note that an optimistic scenario was considered for the ratio $Q_\mathrm{S,Pb}/Q_\mathrm{S,p}$, whose value at FCC energies is a bit uncertain.

\subsubsection{Constraining nuclear parton densities at large $Q^2$}
\label{sec:HI_npdf}

\subsubsubsection{Constraining nPDFs with W and Z production in p--Pb collisions}

The production of on-shell $W$ and $Z$ bosons at hadron colliders probes the PDFs at large $Q^2\sim M_{W,Z}^2\sim 10^4~\gev^2$ and at momentum fractions around $x_{1,2} = \left({M_{\rm W,Z}}/{\sqrt{s}} \right)e^{\pm y}$ such that an order of magnitude increase in $\sqrt{s}$ from the LHC to the FCC extends the small-$x$ reach similarly by an order of magnitude. In addition, the cross sections are larger and the rapidity distributions broader such that these cross sections could be more easily measured (with wide-enough detectors) in a larger phase space than at the LHC. The increase of the production cross sections (computed at NLO by MCFM~\cite{mcfm} using CT10NLO proton PDFs~\cite{Lai:2010vv}) from RHIC energies to the FCC is illustrated for the case of $Z$ production in the left-hand panel of Fig.~\ref{fig:WZ1}. Due to the large FCC energy, the expectations are that within most of the experimental rapidity coverage (here assumed $|y_{\rm FCC} |< 4$) the cross sections are suppressed due to shadowing in nuclear PDFs, whereas at the LHC (here, $|y_{\rm LHC} |< 3$) the probed region is on both sides around the anti-shadowing peak. This is sketched in the right-hand panel of Fig.~\ref{fig:WZ1}, which shows the approximate $x$ intervals probed at the LHC and FCC including the average nuclear modifications of sea and valence quarks according to EPS09 nuclear PDFs~\cite{Eskola:2009uj} at scale $Q^2=(100 \, {\rm GeV})^2$.

Estimates of the rapidity distributions in p--Pb collisions at the FCC with and without nuclear shadowing are shown in Fig.~\ref{fig:WZ2} for $W$ (left-hand panel) and $Z$ production (right-hand panel). As expected, the predictions based on EPS09 nuclear effects show suppression down to $y \sim -3$. The uncertainties on the size of the nuclear effects might be significant. This is caused by the simplified functional forms assumed for small-$x$ distributions (for the lack of experimental high-$Q^2$ data at small $x$), but also by the fact that the partonic channels that contribute at the FCC are more diverse. Indeed, the scatterings between heavier sea quarks (e.g. $c\overline{s}$ and $s\overline{c}$ in $W$ production, $s\overline{s}$ and $c\overline{c}$ in $Z$ production) play a significant role at the FCC energies, while the currently available nuclear PDFs analyses have not extensively tuned e.g. the $s$-quark distribution.

\begin{figure}[t]
\center
\includegraphics[width=0.49\textwidth]{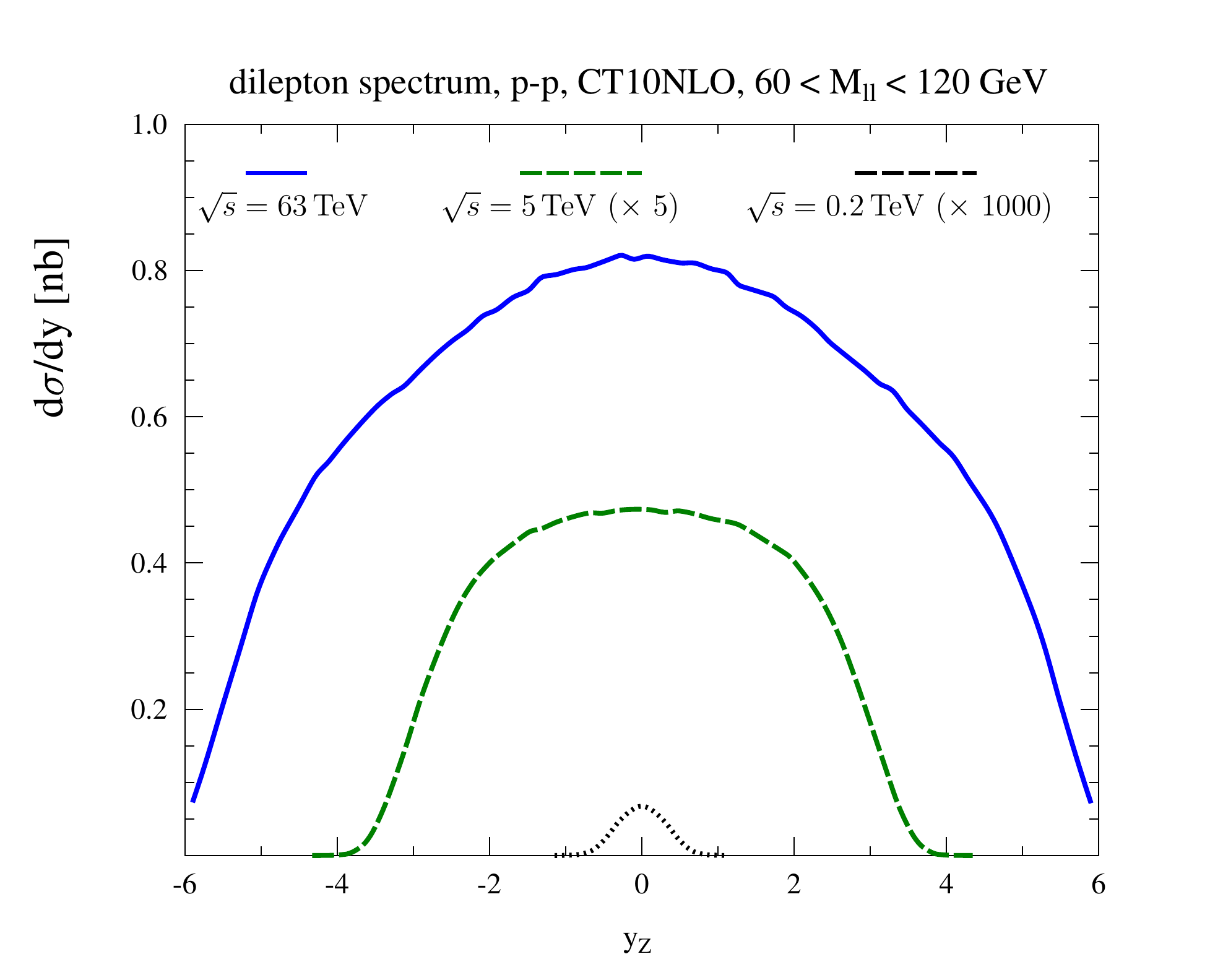}
\includegraphics[width=0.49\textwidth]{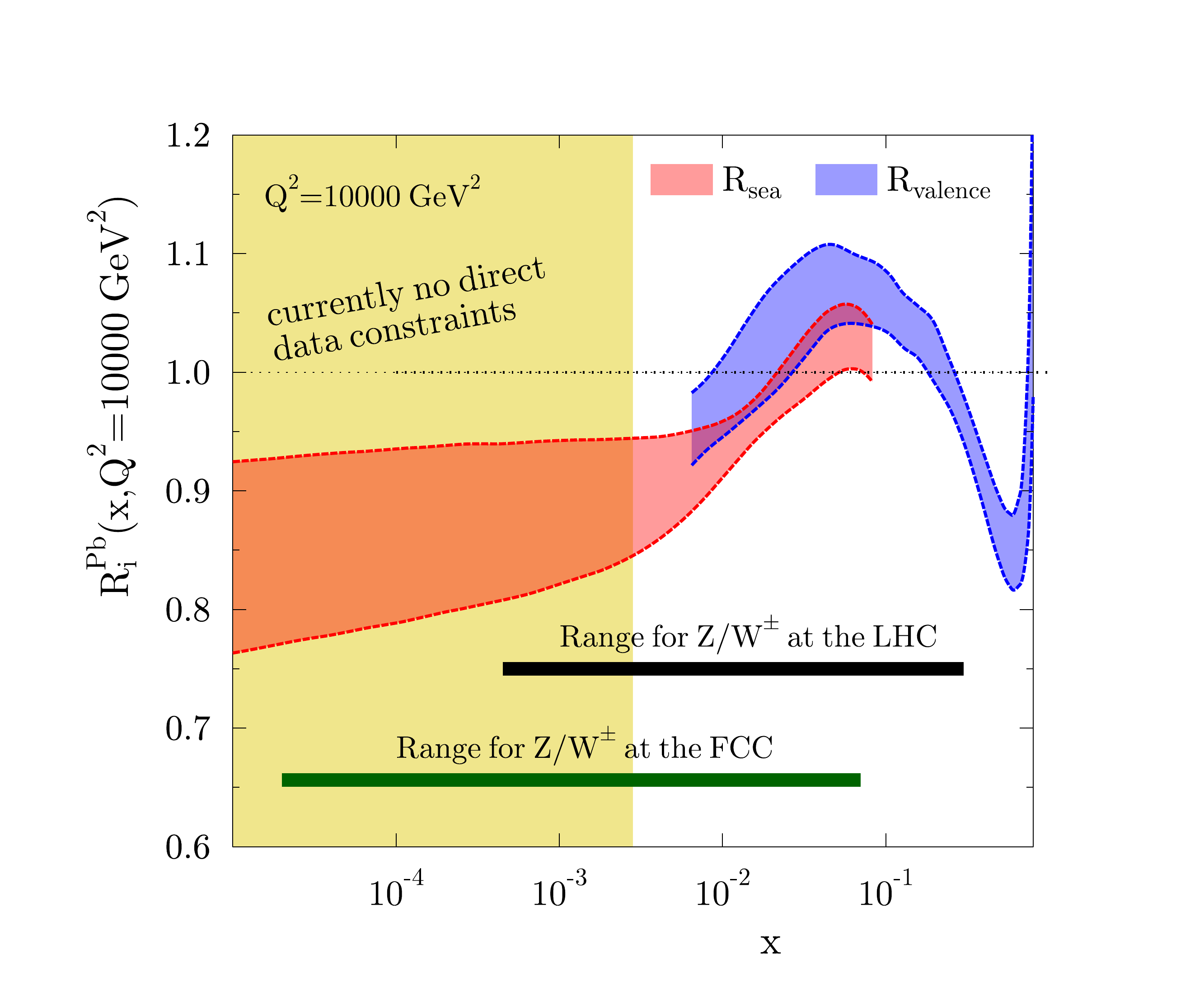}
\caption{Left: Rapidity distributions of high-mass dilepton pairs in pp collisions computed at $\sqrt{s}=200~{\rm GeV}$ (black dotted, multiplied by 1000), $\sqrt{s}=5~{\rm TeV}$ (green dashed, multiplied by 5), and $\sqrt{s}=63~{\rm TeV}$ (blue continuous). 
  Right: The average EPS09 nuclear modifications for valence (violet) and sea (red) quarks at $Q^2=(100~{\rm GeV})^2$. The $x$ regions probed by on-shell $Z$ and $W$ production at the LHC and FCC are indicated.}
\label{fig:WZ1}
\end{figure}

\begin{figure}[t]
\center
\includegraphics[width=0.49\textwidth]{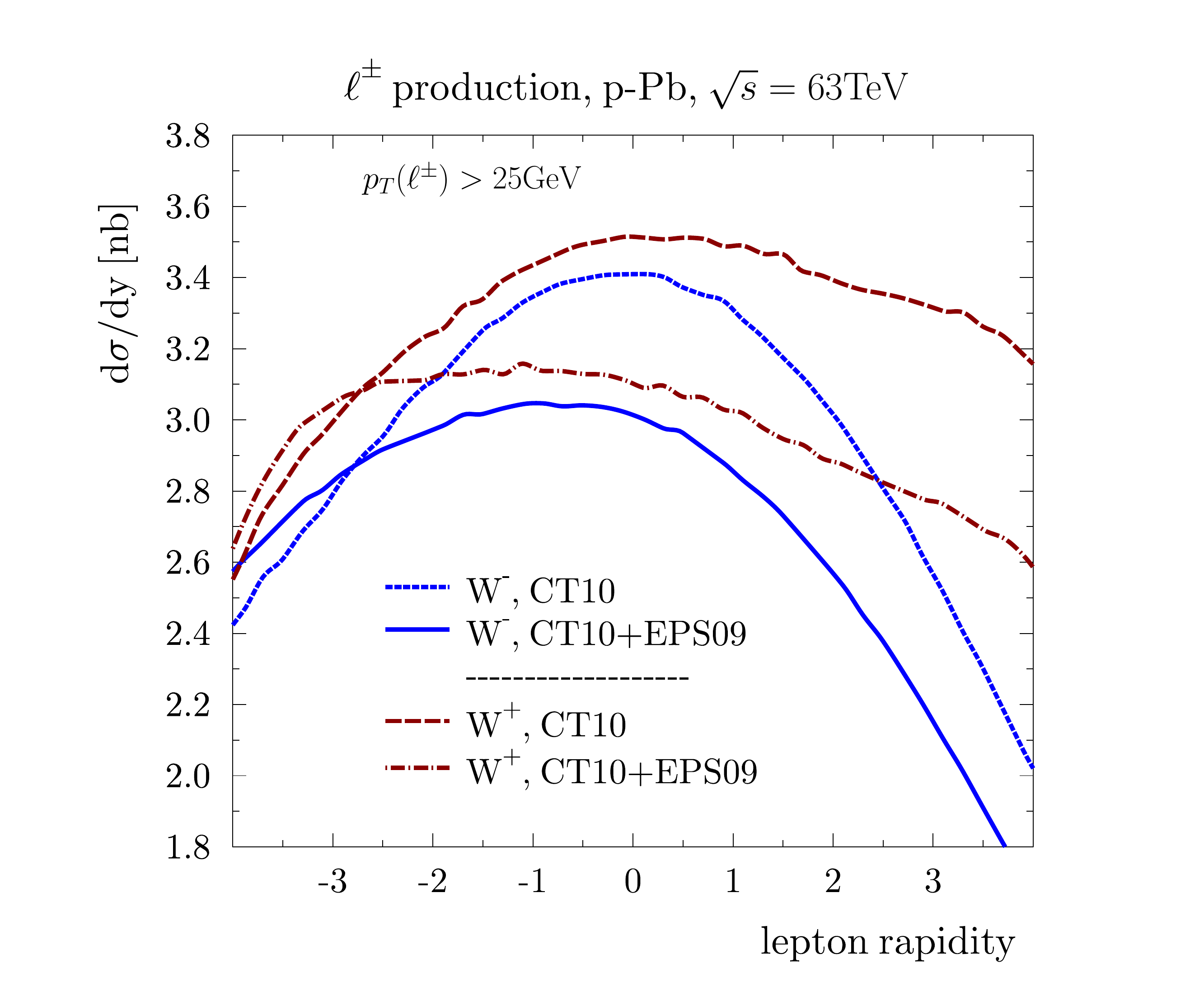}
\includegraphics[width=0.49\textwidth]{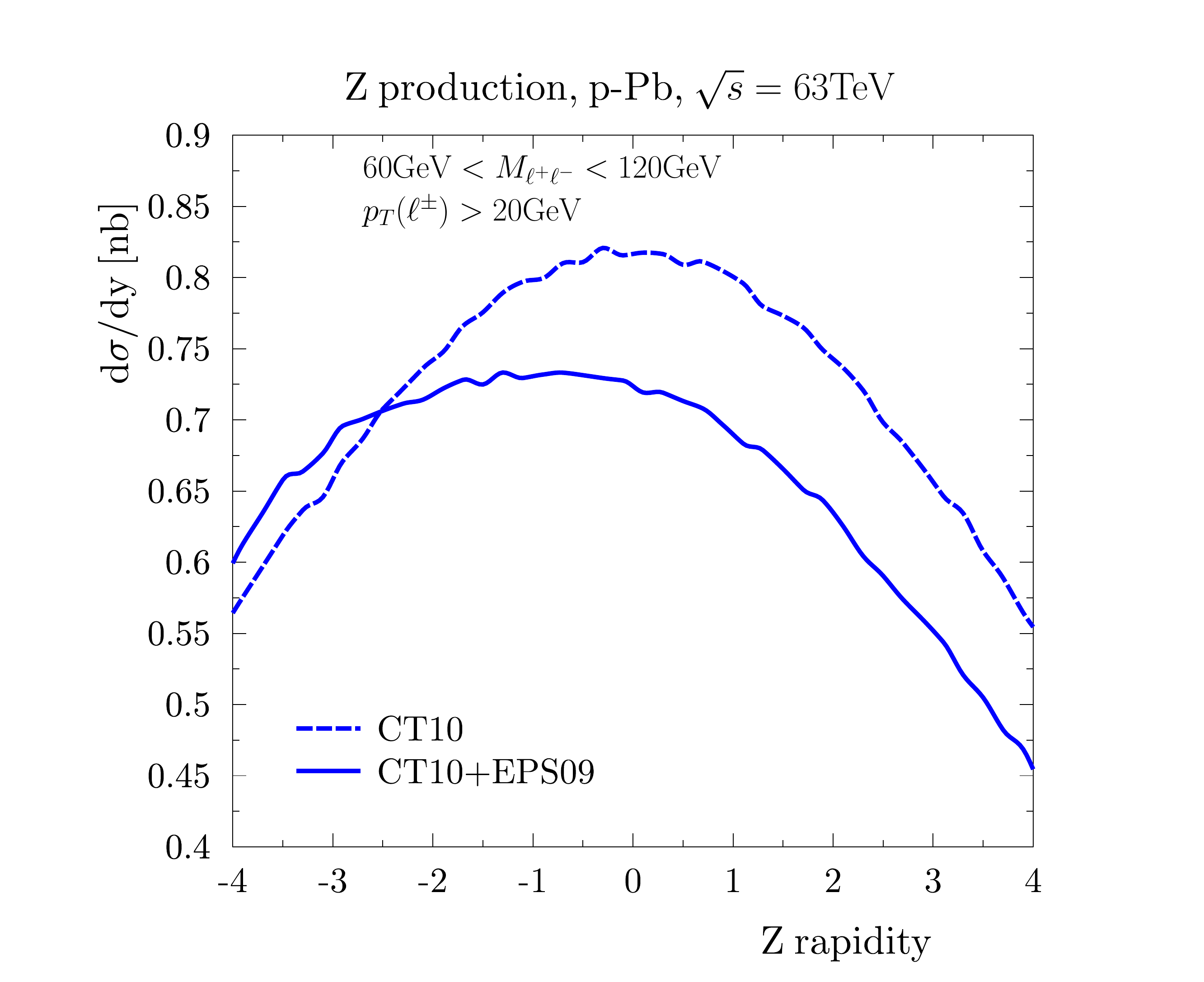}
\caption{Left: The rapidity distributions of charged leptons ($\ell^+$ in dark red, $\ell^-$ in blue) from $W$ decays in p--Pb collisions at $\sqrtsNN = 63 \, {\rm TeV}$ with (continuous, dashed-dotted) and without (short dashed, long dashed) EPS09 nuclear modifications. Right: As the left-hand panel, but for dilepton pairs from $Z$ decays.}
\label{fig:WZ2}
\end{figure}

\subsubsubsection{Constraining nPDFs with top-quark pair-production}

At hadron colliders, top quarks are produced either in pairs, dominantly through the strong interaction, or
singly through the weak interaction. At the energies considered here, the dominant production channels, as
obtained at NLO accuracy~\cite{d'Enterria:2015jna} with the MCFM code~\cite{mcfm}, are 
(Fig.~\ref{fig:txsec} left): (i) gluon-gluon fusion, $g\,g\to t\overline t+X$, 
contributing by 80--95\% to the total pair production (the remaining 5--20\% issuing from quark-antiquark
annihilation), (ii) $t$-channel single-top electroweak production $q\,b\to q'\,t+X$ (the $s$-channel process,
decreasing with energy, amounts to 5--1.5\% of the total single-$t$ cross section), and 
(iii) associated top plus $W$-boson, $g\,b\to W\,t+X$, production (increasing with energy, it amounts to
25--50\% of the $t$-channel process). In pp collisions at the LHC, top-quark production is already being
used as a high-precision tool to constrain the gluon distribution function in the proton at next-to-NLO (NNLO)
accuracy~\cite{Czakon:2013tha}. At the energies attainable in p--Pb and Pb--Pb collisions at the FCC, the same
studies would be feasible also for the nuclear PDFs. The top-quark, the only coloured particle that decays
before its hadronization, decays almost exclusively into a $t\to W\,b$ final-state with a nearly 100\% branching
ratio, and the $W$ can themselves decay leptonically ($t\to W\,b \to \ell\,\nu,b$, one-third of the
times) or hadronically ($t\to W\,b \to \qqbar\,b$, two-thirds of the times). 
Its short lifetime, $\tau_0 = \hbar/\Gamma_{\rm t}\approx 0.1$~fm/$c$, implies that most of the (non-boosted)
top quarks will decay before any significant gluon radiation and before the formation of any
strongly-interacting medium (typical QGP formation times are ${\cal O}$(1~fm/$c$)). To avoid any potential bias
from parton energy loss effects on the top-decay quarks, one can study nPDFs by analysing the
distributions of the (isolated) charged leptons ($\ell$ = $e$, $\mu$) in events clearly identified as containing
top-quarks.
To estimate the impact that the FCC would have on nuclear gluon densities the computed top-pair cross sections in
pp, p--Pb and Pb--Pb with analysis cuts (see discussion for Table.~\ref{tab:tyields}) have been binned in the rapidity $y_\ell$ of the decay leptons. 
In the p--Pb and Pb--Pb cases the calculations include EPS09 nuclear modifications~\cite{Eskola:2009uj} for
PDFs. 

\begin{figure}[t]
\begin{minipage}{\textwidth}
\includegraphics[width=0.49\textwidth]{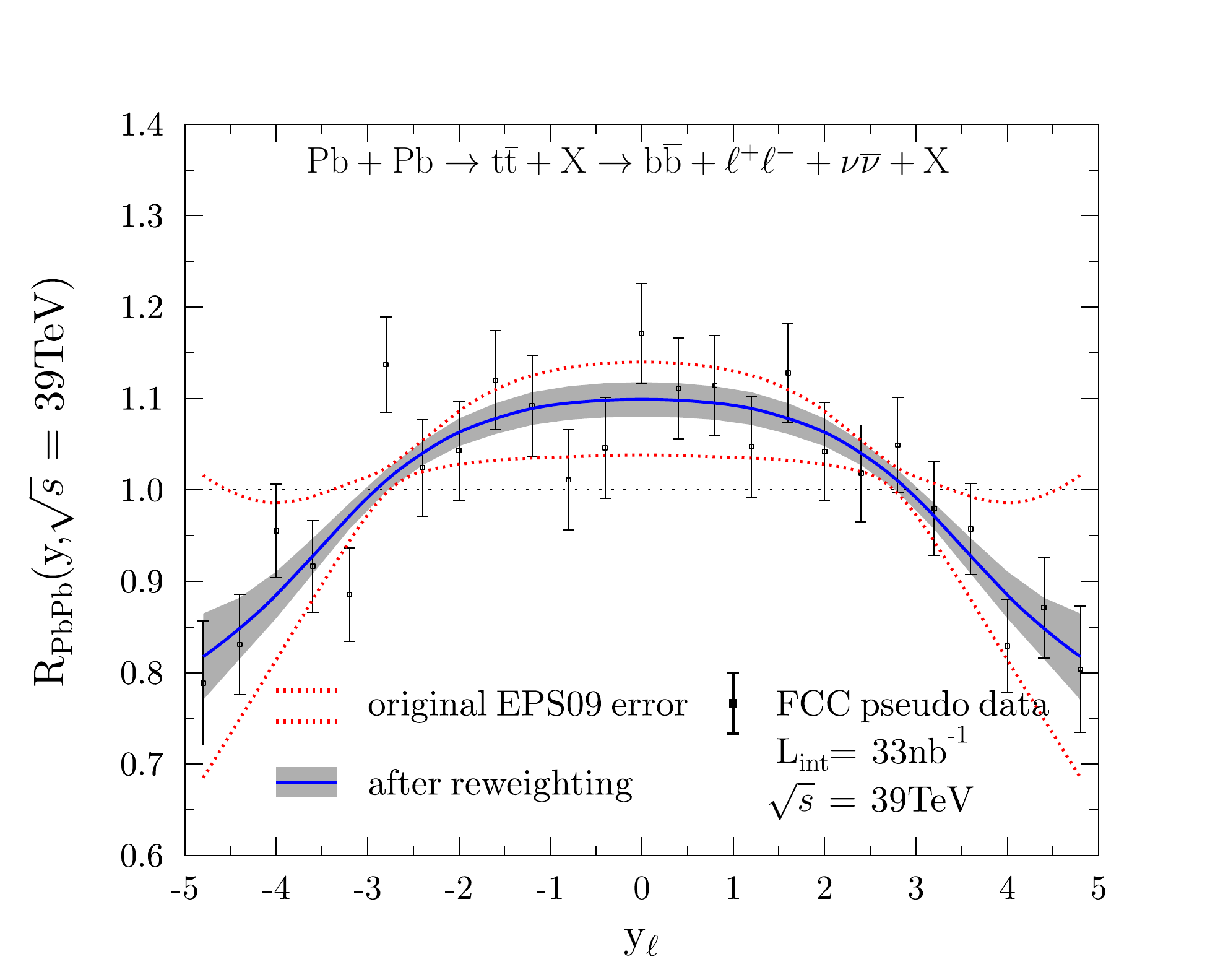}
\includegraphics[width=0.49\textwidth]{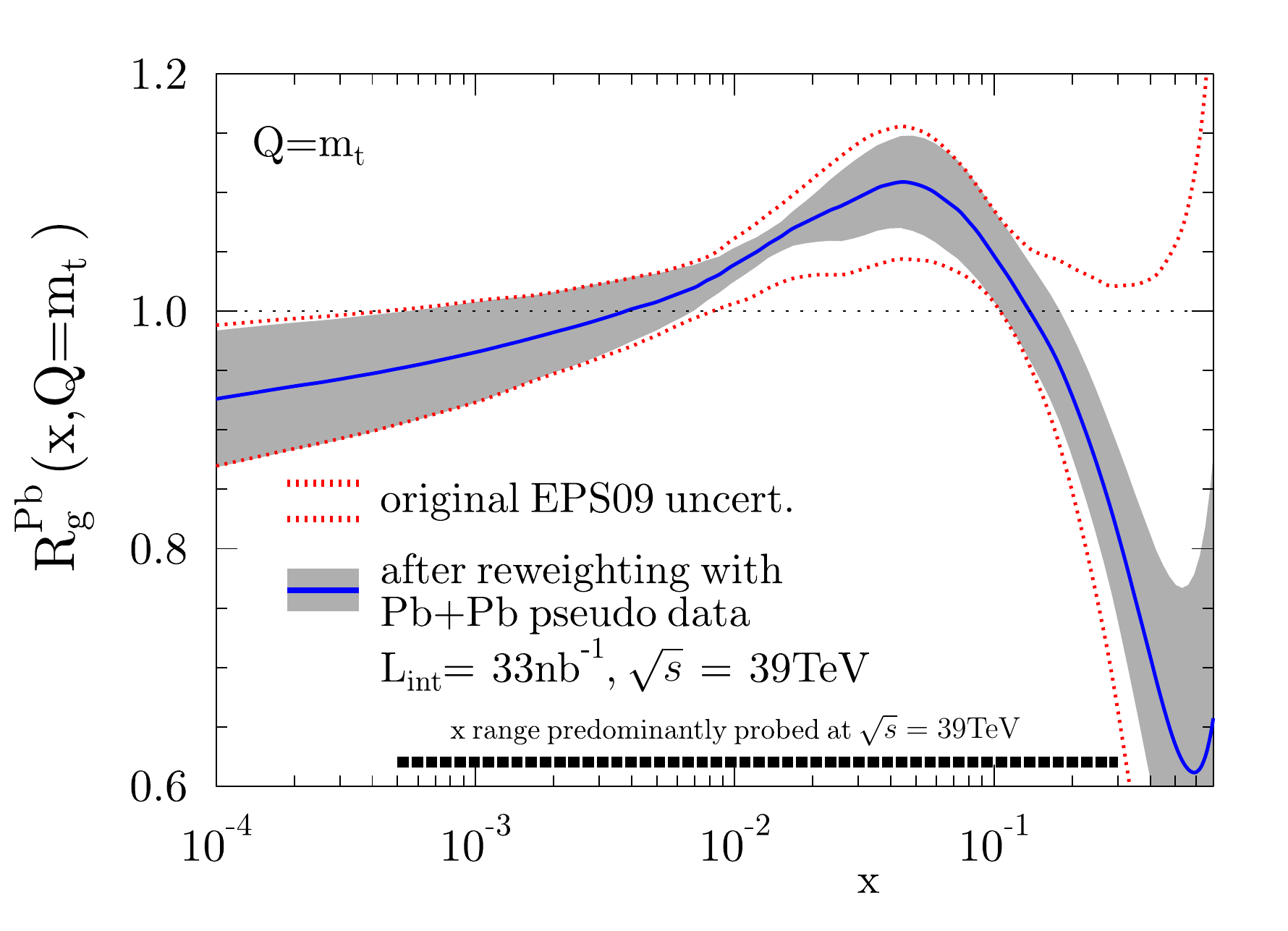}
\includegraphics[width=0.49\textwidth]{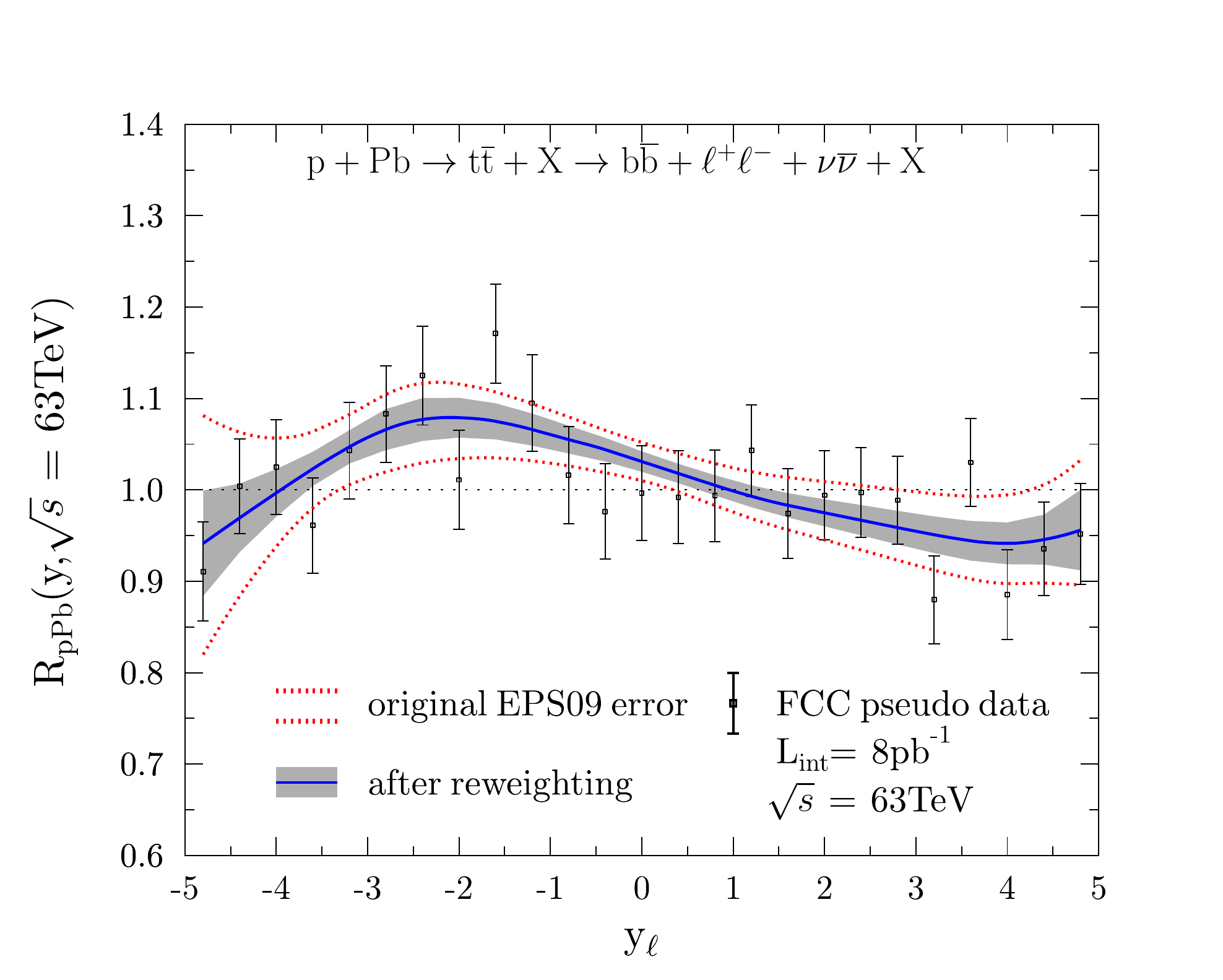}
\includegraphics[width=0.49\textwidth]{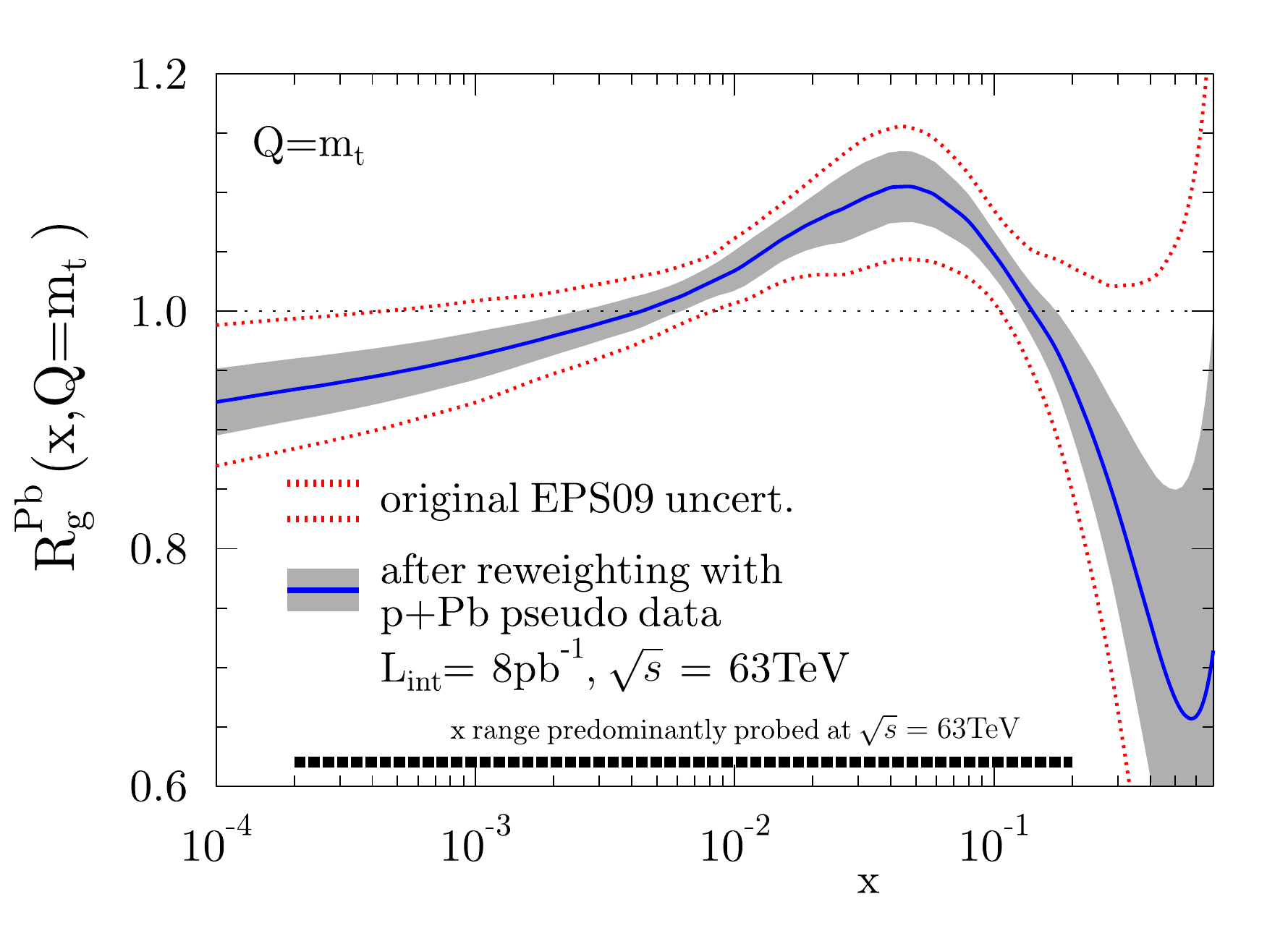}
\caption{Left: FCC pseudodata for nuclear modification factors expected in Pb--Pb (upper panel) p--Pb
  (lower panel). Right: Original EPS09 gluon nuclear modification  at $Q=m_{\rm top}$ and 
  estimated improvement in it obtained by reweighting using the Pb--Pb (upper panel) and p--Pb (lower panel)
  FCC pseudodata. The figures are adapted from Ref.~\cite{d'Enterria:2015jna}.} 
\label{fig:tnPDF}
\end{minipage}
\end{figure}

The left panels of Fig.~\ref{fig:tnPDF} show pseudodata distributions for the expected nuclear modification
factors $R_{\rm pPb}(y_\ell) = \dd\sigma_{\rm pPb}(y_\ell)/(A\,\dd\sigma_{\rm pp}(y_\ell))$ and $R_{\rm PbPb}(y_\ell) =
\dd\sigma_{\rm   PbPb}(y_\ell)/(A^2\,\dd\sigma_{\rm pp}(y_\ell))$ in minimum-bias collisions (Table.~\ref{tab:tyields}). The assigned uncertainties include statistical errors
based on the luminosities of 8~pb$^{-1}$ and 33~nb$^{-1}$ for p--Pb and Pb--Pb, a 5\% uncorrelated systematic uncertainty, and an overall 5\% normalization
error. The effects these pseudodata would have in EPS09 global fit of nuclear PDFs are quantified via the
Hessian reweighting technique~\cite{Paukkunen:2014zia}. The expected impact of these measurements on gluon PDF nuclear modification factor
$R_{\rm g}^{\rm Pb}(x,Q^2) = g^{\rm Pb}(x,Q^2)/g^{\rm p}(x,Q^2)$ at $Q^2=m^2_{\rm top}$ are shown in 
Fig.~\ref{fig:tnPDF} (right). The uncertainties on the nuclear gluon PDF are observed to reduce by more than
50\% in some regions of $x$.

\subsubsubsection{Constraining nPDFs with dijets in p--Pb collisions}

Dijet measurements at LHC have proven to be sensitive to the shadowing and anti-shadowing of quarks in nPDFs. A shift of the average rapidity of the dijet system for jets with $\pt > 100 \GeVc$ was observed in p--Pb collisions~\cite{Chatrchyan:2014hqa}. This shift is consistent with the expectations from (small) modifications due to nuclear parton density functions. Precision measurements of this type can thus improve our knowledge of nuclear PDFs in regions that are not well constrained by existing measurements~\cite{Helenius:2015wda}.

\subsubsection{Relation with the proposed electron--hadron colliders}

\begin{figure}[!b]
\begin{center}
\includegraphics[width=0.6\textwidth]{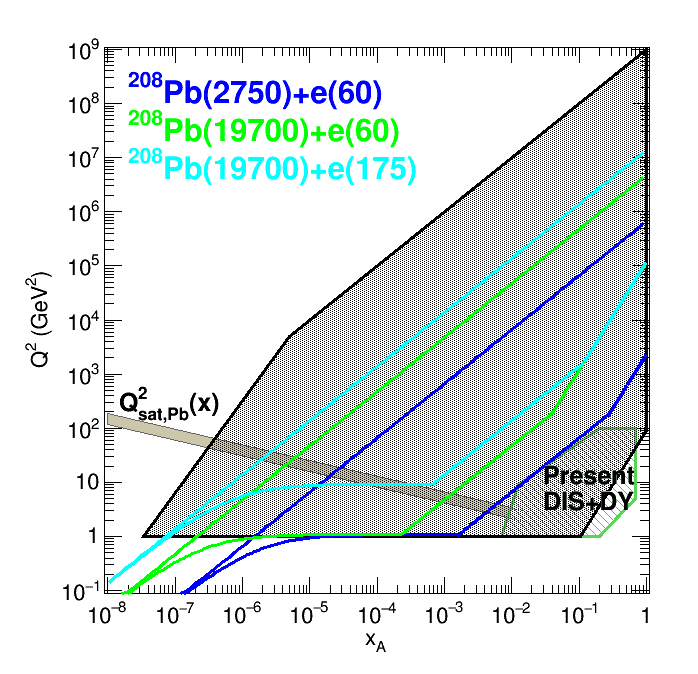}
\end{center}
\caption{\label{fig:kinplane2}Regions in the $x-Q^2$ plane already studied in present DIS and DY experiments (light green), and accessible at the LHeC (dark blue) and different versions of the FCC-he (light green and light blue), and at the FCC p--Pb (black). Values of the saturation scale are shown for illustration.}
\end{figure}

Electron--proton/ion colliders at high energies, while having smaller kinematic coverage with respect to hadronic colliders, offer the advantage of fully constrained kinematics and a much cleaner experimental environment. It is widely recognized that the physics programs at electron--proton/ion colliders and hadron colliders are complementary, and that the precise data obtained from electron--proton/ion colliders can further enhance the physics opportunities of hadron colliders. In the context of the Electron-Ion Collider in the USA~\cite{Accardi:2012qut}, that is already included in the NSAC 2015 Long Range Plan, 
this general argument has been substantiated in detail and the synergies between the ongoing RHIC program and an EIC project are documented. In the context of the LHC hadron collider program, 
the Large Hadron Electron Collider LHeC~\cite{AbelleiraFernandez:2012cc} and the  FCC-he are proposed facilities to provide electron-proton/ion collisions in the TeV regime in the centre-of-mass.
With the larger kinematic reach at the TeV scale, these latter projects are well-positioned to reach conclusive evidence for the existence of a new saturated regime of QCD. 
  For that, the electron beam from an accelerator of about 60 GeV would collide with the LHC or FCC proton or heavy-ion beams. The kinematic coverage of such machines is given in Fig.~\ref{fig:kinplane2}. 
They are clearly complementary with the FCC-hh as they should provide a precise knowledge on the partonic structure of nucleons and nuclei and on the small-$x$ dynamics. Such knowledge will allow QCD studies at hadronic colliders, both in pp and in p--A and A--A, to become rid of several of their most important uncertainties, as PDFs and the values of $\alpha_s$ and of heavy quark masses. They will also set the benchmark for precision factorisation tests.

\subsection{Exclusive photoproduction of heavy quarkonia}
\label{sec:HI_upc}

All charges accelerated at high energies generate electromagnetic fields which, in the equivalent photon
approximation (EPA)~\cite{vonWeizsacker:1934nji,Williams:1934ad,Fermi:1925fq}, can be considered as quasireal $\gamma$ beams of very low virtuality $Q^{2} <
1/R^{2}$, where $R$ is the radius of the charge, \ie\ $Q^2\approx 0.08$~GeV$^2$ for protons
($R\approx 0.7$~fm), and $Q^2<4\cdot 10^{-3}$~GeV$^2$ for nuclei ($R_{\rm A}\approx 1.2\,A^{1/3}$~fm, for mass number
$A>16$). The photon spectra have a typical $E_{\gamma}^{-1}$ power-law fall-off up to energies of the order of
the inverse Lorentz-contracted radius, $\omega_{\rm max}\approx\gamma_{_{\rm L}}/R$, where 
$\gamma_{_{\rm  L}}=\sqrtsnn/(2\,m_{\rm p})$ is the Lorentz factor of the proton or ion. Given that the photon flux
scales with the square of the emitting charge ($Z^2$), the emission of quasireal photons from the Pb-ion is
strongly enhanced compared to that from proton (or electron) beams. The basic characteristics of photon-induced
interactions in ``ultraperipheral'' collisions (UPCs) of proton~\cite{d'Enterria:2008sh} and lead (Pb)
beams~\cite{Baltz:2007kq} ---occurring at impact parameters larger than the sum of their radii and thereby
largely suppressing their hadronic interaction--- at the FCC are listed in Table~\ref{tab:UPCs}.

\begin{table}[b!]
\begin{center}
\caption{Basic characteristics of UPCs at the FCC-hh: 
(i) nucleon-nucleon \cm\ energy, $\sqrtsnn$,
(ii) beam Lorentz factor, $\gamma_{_{\rm L}}$,   
(iii) maximum photon energy in the \cm\ frame, $\omega_{\rm max}$, 
and (iv) maximum $\gamma$-proton, $\gamma$-ion \cm\ energy, $W_{\gamma {\rm p},\gamma A}^{\rm max}$.
}
\label{tab:UPCs}
\vspace{0.4cm}
\begin{tabular}{l|cccc} \hline 
\hspace{-0.4cm} System  \hspace{-0.2cm} & \hspace{-0.1cm} $\sqrtsnn$
\hspace{0.01cm} & 
\hspace{0.01cm} $\gamma_{\rm L}$ \hspace{0.01cm} 
& \hspace{0.01cm} $\omega_{\rm max}$ \hspace{0.01cm} 
& \hspace{0.01cm} $W_{\gamma {\rm p},\gamma A}^{\rm max}$ \hspace{0.01cm} \\
                       & (TeV) &  ($\times 10^3$) & (TeV) & (TeV) \\ \hline
\hspace{-0.3cm} \pp    & 100 &  53.0  & 17.6 & 10.0 \\ 
\hspace{-0.3cm} \pPb   &  63 & 33.5 & 0.95 & 10.0 \\ 
\hspace{-0.3cm} \PbPb  &  39 & 21.0  & 0.60 & 7.0 \\ \hline
\end{tabular}
\end{center}
\end{table}

Exclusive photoproduction of vector mesons in UPCs of protons or ions ---where an exchanged
quasireal photon ``materializes'' into a $\QQbar$ bound state after interacting with the gluon field of the
``target'' proton (ion), without breaking the colour flow and thereby leaving intact the incoming hadrons---
is depicted in Fig.~\ref{fig:gammaPb_QQbar} (left). Since in such processes the gluon couples {\em directly}
to the $c$ or $b$ quarks  and the  cross section is proportional to the gluon density {\em squared}, they
provide a very clean probe of the gluon density in the ``target''
hadron~\cite{Baltz:2008bb,d'Enterria:2007pk,Lansberg:2008zm}, with the large mass of the $\jpsi$ and 
$\Upsilon$ mesons providing a hard scale for pQCD calculations~\cite{Frankfurt:2001,Ryskin:2008rt}.
Exclusive quarkonia photoproduction was measured in electron-proton collisions at
HERA~\cite{ZEUS:1998,H1:2000,ZEUS:2009,ZEUS:2012}, and in ultraperipheral proton--proton and nuclear collisions
at the LHC~\cite{LHCb:2013aa,ALICE:2013aa,Abbas:2013oua,TheALICE:2014dwa,Aaij:2015kea}.

\begin{figure}[t!]
\centering
 \raisebox{3.cm}{\includegraphics[width=0.4\textwidth]{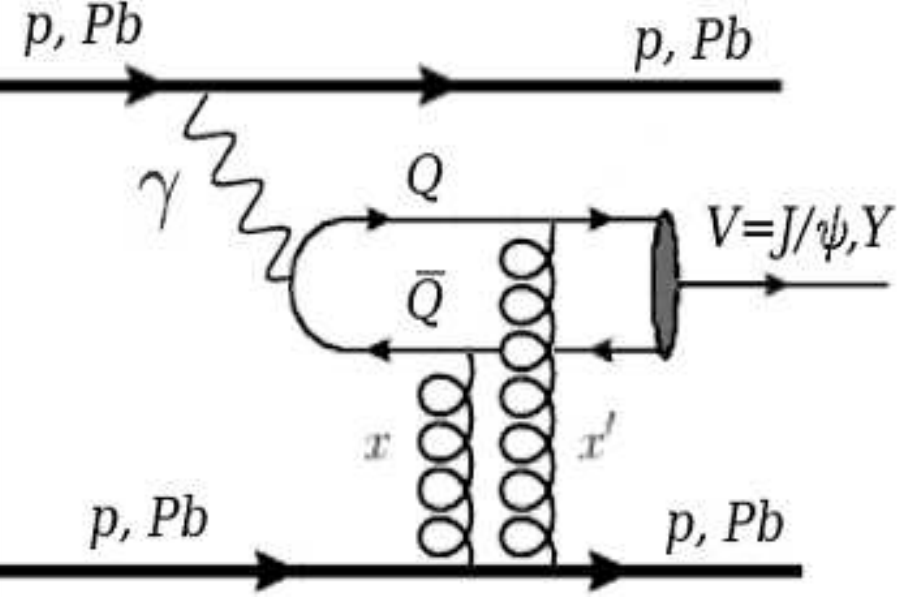}}\hspace{0.25cm}
 \includegraphics[width=0.57\textwidth,height=9.cm]{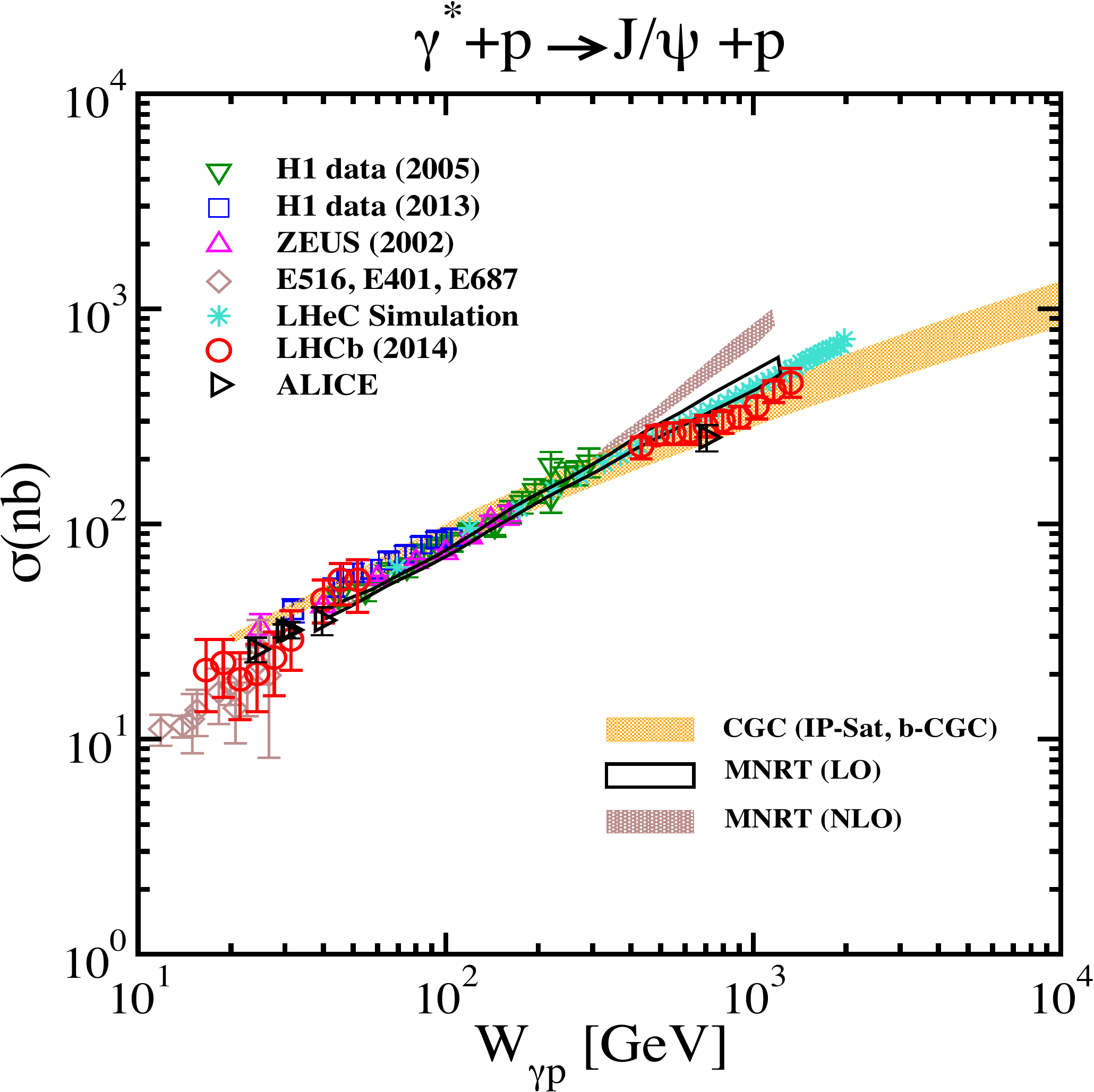}
 \caption{Left: Diagram representing exclusive quarkonia photoproduction in UPCs. Right: Dependence of the
   exclusive $\jpsi$ photoproduction cross section on the photon-hadron \cm\ energy in the regions covered by
   HERA, LHC and future FCC studies~\cite{Armesto:2014sma}.}
\label{fig:gammaPb_QQbar}
\end{figure}

Their measured cross sections rise steeply with photon-hadron
centre-of-mass energy $W_{\gamma {\rm p}}$, following a power-law
dependence $W_{\gamma {\rm p}}^{\delta}$ with
$\delta=0.7$--$1.2$~\cite{H1:2000,ZEUS:2009}, reflecting the steep rise in the gluon density in the hadrons at 
increasingly lower values of parton fractional momentum $x$ (Fig.~\ref{fig:gammaPb_QQbar}, right).
At the FCC, $\jpsi$ and $\Upsilon$ photoproduction will reach photon-hadron \cm\ energies as large as 
$W_{\gp}\approx 10$~TeV, and thereby probe the gluon density in the proton and ion in an
unexplored region values of Bjorken-$x$ as low as $x\approx M_{\jpsi,\Upsilon}^2/W_{\gamma {\rm p}}^2\approx
10^{-7}$, at least two orders of magnitude below the range probed at
the LHC (Fig.~\ref{fig:kinplane}, right). 
As discussed in Section~\ref{sec:HI_smallxintro},
in such a low-$x$ regime,
non-linear (gluon recombination) QCD effects may become important, leading to a saturation of
the PDFs~\cite{Gribov:1983,Mueller:1986,McLerran:1994}. The evolution of the
cross section with energy is very sensitive to the underlying small-$x$ dynamics as shown by the bands in
Fig.~\ref{fig:gammaPb_QQbar} (right) showing different theoretical predictions based on LO and
LO pQCD calculations~\cite{Ryskin:2013jmr}, colour dipole
formalism~\cite{Sampaio:2013aa,Santos:2014zna}, and gluon saturation
approaches (labelled CGC)~\cite{Armesto:2014sma,Rezaeian:2013tka,Rezaeian:2012ji}.

\clearpage 
\section{Contributions to other sectors of high-energy physics
  \footnote{Editors: D.~d'Enterria, J.-P.~Lansberg}}
\label{sec:HI_otherhep}

\subsection{Photon--photon collisions}
\label{sec:HI_gammagamma}

In Section~\ref{sec:HI_upc} we have discussed how the large
electromagnetic fields produced by accelerated protons or ions can be considered as quasireal $\gamma$ beams of very low virtuality.
Photon--photon collisions in UPCs of proton~\cite{d'Enterria:2008sh} 
and lead (Pb) beams~\cite{Baltz:2007kq} have been experimentally observed at the
LHC~\cite{Chatrchyan:2012tv,Chatrchyan:2013foa,Abbas:2013oua,Aad:2015bwa}.
Although the $\gamma$ spectrum is harder for smaller charges --which favours proton over nuclear beams in the
production of heavy diphoton systems-- each photon flux scales with the squared charge of the hadron, $Z^2$, and
thus $\gaga$ luminosities are extremely enhanced for ion beams
($Z^4=5\cdot 10^{7}$ in the case of
\PbPb). Figure~\ref{fig:gamgam_lumi} (left) shows a typical $\gaga$ process in UPCs (light-by-light
scattering, in this particular case), and Table~\ref{tab:upc} summarises the relevant parameters for
ultraperipheral pp, p--Pb, and \PbPb\ collisions at FCC energies.

\begin{figure}[!hb]
\centering
\raisebox{1.35cm}{\includegraphics[width=0.32\columnwidth]{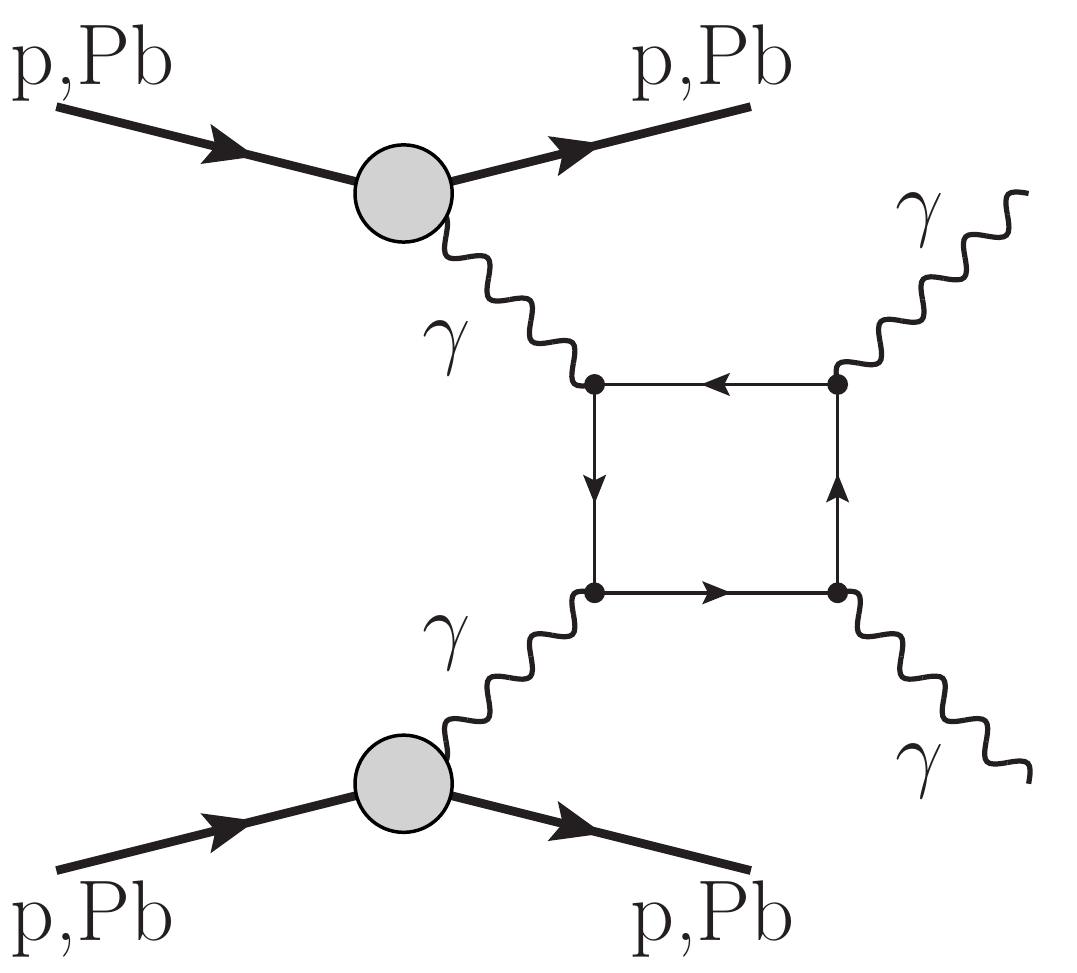}}\hspace{0.25cm}
\includegraphics[width=0.6\columnwidth]{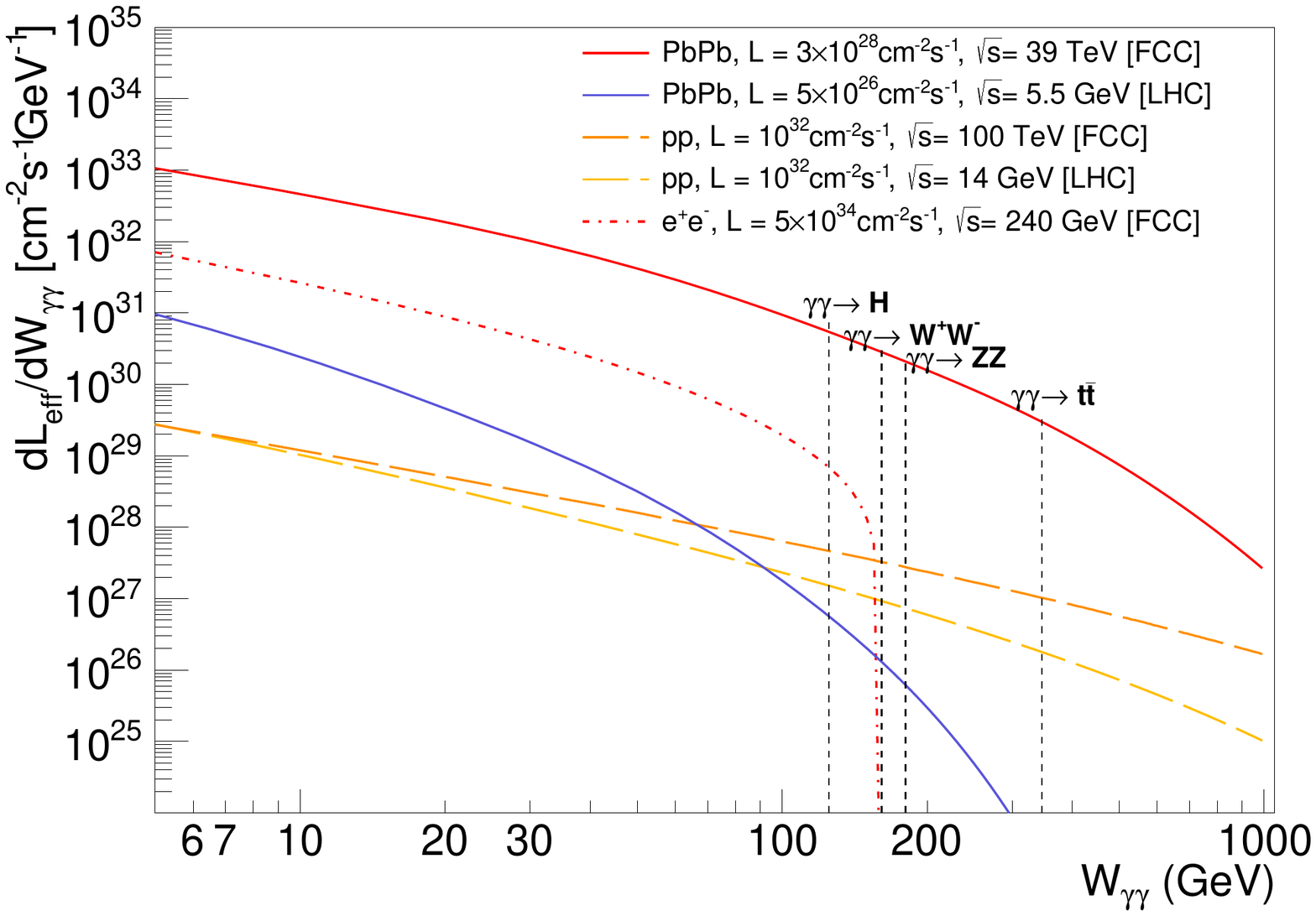}
\caption{Left: Diagram of elastic $\gaga\to\gaga$ scattering in an UPC where the initial-state photons are
  emitted coherently by the protons and/or nuclei which survive the electromagnetic interaction.
  Right: Effective photon--photon luminosities as a function of $\gaga$ \cm\ energy ($W_{\gaga}$) for five
  colliding systems at FCC and LHC energies: Pb--Pb at $\sqrts =39,\,5.5$~TeV (at their corresponding nominal
  beam luminosities); pp at $\sqrts=100,\, 14$~TeV (corresponding to 1~fb$^{-1}$ integrated luminosities); 
  and $\epem$ at $\sqrts=240$~GeV (FCC-ee nominal luminosity per IP). The vertical dashed lines
  indicate the energy thresholds for Higgs, $W^+W^-$, $Z\,Z$, and $\ttbar$ production.}
\label{fig:gamgam_lumi}
\end{figure}

\begin{table}[htpb!]
\begin{center}
\caption[]{Characteristics of $\gaga\to\gaga$ processes at the FCC-hh: 
(i) nucleon-nucleon \cm\ energy, $\sqrtsnn$, (ii) integrated luminosity $\mathcal{L}_{\rm AB}\cdot\Delta t$ 
($\mathcal{L}_{\rm AB}$ are beam luminosities --for low pileup in the pp case-- and a ``year''
is $\Delta t=10^{7}$~s for pp, and $10^{6}$~s in the ion mode), 
(iii) beam Lorentz factor, $\gamma_{\rm L}$, 
(iv) maximum photon energy in the \cm\ frame, $\omega_{\rm max}$, and
(v) maximum photon-photon \cm\ energy, $\sqrt{s_{\gaga}^{\rm max}}$. The last two columns are
(vi) cross sections, and (vii) expected number of counts/year after selection cuts, for the
exclusive $\gaga\to\gaga$ process at masses above 5~GeV.}
\label{tab:upc}
\vspace{0.4cm}
\begin{tabular}{l|ccccc|cc} \hline 
\hspace{-0.4cm} System  \hspace{-0.2cm} & \hspace{-0.1cm} $\sqrtsnn$ \hspace{0.01cm} & \hspace{0.01cm}$\mathcal{L}_{\rm AB}\cdot\Delta t$ \hspace{0.01cm} 
& \hspace{0.01cm} $\gamma_{\rm L}$ \hspace{0.01cm} 
& \hspace{0.01cm} $\omega_{\rm max}$ \hspace{0.01cm} 
& \hspace{0.01cm} $\sqrt{s_{\gaga}^{\rm max}}$ \hspace{0.01cm} 
& \hspace{0.01cm} $\sigmagg^{\rm excl}$\hspace{0.01cm} 
& \hspace{0.01cm} $N_{\gaga}^{^{\rm cuts}}$ \hspace{0.01cm} \\
       & (TeV) &  (per year) &  ($\times 10^3$) & (TeV) & (TeV) &      \multicolumn{2}{c}{$[\mgg>5~{\rm GeV}]$} \\ 
\hline
\hspace{-0.3cm} pp    & 100 &  1~fb$^{-1}$  & 53.0  & 17.6 & 35.2 & 240 $\pm$ 24 fb &  50 \\ 
\hspace{-0.3cm} \pPb   &  63 &  8~pb$^{-1}$  & 33.5 & 0.95 & 1.9  & 780 $\pm$ 78 pb & 1\,200 \\ 
\hspace{-0.3cm} \PbPb  &  39 & 33~nb$^{-1}$  & 21.0  & 0.60 & 1.2  &
1.85$\pm$ 0.37 $\mu$b & 16\,500\\ 
\hline
\end{tabular}
\end{center}
\end{table}

The figure of merit for UPC $\gaga$ processes is the effective $\gaga$ luminosity, 
$\dd\mathcal{L}_{\rm eff}/\dd W_{\gaga} \equiv
\mathcal{L}_{AB}\,\dd\mathcal{L}_{\gaga}/\dd W_{\gaga}$, where ${\cal L}_{\rm AB}$ is the
collider luminosity for the $A$\,$B$ system and
$\dd\mathcal{L}_{\gaga}/\dd W_{\gaga}$ is the photon--photon luminosity
as a function of the $\gaga$ centre-of-mass energy $W_{\gaga}$, obtained integrating the two photon fluxes over all
rapidities $y$, \ie\ $\dd^2\mathcal{L}_{\gaga}/\dd W_{\gaga}\dd y=(2/W_{\gaga})f_{\gamma/A}(W_{\gaga}/2
e^{y})f_{\gamma/B}(W_{\gaga}/2 e^{-y})$.  
Figure~\ref{fig:gamgam_lumi} (right) shows a comparison of the $\dd\mathcal{L}_{\rm eff}/\dd W_{\gaga}$ reachable as a
function of $W_{\gaga}$ for five different colliding systems at LHC and FCC energies. Two-photon centre-of-mass
energies at the FCC will reach for the first time the range beyond 1~TeV. Clearly, Pb--Pb at $\sqrtsNN=39$~TeV provides
the largest two-photon luminosities of all colliding systems. The effective luminosities are very high up 
to large diphoton masses (in the next section we present a case study for the measurement of
light-by-light scattering above $\mgg=5$~GeV that profits from the large photon fluxes available at FCC). The
vertical lines in Fig.~\ref{fig:gamgam_lumi} show the thresholds for photon-fusion production of Higgs,
$W^+W^-$, $Z\,Z$, and $\ttbar$. All such processes, sensitive to different tests of the electroweak sector of
the Standard Model (SM)~\cite{Teles:2015xua}, such as anomalous quartic-gauge couplings and top-electroweak
moments, should have visible counts at the FCC(Pb--Pb) although dedicated studies need to be carried out in
order to estimate the significance of the corresponding observations.

\subsubsection*{Case study: light-by-light scattering}

It has been demonstrated that one can detect
the very rare elastic scattering of two photons in vacuum, $\gaga\to\gaga$, a simple yet unobserved process so
far (Fig.~\ref{fig:gamgam_lumi}, left), using the large quasireal photon fluxes of the ions accelerated at TeV
energies at LHC and FCC
energies~\cite{d'Enterria:2013yra,d'Enterria:2016qsv}. 
Such a measurement at the LHC will constitute the first-ever observation of
a fundamental quantum mechanical process in the lab, whereas at the FCC, due to the higher diphoton masses
reached, one may be sensitive to physics beyond the SM through new heavy charged particles contributing to the
virtual loop in Fig.~\ref{fig:gamgam_lumi} (left) such as, \eg\ from SUSY
particles~\cite{Gounaris:1999gh}. Light-by-light (LbyL) scattering has also been proposed as a tool to
search for monopoles~\cite{Ginzburg:1998vb}, axions~\cite{Bernard:1997kj}, unparticles~\cite{Kikuchi:2008pr}, 
low-scale gravity effects~\cite{Cheung:1999ja}, and non-commutative interactions~\cite{Hewett:2000zp}.

We update here the results of Ref.~\cite{d'Enterria:2016qsv} to
account for factors $\times 6.5$
and $\times 8$ increase in the FCC luminosities considered now for the
\PbPb\ and \pPb\ systems (see Section~\ref{sec:HI_machine}). The UPC final-state
signature is the exclusive production of two photons, \ABgaga, with the diphoton final-state measured in the
central detector, and A,B\,=\,p,Pb surviving the electromagnetic interaction scattered at very low
angles with respect to the beam. In the equivalent photon
approximation (EPA), the elastic $\gaga$ production cross section in UPCs of hadrons A
and B factorizes into the product of the elementary $\gaga\rightarrow \gaga$ cross section at $\sqrtsgg$,
convoluted with the photon fluxes $f_{\rm \gamma/A,B}(\omega)$ of the two colliding beams: 
\begin{equation}
\sigmagg^{\rm excl}=\sigma(\rm{A} \rm{B} \xrightarrow{\gaga} \rm{A} \gaga \rm{B})=
\int d\omega_1 d\omega_2  \frac{f_{\rm \gamma/A}(\omega_1)}{\omega_1} \frac{f_{\rm \gamma/B}(\omega_2)}{\omega_2} \sigmagg(\sqrtsgg)\,,
\label{eq:two-photon}
\end{equation}
where $\omega_1$ and $\omega_2$ are the energies of the photons emitted by A and B, 
$f_{\rm \gamma/p}(\omega)$ is the spectrum derived from the proton elastic form factor~\cite{Budnev:1974de},
and $f_{\rm \gamma/A}(\omega)$ the impact-parameter dependent expression for the ion 
spectrum~\cite{Bertulani:1987tz} including a correction equivalent to ensuring that all collisions are purely
exclusive, \ie\ without hadronic overlap and breakup of the colliding beams~\cite{Cahn:1990jk}.
The \madgraph\ v.5 Monte Carlo (MC)~\cite{Alwall:2007st} framework is used to convolute the $\gamma$ fluxes with the
LO expression for the $\sigmagg$ cross section~\cite{Bern:2001dg} including all quark and lepton loops, but
omitting the $W^\pm$ contributions which are only important at
$\mgg\gtrsim 200$~GeV. Inclusion of
next-to-leading-order QCD and QED corrections increases $\sigmagg$ by
a few percent only~\cite{Bern:2001dg}.
Propagated
uncertainties to the final cross sections are of order $\pm 10\%$ ($\pm
20\%$) for pp and \pPb\ (\PbPb)
collisions, covering different form-factors parametrizations and the convolution of the nuclear photon
fluxes. The obtained signal cross sections $\sigmagg^{\rm excl}$ are listed in Table~\ref{tab:upc}, and
plotted as a function of \cm\ energies in the range $\sqrtsnn =1$--$100$~TeV in Fig.~\ref{fig:LbyL} (left). 
The increase in cross sections from LHC to FCC is of $\mathcal{O}$(2--5). The cross sections are in at the
$\mu$b level at the FCC, clearly showing the importance of the $Z^4$-enhanced photon-flux for ions compared
to protons.

\begin{figure}[!t]
\centering
\includegraphics[width=0.49\columnwidth,height=7.cm]{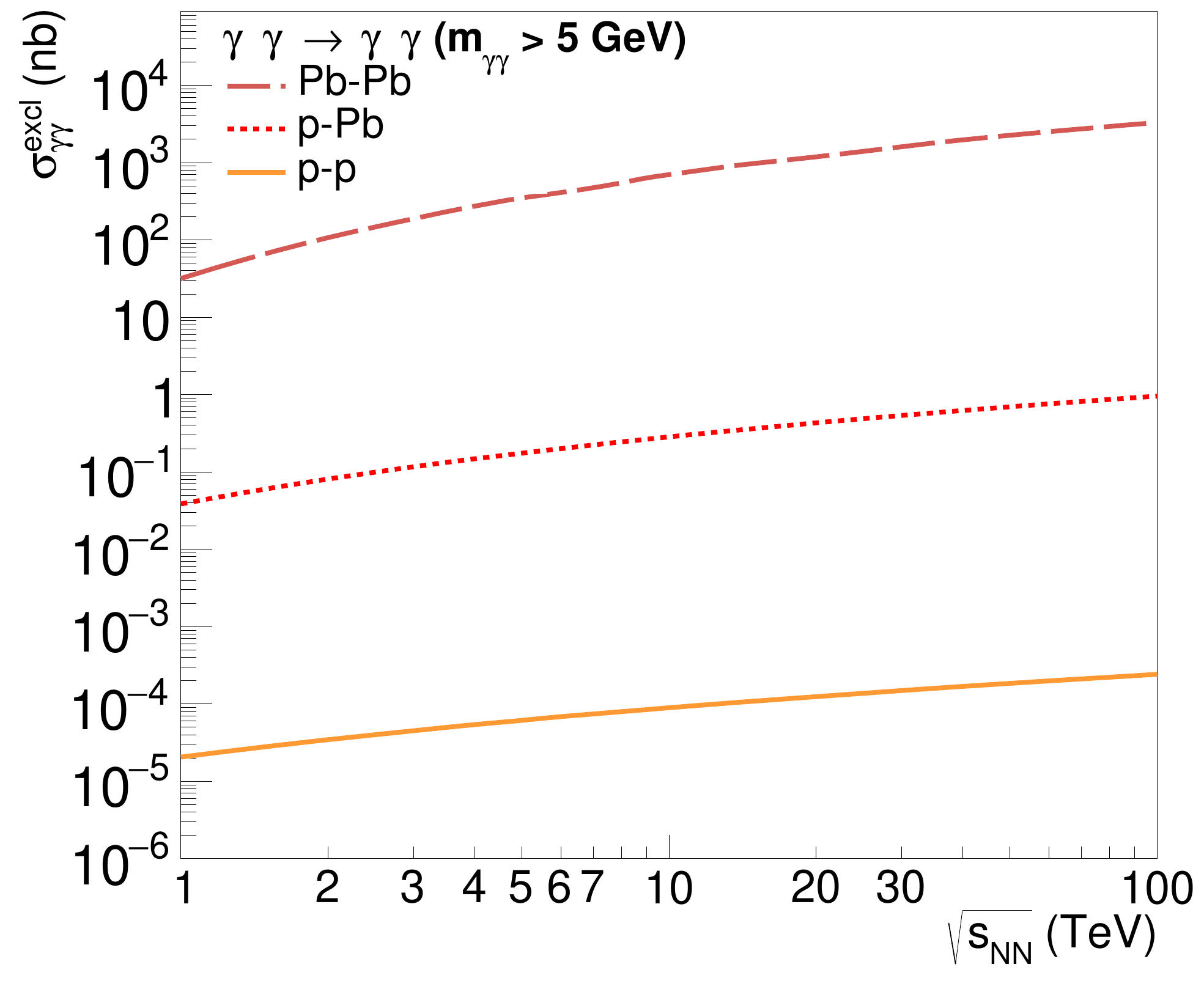}
\includegraphics[width=0.49\columnwidth,height=7.cm]{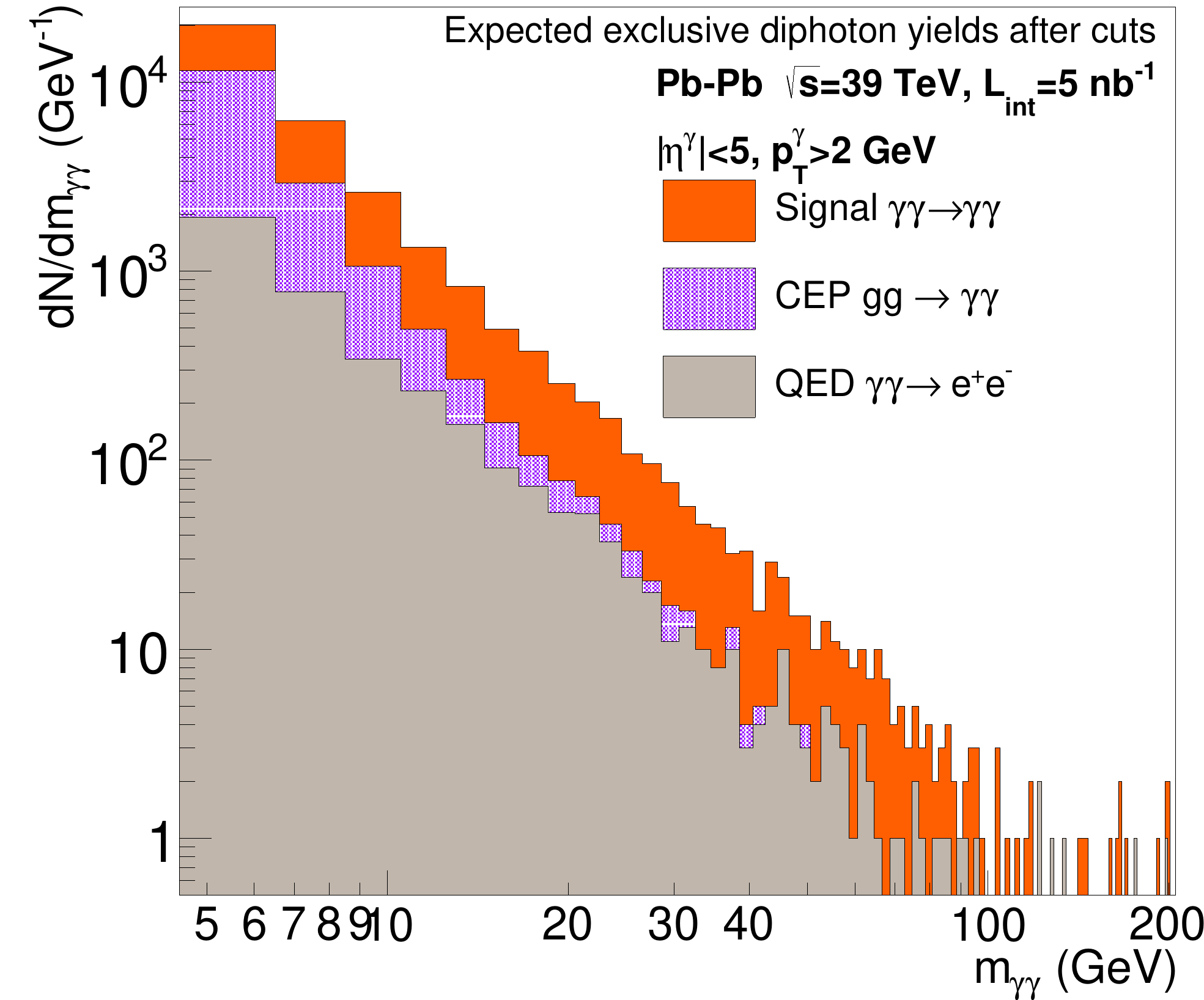}
\caption{
Left: Cross sections for $\gaga\to\gaga$, with pair masses above 5~GeV, in ultraperipheral \PbPb\
(top curve), \pPb\ (middle) and pp (bottom) collisions as a function of \cm\ energy~\cite{d'Enterria:2016qsv}.
Right: Yields as a function of diphoton invariant mass for elastic $\gaga$, plus CEP-$\gaga$ and
QED backgrounds, expected in \PbPb\ at FCC after analysis cuts.}
\label{fig:LbyL}
\end{figure}

The detectable number of $\gaga\to\gaga$ events is estimated by considering nominal luminosities for each
system, geometric detector acceptance, and reconstruction efficiencies. 
Standard trigger+acceptance requirements (both photons with $\pT^\gamma >2$~GeV/$c$ within $|\eta^\gamma|<5$)
reduce the yields by $\varepsilon_{\rm acc}\approx 0.3$ (pp and \pPb), and 
$0.4$ (\PbPb). Accounting for typical offline $\gamma$ reconstruction and identification efficiencies
($\varepsilon_{\rm rec,id\,\gamma}\approx 0.8$ in the photon energy range of interest) results in final
combined signal efficiencies of
$\varepsilon_{\gaga}=\varepsilon_{\rm acc}\cdot\varepsilon_{\rm
  rec,id\,\gamma}^2\approx 20\%$ (pp, \pPb) and
$26\%$ (\PbPb). The number of events expected per year 
are obtained via  $N_{\gaga}^{\rm excl}=\varepsilon_{\gaga}\cdot\sigma_{\gaga}^{\rm excl}\cdot\mathcal{L}_{\rm AB}\cdot\Delta t$ (Table~\ref{tab:upc}). 
The nominal \pPb\ and \PbPb\ luminosities are low enough to keep the number of simultaneous collisions 
well below one, but  pileup in
pp is very  high and only $\cal{O}$(1~fb$^{-1}$/year) can be collected under conditions that preserve the rapidity gaps
adjacent to the central $\gaga$ system (unless one can tag the outgoing quasielastically-scattered protons~\cite{Fichet:2015nia}). 
Clearly, \PbPb\ provides the best signal counting rates, with statistical uncertainties of order $1/\sqrt{N_{\gaga}^{\rm excl}} \sim 1\%$.


Three potential backgrounds share the same (or very similar) final-state signature as $\gaga\to\gaga$: 
(i) diphoton production through a quark-loop in the color-singlet exchange of two gluons, \ABglgl\ (``central
exclusive production'', CEP), 
(ii) QED $\gaga\to\epem$ events, with both $e^{\pm}$ misidentified as photons, and
(iii) diffractive Pomeron-induced ($\Pom\Pom$, or $\gamma\Pom$) processes with final-states containing two
photons plus rapidity gaps. The latter diffractive and $\gamma$-induced final-states have larger 
$\pT^{\gaga}$ and diphoton acoplanarities than $\gaga\to\gaga$, and can be efficiently removed. However,
the CEP $gg\to\gaga$ background (observed at Tevatron~\cite{Aaltonen:2011hi} and
theoretically described by the \superchic~\cite{HarlandLang:2010ep} MC) scales with the fourth power of the gluon
density and is a large potential background. In \PbPb\ at FCC(39~TeV), the CEP cross section within
$|\eta|<5$ is indeed very large: $\sigma_{gg\to\gaga}^{\rm CEP}[m_{\gaga}>5$~GeV] = $1.3$~nb~$\times
208^2 \times (R_g^{^{\rm Pb/p}})^4$ $\approx 14$~$\mu$b (with a factor
of $\sim 3$ uncertainty) as obtained
with \superchic~2.02~\cite{Harland-Lang:2015cta} and the MMHT2014
PDFs~\cite{Harland-Lang:2014zoa}, where $R_g^{^{\rm Pb/p}}\approx 0.7$
is the Pb gluon shadowing according to the EPS09 nuclear PDF modifications~\cite{Eskola:2009uj}. Typical CEP photon pairs peak at 
$\pT^{\gaga}\approx 0.5$~GeV/$c$ and have moderate tails in their azimuthal acoplanarity $\Delta\phi_{\gaga}$,
whereas photon-fusion systems are produced almost at rest. By imposing very tight cuts in the pair momentum,
$\pT^{\gaga}\lesssim 0.1$~GeV/$c$ and acoplanarity
$\Delta\phi_{\gaga}-\pi\lesssim 0.04$, the CEP $\gaga$ can be
reduced  to $\sim 400$~nb. 
The very large exclusive \PbPb$\xrightarrow{\gaga}\epem$ QED cross section, 
$\sigma_{\rm \gaga\to \epem}^{\rm QED}[m_{\rm \epem}>5$~GeV$]=26$~mb according to 
\textsc{Starlight}~\cite{Nystrand:2004vn}, similarly enhanced by the $Z^4$ factor as the LbyL signal,
can be of concern if neither $e^\pm$ track is reconstructed or if both $e^\pm$ undergo hard
bremsstrahlung. Requiring both $e^\pm$ to fall within the central acceptance and be singly misidentified as
photons with probability $f_{\rm e\to\gamma}\approx 0.5\%$, results in a residual
$\gaga\to\gamma_{(e^+)}\,\gamma_{(e^-)}$ cross section of $\sim 120$~nb. After cuts, both backgrounds 
are thereby smaller than the expected visible LbyL cross section of
$\sim 500$~nb.
Figure~\ref{fig:LbyL} (right) shows the $\gaga$ invariant mass distributions for signal and CEP and QED
backgrounds (after cuts) in one \PbPb\ run at the FCC. We expect about $N_{\gaga}^{\rm excl}\approx$~16\,500
signal counts (reaching diphoton masses above 100~GeV) compared to
$\sim 13\,000$ and $\sim 4\,000$ CEP and QED
counts respectively. The overall (profile likelihood) significance of the signal in the integrated yields is
${\cal S}\approx 35$, considering $20\%$ and $50\%$ theoretical uncertainties on LbyL and CEP yields respectively
(the QED background can be easily well-measured beforehand).  

In summary, light-by-light scattering, a rare fundamental quantum-mechanical process that has escaped
experimental observation so far, can be measured at the LHC~\cite{d'Enterria:2013yra} and
FCC~\cite{d'Enterria:2016qsv} exploiting the large quasireal photon fluxes in electromagnetic interactions of
protons and ions accelerated at TeV energies. 
The increase in $\gaga\to\gaga$ yields from LHC to FCC is of $\mathcal{O}(200)$ thanks to factors
of $\times 30$ larger cross sections times luminosities, and $\times 2$ in the experimental acceptance.
The measurement of elastic $\gaga$ scattering at the LHC will be the first-ever 
observation of such fundamental quantum mechanical process in the lab. At the FCC, 
the higher-masses of the produced diphoton system may be sensitive to new-physics effects
predicted in various SM extensions.

\subsection{Fixed-target collisions using the FCC proton and lead beams}
\label{sec:HI_fixedtarget}
 
\newcommand{\glabcms}{\gamma^{\rm lab}_{\rm c.m.s.}}
\newcommand{\blabcms}{\beta^{\rm lab}_{\rm c.m.s.}}
\newcommand{\dylabcms}{\Delta y^{\rm lab}_{\rm c.m.s.}}

Fixed-target experiments have brought decisive contributions to particle and nuclear physics. They 
have led to particle discoveries such as those of $\Omega^-$, $J/\psi$, $\Upsilon$, etc., as well as
evidence for the novel dynamics of quarks and gluons in heavy-ion collisions. 
In accessing the high Feynman $x_F$ domain\footnote{$x_F$ is defined
  as the difference on the Bjorken $x$ values for the two partons that
enter the hard scattering process, $x_F=x_1-x_2$.} and in offering a number of options 
for polarised and unpolarised proton and nuclear targets, they have also led to the observation 
of surprising QCD phenomena: the 
breakdown of the Lam-Tung relation,  colour transparency, higher-twist effects at high 
$x_F$, anomalously large single- and double-spin correlations, 
and the breakdown of factorisation in $J/\psi$ hadroproduction at high 
$x_F$ in proton-nucleus collisions (see~\cite{Brodsky:2012vg} and references therein).
The fixed-target mode indeed offers critical advantages that remain still nowadays difficult to challenge by 
collider experiments, and hence their complementarity.  Let us emphasise four of key assets,
among others:  accessing the high Feynman $x_F$ domain, achieving high 
luminosities with dense targets, varying the atomic mass of the target almost at will, 
and polarising the target. It is therefore legitimate to investigate the physics opportunities which are offered by the 
ultra-high energy proton and lead beams of the FCC-hh impinging on a fixed target\footnote{A  list 
of physics opportunities offered by the use of  the multi-TeV proton and lead LHC beams on a 
fixed target can be found in Ref.~\cite{Brodsky:2012vg}. We refer to Ref.~\cite{Lansberg:2012kf} for the 
 specific case of quarkonium studies, for spin physics to 
Refs.~\cite{Lorce:2012rn,Rakotozafindrabe:2013au,Lansberg:2014myg,Massacrier:2015nsm,Lansberg:2016urh} and for 
heavy-ion physics with lead beam to Refs.~\cite{Rakotozafindrabe:2012ei,Lansberg:2012kf,Lansberg:2013wpx}. 
}. We will refer in the following to such a set-up as to AFTER@FCC.

\subsubsection{Colliding ultra-high-energy protons and ions on fixed targets}

\subsubsubsection{Fixed-target kinematics}

Contrary to the case of colliding beams of equal energies for which 
the c.m.s. frame obviously corresponds to the laboratory frame,
one has to account for the boost ($\glabcms$)   and the rapidity shift ($\dylabcms$) 
between the c.m.s. frame of the fixed-target collision and the laboratory frame. 
These  are respectively
$\glabcms=\sqrt{s}/(2m_{\rm p})$ and $\dylabcms =\ln(\glabcms+\sqrt{(\glabcms)^2-1}) $. 
 Consequently, the available c.m.s. energy is much lower than in the collider mode, on the order
of $200$--$300$~GeV in the FCC case depending on the beam energy (see Table~\ref{table:BeamParameters}). 

The region of central c.m.s. rapidities, $y_{\rm c.m.s.}\simeq 0$, is thus highly 
boosted at an angle with respect to the beam axis of about one degree 
in the laboratory frame. The entire backward hemisphere, $y_{\rm c.m.s.}<0$, is thus 
easily accessible with standard experimental techniques. With the FCC, the rapidity 
shift is on the order of 5--6, whereas it is 4.8 
for the 7 TeV LHC beams (see also Table~\ref{table:BeamParameters}). A detector covering $2<\eta_{\rm lab} <6$ would thus cover 
nearly half of the physical phase space of the fixed-target mode. In terms of kinematics, 
the advantage of such a mode running at ultra high beam energies is that particles nearly at the end of 
the phase space at backward c.m.s. rapidities, which would not be detectable in the collider mode, 
are at large angle and can be detected since they do not fly in the beam pipe.

\begin{table*}[!ht]
\caption{Comparison between the beam and kinematics parameters for the proposed FCC with 
proton and lead beams~\cite{Schaumann:2015fsa} and that of the LHC, including the fixed-target 
energies, the boost ($\glabcms$) and the rapidity shift ($\dylabcms$) between 
the c.m.s. of the fixed-target collision and the laboratory frame.}
              \label{table:BeamParameters}
\begin{center}\renewcommand{\arraystretch}{1.3}\footnotesize
    \begin{tabularx}{16cm}{ p{5.5cm} | c | c |c |p{3.25cm} }
     &  p@LHC & Pb@LHC & p@FCC & Pb@FCC  \\
&&&(25 and 5 ns spacing)& \\ \hline
      Beam Energy ($E_b=\frac{Z}{A}E_{\rm p}$) [TeV] & 7 & 2.76 & 50 & 19.71 \\ 

     Number of bunches stored  & 2808 & 592  & 10600/53000  & 2072 \\ 
     Number of particles ($N_{\rm p}$) per bunch [$10^{8}$] & 1150 & 0.7 &  1000/200 & 2.0 \\ 
Circumference [km] & 26.659 & 26.659 & 100 & 100 \\  
Revolution frequency [kHz] & 11.25 & 11.25 & 3 & 3 \\
Particle flux [s$^{-1}$] & $3.6\times 10^{18}$ & $4.6\times 10^{14}$ &$3.2\times 10^{18}$ & $1.2\times 10^{15}$\\\hline
     Nucleon--Nucleon c.m.s. energy ($\sqrt{s_{\rm NN}}=\sqrt{2E_b m_N}$) [GeV] & 114.6 & 72.0 & 306.6 & 192.5 \\ 
Lorentz factor ($\glabcms=\sqrt{\frac{E_b}{2m_{\rm p}}}$) &  61.0  & 38.3   & 163.1  & 102.4   \\
$\dylabcms =\ln(\glabcms+\sqrt{(\glabcms)^2-1})$ &  4.80  & 4.33   & 5.79  & 5.32   \\
    \end{tabularx}
\end{center}
\end{table*}

\subsubsubsection{Beam extraction vs. internal gaseous target}

There are two promising techniques to achieve the fixed-target mode 
with ultra-high energy beams: the slow extraction by a bent crystal or an internal gas target. Both of them are currently being investigated on the LHC beams.

For what concerns the slow extraction by a bent crystal, a first computation  of the approximate 
deflection efficiency as a function of crystal length, along the lines of~\cite{Baurichter:2000wk} for a single pass, was presented for the FCC in~\cite{Lansberg:2015pra}.
It was found that the efficiency (excluding surface transmission) in 
Si (110) is 84\% for a 50~TeV beam  at a deflection 
angle of 0.5 mrad, as approximately required for the passage of a septum blade 
downstream which is required for further extraction.
This  efficiency corresponds to an optimal crystal length of $L/L_D=0.085$, 
thus to a length of $1.6$~m. It was, however, stressed that such a length
is probably overestimated since it was evaluated for a single pass only. 
Results of the order $20$--$30$~cm are certainly not unrealistic.

The beam extraction may however not be the only way to perform fixed-target experiments at colliders. 
The injection of a small amount of gas into the detector region of a running machine
is sufficient to  increase the probability of beam-gas interactions 
such as to reach large luminosities -- yet at essentially zero cost in terms of 
the beam lifetime. In fact, the LHCb experiment running at the LHC has implemented this idea. 
The initial objective of their system, referred to as SMOG for  System for Measuring 
Overlap with Gas (SMOG), was to monitor the beam 
luminosity~\cite{FerroLuzzi:2005em}. 
SMOG has so far proven to be functioning well,  while not disturbing the primary beam\footnote{
LHCb took data from proton-neon and lead-neon collisions over short periods during beam tests in 2012 and 2013, 
as well as in 2015 from collisions of proton-neon (12 hours), proton-helium (8 hours) and proton-argon (3 days) 
at 110.4 GeV and lead-argon (1 week) and proton-argon (a few hours) at 68.6 GeV.}.

 In order to derive the luminosities that can be expected in the 
fixed-target mode with the beams of a future collider like the FCC, we 
need to account for some specific parameters (see Table~\ref{table:BeamParameters}), 
some of them like the beam loss are not yet well known. Indeed, in the case of
the extraction with the LHC beam with a bent crystal positioned in the beam halo, an extraction flux corresponding to half
of the beam loss has been assumed~\cite{Uggerhoj:2005xz,Brodsky:2012vg}. Lacking such numbers in the FCC case, 
we will take as a working hypothesis that such a parasitic mode is reached with an extracted flux corresponding to 
$5\%$ of the protons stored in the FCC per fill.

Table~\ref{table:luminosities-FTE} summarises the estimated
luminosities for the two modes (slow extraction and internal target), 
for the proton and Pb beams at the FCC. Two options for the internal target case are considered: one is similar to
the LHCb SMOG system with a slightly higher pressure; the other is inspired by the HERMES target system used
at DESY~\cite{Airapetian:2004yf}, which offers the possibility to
have polarised hydrogen, deuterium or helium targets. 
The injection of heavier inert gases is also possible. In this case,
the limit on the target density is determied by the maximum acceptable impact on
the collider-mode luminisity and by the detector readout capabilities.
Yearly luminosities, as for AFTER@LHC, for the FCC proton beam, range from $1$~fb$^{-1}$yr$^{-1}$  up to $60$~fb$^{-1}$yr$^{-1}$  with light targets and, for the Pb beam,  from $40$~nb$^{-1}$yr$^{-1}$  up to $2$~pb$^{-1}$yr$^{-1}$  with a long hydrogen target.

\begin{table*}[!ht]

             \caption{For the FCC extracted beam, the extracted flux is calculated by assuming that 5\% of the beam is used 
per fill of 10 hours and the luminosities are calculated for the case of targets that are 1 cm thick for Be and W and 1m long for liquid hydrogen.
The values for the internal gas-target (\`a la SMOG)  are calculated using the same parameters as in~\cite{Massacrier:2015nsm} 
for an ideal gas at a pressure of $10^{-6}$ mbar in a zone of 100 cm. Those for an internal gas storage-cell target (\`a la HERMES) 
are computed for a target areal density of $2.5\cdot  10^{14}$~cm$^{-2}$ (H$_2$), $3.2\cdot  10^{14}$~cm$^{-2}$ (D$_2$) and $3.8\cdot  10^{13}$~cm$^{-2}$ (Xe)~\cite{Barschel:2015mka}. 
A year is assumed to be $10^7$ s for p and $10^6$ s for Pb for both FCC and LHC cases.}
              \label{table:luminosities-FTE}
\begin{center}\renewcommand{\arraystretch}{1.2}\footnotesize
    \begin{tabularx}{16cm}{  p{4cm} | c  | c | c |c }
     & p@LHC & Pb@LHC & p@FCC & Pb@FCC  \\ \hline
{\it Extracted beam on an external target} &liq. H / Be / W &liq. H / Be / W &liq. H / Be / W &liq. H / Be / W \\

     Extracted flux [s$^{-1}$] & $5\cdot10^{8}$ & $2\cdot10^{5}$ & $1.5\cdot  10^9$ & $5.8\cdot  10^5$  \\ 
     $\mathcal{L} (\mu $b$^{-1}$s$^{-1})$ & 2000~/~62~/~31 & 0.8~/~0.025~/~0.013 &  6000~/~190~/~93 & 2.32~/~0.072~/~0.036 \\ 
     $\int dt \mathcal{L}($pb$^{-1}$yr$^{-1})$ & 20000 ~/~ 620 ~/~310 & 0.8~/~0.025~/~0.013 &  60000~/~1900~/~930 &  2.32~/~0.072~/~0.036 \\ 
   \hline
{\it   Internal gas target} (SMOG type)  &ideal gas &ideal gas&ideal gas&ideal gas\\
$\mathcal{L}(\mu $b$^{-1}s^{-1})$& 10 &0.001&8.9& $3.3\cdot  10^{-3}$\\
     $\int dt \mathcal{L}($pb$^{-1}$yr$^{-1})$ & 100 & 0.001  & 89 & $3.3\cdot  10^{-3}$ \\
{\it   Internal gas storage-cell target} ~~~~~~(HERMES type) &H$_2$~/~D$_2$~/~Xe&H$_2$~/~D$_2$~/~Xe&H$_2$~/~D$_2$~/~Xe&H$_2$~/~D$_2$~/~Xe\\
$\mathcal{L}(\mu $b$^{-1}s^{-1})$& 900~/~1200~/~140 &0.12~/~0.15~/~0.02&800~/~1100~/~120&0.3~/~0.4~/~0.05\\
     $\int dt \mathcal{L}($pb$^{-1}$yr$^{-1})$ & 9000 ~/~ 12000~/~1400 & 0.12~/~0.15~/~0.02  &8000~/~11000~/~1200  & 0.3~/~0.4~/~0.05 \\
        \end{tabularx}
\end{center}
\end{table*}

\subsubsection{Physics opportunities}

It is of course impossible to make an exhaustive list of the original measurements 
which can be carried out with AFTER@FCC.  We will limit our discussion here to some highlights. The list below 
would certainly evolve with time with the future RHIC and LHC data. The c.m.s. energy  range (per nucleon-nucleon collisions) to be covered by such a setup 
would obviously not be new since it is that of RHIC studies\footnote{Not considering the so-called beam-energy-scan 
studies with limited luminosities.} in pp collision from 200 to 500~GeV and d--A and A--A collisions at 200 GeV. 
Yet, the key asset of the fixed-target mode, beside the much larger luminosities for the p--A systems than at RHIC, is 
the extensive access towards very backward rapidities. This is crucial since it allows to study
nuclear effects in a wide rapidity range, thus to scan very different boosts between the probe and the nuclear matter, 
be it hot (A--A collisions) or cold (p--A collisions). This may happen to be essential for instance to disentangle the different 
processes involved in the quarkonium production in heavy-ion collisions. An extensive rapidity coverage down to the target-ion 
rapidity can also help to study the origin of azimuthal asymmetries, 
like the elliptic flow. Given the similarities with a setup like AFTER@LHC, we also guide the readers to 
Refs.~\cite{Brodsky:2012vg,Lansberg:2012kf,Lorce:2012rn,Rakotozafindrabe:2013au,Lansberg:2014myg,Massacrier:2015nsm,Lansberg:2016urh,Rakotozafindrabe:2012ei,Lansberg:2012kf,Lansberg:2013wpx}. It has to be noted that measurements in the backward c.m.s. region may require to look at probes with transverse momenta
 down to a few GeV. This is not an issue since most of the particles released in the collisions fly forward; the 
multiplicities and the detector occupancies in this region are expected to be easily tractable. It however remains to be studied
if this can be done with a detector also running in the collider mode with very high $\pt$ thresholds.

As it was discussed in~\cite{Lansberg:2015pra}, RHIC luminosities in pp collisions at 200 GeV are limited and
could not allow for the study of vector boson production close to threshold which can help us probe
the large $x$ content in the proton and nucleus, 0.7 and above. These measurements are potentially sensitive to 
threshold effects~\cite{Mukherjee:2006uu} which can also be relevant for the production of heavy BSM particles at colliders.
The production of $W$ and $Z$ boson near thereshold could also enable
the study of  decay modes that cannot be analysed at higher
c.m.s. energies. 
Higgs boson production near threshold in p~A collisions is very challenging since gluons with enough energy
to produce the Higgs boson are suppressed at high $x$, and quark-induced reactions via vector boson
fusion are disfavoured compared to gluon-gluon production.

With a longitudinally polarised target, the study of vector boson production opens the possibilities
to study (anti)quark helicity distributions in the proton at very large $x$. With deuterium and helium targets, measurements 
can also be carried out on the neutron. Using a transversely polarised target allows one
to access transverse-momentum dependent distributions (TMDs) which are connected to the orbital angular momentum carried
by the partons. See~\cite{Lansberg:2016urh} for a recent discussion for AFTER@LHC. The increase in energy with AFTER@FCC,
which is nearly three times higher, 
will allow one for studies of systems with large scales where the applicability of the TMD factorisation will probably be 
even safer with more event counts.

In summary, a setup as AFTER@FCC could be considered as a facility by itself opening a new realm of investigations
that would complement measurements carried out at RHIC and at lower energy facilities.



\clearpage
\section*{Acknowledgements}

J.L.\,Albacete is supported by the
FP7-PEOPLE- 2013-CIG Grant of the European Commission, reference QCDense/631558, and by Ramon y Cajal and MINECO projects reference RYC-2011-09010 and FPA2013-47836. 
N.\,Armesto, H.\,Paukkunen and C.A.\,Salgado acknowledge support from European Research Council grant HotLHC ERC-2011- StG-279579; Ministerio de Ciencia e Innovacion of Spain under project FPA2014-58293-C2-1-P; Xunta de Galicia (Conseller\'a de Educaci\'on) -- the group is part of the Strategic Unit AGRUP2015/11. 
N.\,Armesto is also supported by
the People Programme (Marie Curie Actions) of the European Union's Seventh Framework Programme FP7/2007-2013/ under REA grant agreement no.\,318921.
The work of C.-M.\,Ko is supported by the US Department of Energy under Contract No. DE-SC0015266 and the Welch Foundation under Grant No. A-1358.
The work of J.G.\,Milhano and L.\,Apolin\'ario is supported by Funda\c c\~ao para a Ci\^encia e a Tecnologia of Portugalunder project CERN/FIS-NUC/0049/2015 and contracts ``Investigador FCT - Development Grant'' (JGM) and SFRH/BD/64543/2009 (LA).
C.\,Ratti is supported by the National Science Foundation under grant no.\,PHY-1513864.
The work of A.H.\,Rezaeian is supported in part by Fondecyt grant 1150135, Ecos-Conicyt C14E01, Anillo ACT1406 and Basal FB0821.  
M.\,van\,Leeuwen acknowledges support from the  Stichting voor Fundamenteel Onderzoek der Materie (FOM) and the Nederlandse Organisatie voor Wetenschappelijk Onderzoek (NWO), Netherlands.

\newpage
\bibliographystyle{report}
\bibliography{01-introduction/intro,02-machine/machine,04-qgphard/heavyflavour,04-qgphard/top,04-qgphard/hardxsections,04-qgphard/quenching,05-smallx/smallxhadro,05-smallx/upc,03-qgpsoft/soft,06-otherhep/fixedtarget,06-otherhep/gammagamma}

\end{document}